\documentclass[longauth]{aa}  

\usepackage{graphicx}
\usepackage{txfonts}
\usepackage[colorlinks=true,citecolor=blue]{hyperref}

\newcommand{\eg}{e.g.,\ }

\newcommand{\Msun}{$M_{\odot}$}

\newcommand{\kms}{km~s$^{-1}$}

\newcommand{\OI}{O~{\sc i}}

\newcommand{\CII}{C~{\sc ii}}

\newcommand{\alII}{Al~{\sc ii}}

\newcommand{\MgII}{Mg~{\sc ii}}
\newcommand{\MgI}{Mg~{\sc i}}

\newcommand{\SiII}{Si~{\sc ii}}
\newcommand{\SiIII}{Si~{\sc iii}}

\newcommand{\SII}{S~{\sc ii}}
\newcommand{\CaII}{Ca~{\sc ii}}
\newcommand{\TiII}{Ti~{\sc ii}}

\newcommand{\FeII}{Fe~{\sc ii}}

\newcommand{\CoI}{Co~{\sc i}}
\newcommand{\CoII}{Co~{\sc ii}}

\newcommand{\Nifs}{$^{56}$Ni}

\newcommand{\vph}{v_{\rm ph}}
\newcommand{\Dm}{\Delta{\rm m_{15}(B)}}

\newcommand{\ab}{$\sim$}

\begin{document} 

\title{On the Ca-strong 1991bg-like type Ia supernova 2016hnk: \\ evidence for a Chandrasekhar-mass explosion}
\titlerunning{\mbox{SN~2016hnk}}

\author{Llu\'is Galbany\inst{1,2}\fnmsep\thanks{E-mail: llgalbany@pitt.edu}
\mbox{Chris Ashall\inst{3}},
\mbox{Peter H\"oflich\inst{3}},
\mbox{Santiago Gonz\'alez-Gait\'an\inst{4}},
\mbox{Stefan Taubenberger\inst{5,6}},
\mbox{Maximilian Stritzinger\inst{7}},
\mbox{Eric Y. Hsiao\inst{3}},
\mbox{Paolo Mazzali\inst{8,6}},
\mbox{Eddie Baron\inst{9}},
\mbox{St\'ephane Blondin\inst{10,11}},
\mbox{Subhash Bose\inst{12}}
\mbox{Mattia Bulla\inst{13}},
\mbox{Jamison F. Burke\inst{14,15}},
\mbox{Christopher R. Burns\inst{16}},
\mbox{R\'egis Cartier\inst{17}},
\mbox{Ping Chen\inst{12,18}} 
\mbox{Massimo Della Valle\inst{19,5}},
\mbox{Tiara R. Diamond\inst{20}},
\mbox{Claudia P. Guti\'errez\inst{21}},
\mbox{Jussi Harmanen\inst{22}},
\mbox{Daichi Hiramatsu\inst{14,15}},
\mbox{T.~W.-S.~Holoien\inst{16}},
\mbox{Griffin Hosseinzadeh\inst{23}},
\mbox{D. Andrew Howell\inst{14,15}},
\mbox{Yiwen Huang \inst{24}},
\mbox{Cosimo Inserra\inst{25}},
\mbox{Thomas de Jaeger\inst{26}},
\mbox{Saurabh W. Jha\inst{27}},
\mbox{Tuomas Kangas\inst{28}},
\mbox{Markus Kromer\inst{29,30}},
\mbox{Joseph D. Lyman\inst{31}},
\mbox{Kate Maguire\inst{32,33}},
\mbox{George Howie Marion\inst{34}},
\mbox{Dan Milisavljevic\inst{35}},
\mbox{Simon J. Prentice\inst{32}},
\mbox{Alessandro Razza\inst{36,37}},
\mbox{Thomas M Reynolds\inst{21}},
\mbox{David J. Sand\inst{38}},
\mbox{Benjamin J. Shappee\inst{39}},
\mbox{Rohit Shekhar\inst{40,13}},
\mbox{Stephen J. Smartt\inst{32}},
\mbox{Keivan G. Stassun\inst{41}},
\mbox{Mark Sullivan\inst{21}},
\mbox{Stefano Valenti\inst{42}},
\mbox{Steven Villanueva,\inst{43,44}},
\mbox{Xiaofeng Wang\inst{45}},
\mbox{J. Craig Wheeler\inst{46}},
\mbox{Qian Zhai\inst{47,48}},
\mbox{Jujia Zhang\inst{47,48}}
}
\authorrunning{Galbany, Ashall, H\"oflich, et al.}

\institute{
\begin{scriptsize}
PITT PACC, Department of Physics and Astronomy, University of Pittsburgh, Pittsburgh, PA 15260, USA.\and
Departamento de F\'isica Te\'orica y del Cosmos, Universidad de Granada, E-18071 Granada, Spain.\and
Department of Physics, Florida State University, Tallahassee, FL 32306, USA.\and
CENTRA/COSTAR, Instituto Superior T\'ecnico, Universidade de Lisboa, Av. Rovisco Pais 1, 1049-001 Lisboa, Portugal.\and
European Southern Observatory, Karl-Schwarzschild-Stra$\beta$e 2, D-85748 Garching bei M\"unchen, Germany.\and
Max-Planck-Institut f\"ur Astrophysik, Karl-Schwarzschild-Stra$\beta$e 1, D-85748 Garching bei M\"unchen, Germany.\and
Department of Physics and Astronomy, Aarhus University, Ny Munkegade 120, DK-8000 Aarhus C, Denmark.\and
Astrophysics Research Institute, Liverpool John Moores University, IC2, Liverpool Science Park, 146 Brownlow Hill, Liverpool L3 5RF, UK\and
Homer L. Dodge Department of Physics and Astronomy, University of Oklahoma, Norman, OK, USA.\and
Aix Marseille Univ, CNRS, CNES, Laboratoire d'Astrophysique de Marseille (LAM), Marseille, France.\and
Unidad Mixta Internacional Franco-Chilena de Astronom\'ia, CNRS/INSU UMI 3386, and Departamento de Astronom\'ia, Universidad de Chile, Camino El Observatorio 1515, Las Condes, Santiago, Chile.\and
Kavli Institute for Astronomy and Astrophysics, Peking University, Yi He Yuan Road 5, Hai Dian District, Beijing 100871, People's Republic of China \and
The Oskar Klein Centre, Department of Physics, Stockholm University, AlbaNova, SE-106 91 Stockholm, Sweden.\and
Las Cumbres Observatory, 6740 Cortona Dr. Suite 102, Goleta, CA, USA 93117.\and
University of California, Santa Barbara, Department of Physics, Broida Hall, Santa Barbara, CA, USA 93111.\and
The Observatories of the Carnegie Institution for Science, 813 Santa Barbara St., Pasadena, CA 91101, USA.\and
Cerro Tololo Inter-American Observatory, National Optical Astronomy Observatory, Casilla 603, La Serena, Chile.\and
Department of Astronomy, School of Physics, Peking University, Yi He Yuan Road 5, Hai Dian District, Beijing 100871, China. \and
Capodimonte Observatory, INAF-Naples , Salita Moiariello 16, 80131-Naples, Italy.\and
Astrophysics Science Division, NASA Goddard Space Flight Center, Greenbelt, MD 20771, USA.\and
School of Physics and Astronomy, University of Southampton, Southampton, SO17 1BJ, UK.\and
Tuorla Observatory, Department of Physics and Astronomy, FI-20014 University of Turku, Finland.\and
Harvard-Smithsonian Center for Astrohysics, 60 Garden Street, Cambridge, MA 02138.\and
University of California, San Diego, La Jolla, CA 92093.\and
School of Physics \& Astronomy, Cardiff University, Queens Buildings, The Parade, Cardiff, CF24 3AA, UK.\and
Department of Astronomy, University of California, Berkeley, CA 94720-3411, USA.\and
Department of Physics and Astronomy, Rutgers the State University of New Jersey, 136 Frelinghuysen Road, Piscataway, NJ 08854 USA.\and
Space Telescope Science Institute, 3700 San Martin Drive, Baltimore, MD 21218, USA.\and
Zentrum f\"ur Astronomie der Universit\"at Heidelberg, Institut f\"ur Theoretische Astrophysik, Philosophenweg 12, 69120, Heidelberg, Germany.\and
Heidelberger Institut f\"ur Theoretische Studien, Schloss-Wolfsbrunnenweg 35, 69118 Heidelberg, Germany.\and
Department of Physics, University of Warwick, Coventry, CV4 7AL, UK.\and
Astrophysics Research Centre, School of Mathematics and Physics, Queen's University Belfast, BT7 1NN, UK.\and
School of Physics, Trinity College Dublin, Dublin 2, Ireland.\and
University of Texas at Austin, 1 University Station C1400, Austin, TX, 78712-0259, USA.\and
Department of Physics and Astronomy, Purdue University, 525 Northwestern Avenue, West Lafayette, IN, 47907, USA.\and
European Southern Observatory, Alonso de C\'ordova 3107, Casilla 19, Santiago, Chile.\and
Departamento de Astronom\'ia, Universidad de Chile, Camino El Observatorio 1515, Las Condes, Santiago, Chile. \and
Steward Observatory, University of Arizona, 933 North Cherry Avenue, Tucson, AZ 85721-0065, USA.\and
Institute for Astronomy, University of Hawai'i, 2680 Woodlawn Drive, Honolulu, HI 96822, USA. \and
Department of Physics, BITS Pilani K.K. Birla Goa Campus.\and
Vanderbilt University, Department of Physics \& Astronomy, Nashville, TN 37235, USA.\and
Department of Physics, University of California, Davis, CA 95616, USA.\and
Massachusetts Institute of Technology, Cambridge, MA 02139, USA.\and
Pappalardo Fellow.\and
Physics Department/Tsinghua Center for Astrophysics, Tsinghua University, Beijing, 100084, China.\and
Department of Astronomy, University of Texas at Austin, Austin, TX 78712, USA.\and
Yunnan Observatories, Chinese Academy of Sciences, Kunming 650216, China.\and
Key Laboratory for the Structure and Evolution of Celestial Objects, Chinese Academy of Sciences, Kunming 650216, China.
\end{scriptsize}
}

\date{Received April 5, 2019; accepted ---}

\abstract
{}
{We present a comprehensive dataset of optical and near-infrared photometry and spectroscopy of type~Ia supernova (SN) 2016hnk, combined with integral field spectroscopy (IFS) of its host galaxy, \mbox{MCG~-01-06-070}, and nearby environment. Our goal with this complete dataset is to understand the nature of this peculiar object.}
{Properties of the SN local environment are characterized by means of single stellar population synthesis applied to IFS observations taken two years after the SN exploded. We perform detailed analyses of SN photometric data by studying its peculiar light and color curves. SN~2016hnk spectra are compared to other 1991bg-like SNe Ia, 2002es-like SNe Ia, and Ca-rich transients. In addition, abundance stratification modelling is used to identify the various spectral features in the early phase spectral sequence and the dataset is also compared to a modified non-LTE model previously produced for the sublumnious SN~1999by.}
{\mbox{SN~2016hnk} is consistent with being a sub-luminous (M$_{\rm B}=-16.7$ mag, s$_{\rm BV}$=0.43$\pm$0.03), highly reddened object. IFS of its host galaxy reveals both a significant amount of dust at the SN location, as well as residual star formation and a high proportion of old stellar populations in the local environment compared to other locations in the galaxy, which favours an old progenitor for \mbox{SN~2016hnk}. Inspection of a nebular spectrum obtained one year after maximum contains two narrow emission lines attributed to the forbidden [\ion{Ca}{ii}] $\lambda\lambda$7291,7324 doublet with a Doppler shift of 700 km s$^{-1}$. Based on various observational diagnostics, we argue that the progenitor of \mbox{SN~2016hnk} was likely a near Chandrasekhar-mass ($M_{\rm Ch}$) carbon-oxygen white dwarf that produced 0.108 $M_\odot$ of $^{56}$Ni. Our modeling suggests that the narrow [\ion{Ca}{ii}] features observed in the nebular spectrum are associated with $^{48}$Ca from electron capture during the explosion, which is expected to occur only in white dwarfs that explode near or at the $M_{\rm Ch}$ limit.}
{}


\maketitle


\section{Introduction}

Type Ia supernovae (SNe Ia) are among the brightest transient objects in the Universe, typically reaching peak absolute magnitudes of $\lesssim -19$ mag.
They are also very homogeneous, with peak brightness varying $\sim$1$-$2 mag (\eg \citealt{1996AJ....112.2391H}). However, through empirical calibrations, such as the relations between absolute peak magnitude and both the post-maximum brightness decline (e.g. $\Delta$m$_{15}(B)$, the parameter accounting for the magnitude decay after 15 days post-max in the $B$ band; \citealt{1993ApJ...413L.105P}) and the peak color \citep{1996ApJ...473...88R,1998A&A...331..815T}, this scatter around peak brightness is reduced to $\sim$0.1 mag.
Due to their standardizability, and because they can be discovered from the ground up to high redshifts (z $\lesssim$ 1.0), SNe Ia have been widely used in the last decades to put constraints on cosmological parameters and the Hubble constant (e.g. \citealt{2011NatCo...2E.350H,2018ApJ...869...56B}), 
they were the key to the discovery of the accelerated expansion of the Universe \citep{1998AJ....116.1009R,1999ApJ...517..565P} and remain one of the key probes of dark energy (e.g. \citealt{2014A&A...568A..22B,2018arXiv181102374D}). 

However, their success as cosmological probes is not matched by our understanding of their progenitor systems,  the details on how they explode, and the underlying physical interpretation of the known empirical relations that enable them to provide precise cosmological distances.
Further improvement requires a much better understanding of the physical model of the explosion, observational constraints on which kind of progenitors can produce SNe Ia, a better control of reddening, and a complete characterization of peculiar (non-standardizable) types.

The fact that SNe Ia are discovered in all types of galaxies, including passive galaxies with no signs of star formation, points to old stars being the most probable progenitors ($>$100s Myr; \citealt{2006MNRAS.370..773M}).
At the same time, their uniform brightness suggests that their progenitor systems form a homogeneous family. 
The consensus picture is that a SN Ia results from a degenerate (held by electron degeneracy pressure) 
carbon-oxygen (C-O) white dwarf (WD) undergoing a thermonuclear runaway \citep{1960ApJ...132..565H}, and that they originate from close binary systems.
However, the nature of the system and the explosion mechanism is still under debate.
Potential progenitor systems may either consist of two WDs, called a double degenerate (DD) system, or a single WD with a non-degenerate companion, called a single degenerate (SD) system. 
Candidate progenitor systems have been inferred for both SD and DD systems. 
Although the lack of donor stars surviving the explosion may favor DD systems \citep{2019arXiv190305115T}, other examples would favor SD progenitors, such as the potential He donor star identified for the well-observed 2002cx-like type Ia SN~2012Z \citep{2014Natur.512...54M,2015A&A...573A...2S} and the massive asymptotic-giant-branch star inferred for SN~2002ic \citep{2003Natur.424..651H}.

Within these progenitor scenarios four classes of explosion scenarios are most favored:
(1) the dynamical merging of two C-O WDs in a binary system after losing angular momentum via gravitational radiation, where the thermonuclear explosion is triggered by the heat of the merging process \citep{2017hsn..book.1237G};
(2) the explosion of a C-O WD with a mass close to M$_{\rm Ch}$ triggered by compressional heating near the WD center, caused by accretion from the donor star, which may be a red giant, a main sequence star of mass less than 7-8\Msun, a He star,
or the tidally disrupted WD from a DD system \citep{2004MNRAS.353..243P};
(3) the explosion of a sub-M$_{\rm Ch}$ C-O WD triggered by detonating a thin surface He layer on the WD (double detonations; \citealt{2010ApJ...719.1067K,2014ApJ...797...46S}); and 
(4) direct collision of two WDs in tertiary systems, where the third star induces oscillations in the eccentricity of the other two increasing gravitational-wave losses and finishing in a head-on collision
(e.g. \citealt{2011ApJ...741...82T,Mazzali18}). 
The debate about dynamical mergers, M$_{\rm Ch}$ explosions, and double-detonations is ongoing, but all channels may contribute to the SNe~Ia population.

\begin{table}\scriptsize
\centering
\caption{Properties of \mbox{SN~2016hnk} and its host galaxy.}
\label{table:sn_parameters}
\begin{tabular}{lll} 
\hline\hline
\textbf{Parameters} & \textbf{Values} & \textbf{Refs} \\
\hline
\textbf{\mbox{SN~2016hnk}:} & & \\
RA (J2000) & 02$^{\rm h}$13$^{\rm m}$16\fs63 & (1) \\
 & 33.319312$^{\circ}$ & (1) \\
DEC (J2000) & $-$07$^{\circ}$39\arcmin40\farcs80 & (1) \\
 & $-$7.661332$^{\circ}$ & (1) \\
Date of explosion (JD) & 2,457,676.54 $\pm$ 4.49 & This work\\
Date of maximum (JD) & 2,457,689.98 $\pm$ 3.27 & This work\\
Projected offset &   11.55 arcsec               &This work \\
     & 	3.71 kpc &This work \\
Deprojected offset &  14.23 arcsec &This work \\
     & 	4.57 kpc &This work \\
Milky Way extinction $E(B-V)$$_{\rm MW}$ & 0.0224 $\pm$ 0.0008 mag & (4)\\
Host galaxy extinction $E(B-V)$$_{\rm HG}$ & 0.45 $\pm$ 0.06 mag & This work \\
$\Delta$m$_{15}(B)$$_{\rm obs}$ & 1.297 $\pm$ 0.071 mag & This work \\
$\Delta$m$_{15}(B)$$_{\rm true}$ & 1.324 $\pm$ 0.096 mag & This work \\
$\Delta$m$_{15,s}$(B) & 1.803 $\pm$ 0.200 mag & This work \\
$s_{BV}$ & 0.438 $\pm$ 0.030  & This work \\
(m, M)$_{\rm B,max}^{\rm corrected}$  & (17.539,$-$16.656) & This work\\ 
\hline
\textbf{Host galaxy:} & & \\
Name & \mbox{MCG~-01-06-070}  \\
Type & (R')SB(rs)a & (3)\\
RA (J2000) & 02$^{\rm h}$13$^{\rm m}$15\fs79 & (3) \\
 & 33.315792$^{\circ}$ & (3) \\
DEC (J2000) & $-$07$^{\circ}$39\arcmin42\farcs70 & (3) \\
 & $-$7.661861$^{\circ}$ & (3) \\
Recession velocity, v & 4780 $\pm$ 45 km s$^{-1}$ & This work \\
Heiocentric redshift, z$_{\rm helio}$ & 0.01610 $\pm$ 0.00015& This work \\
Projection angle (N$\rightarrow$E) & 308.84 $\pm$ 4.09 deg & This work \\
Inclination & 60.17 $\pm$ 2.84 deg & This work\\ 
Luminosity distance & 68.35 $\pm$ 0.04 Mpc & This work \\
m$_B$ &  14.89 $\pm$ 0.02 mag & (5,6) \\
Stellar mass & 10.78$^{+0.12}_{-0.21}$ dex & This work \\
\hline
\end{tabular}
$ $\\
Notes: 
(1) ATLAS survey; 
(2) \citet{2011ApJ...737..103S}; 
(3) NASA/IPAC Extragalactic Database (\url{http://ned.ipac.caltech.edu/}); 
(4) NASA/IPAC Infrared science archive (\url{https://irsa.ipac.caltech.edu}); 
(5) SDSS DR14 \citep{2018ApJS..235...42A}; 
(6) \cite{2005AJ....130..873J}.
\end{table}

The connection of these progenitor scenarios and explosion mechanisms with SN Ia subtypes is still not well-established.
More extreme sub- or over-luminous SNe Ia may require progenitors deviating from those discussed above. 
Clarifying which progenitor configuration and explosion mechanism lead to SNe Ia with particular properties, and with what rates, is of utmost importance for a number of fields including stellar evolution and SN cosmology.
The presence of these peculiar over- and sub-luminous SNe Ia, and other heterogeneous objects that do not follow the empirical relations \citep{2017arXiv170300528T}, may challenge the picture of all SNe Ia coming from the same family of progenitors.

In particular, subluminous SN 1991bg-like objects seem to break such relations.
Photometrically, they are characterized by being fainter at peak and by having faster light curve rise and decline rates, putting them at the faint and fast extreme of the light curve width-luminosity relation ($\Delta$m$_{15}$ $\gtrsim$ 1.7 mag).
The maxima in the near-infrared (NIR) bands occur after the maximum in optical bands, but one of the main peculiarities is the lack of either a shoulder or a secondary maximum in the NIR light curves.
The latter can be explained by the recombination of \ion{Fe}{iii} to \ion{Fe}{ii} occurring earlier: the \ion{Fe}{ii} line blanketing absorbs flux in the blue that is re-emitted at longer wavelengths, causing the first and second NIR peaks to merge to form a single, slightly delayed maximum \citep{2007ApJ...656..661K,2017ApJ...846...58H}.
As a result, their colors are redder at maximum ($B-V$)$_{B_{\rm max}}\gtrsim$ 0.5 mag and their $B-V$ color curves peak earlier ($\sim$15 days from peak maximum in $B$-band) compared to normal SNe Ia ($\sim$30 days).
Spectroscopically, their main characteristics are the presence of prominent \ion{Ti}{ii} and strong O I $\lambda$7744 lines, indicating low ionization and temperature, deeper \ion{Si}{ii} $\lambda$5972 than in normal SNe Ia, and relatively low ejecta velocities with fast decline and no high-velocity features.
To first order, the luminosity of a supernova is dependent on the mass of radioactive $^{56}$Ni synthesized in the explosion, where less luminous objects produce smaller amounts of $^{56}$Ni \citep{1982ApJ...253..785A,2000ApJ...530..744P,Mazzali01}, though there is some evidence that a variation of the total ejecta mass may also be required \citep{2006A&A...460..793S,2014MNRAS.440.1498S,2017MNRAS.470..157B}. Given their low luminosity, 1991bg-like SNe synthesize rather low amounts ($\lesssim$0.1 \Msun) of radioactive $^{56}$Ni \citep{2006A&A...460..793S}.
In addition, most 1991bg-like SNe Ia are found in elliptical galaxies with rather old stellar populations and no ongoing or recent star formation, although the well-observed SN 1999by occurred in the spiral Sa-type galaxy NGC 2841 \citep{2002ApJ...568..791H,2004ApJ...613.1120G}. 

In this paper, we present a comprehensive dataset collected for the peculiar \mbox{SN~2016hnk} (see also \citealt{2017arXiv171202799S}), which exploded in the Sb-type galaxy \mbox{MCG~-01-06-070}. 
\mbox{SN~2016hnk} showed many similarities with 1991bg-like SNe~Ia, including the lack of hydrogen features, the typical deep \ion{Si}{ii} $\lambda$6355 absorption, other typical SN Ia features with absorption velocities within 6,000 $<$ v $<$ 10,000 km s$^{-1}$, and very strong calcium lines in the optical.
With an ATLAS-c absolute magnitude at discovery of $-$16.1, it was also reported to be similar to the unusual Ca-rich transient PTF09dav \citep{2016ATel.9705....1P}.
Our goal with this complete dataset is to understand the nature of this peculiar object. 
The paper is structured as follows. In Section \ref{sec:obs} we describe the discovery and the photometric and spectroscopic follow up campaign. In Section \ref{sec:host}, integral field spectroscopy of the SN host galaxy from the PISCO survey \citep{2018ApJ...855..107G} is used to study the SN local environment. Later we present the observational analysis in two sections: the photometric view in Section \ref{sec:photanal} and the spectroscopic view in Section \ref{sec:specanal}, including an observational characterization within the 1991bg-like group. We make a comparison with theoretical models from the abundance stratification technique and detailed non-LTE modified models of SN 1999by in Section \ref{sec:mod}. Finally, we discuss the origin of this object in Section \ref{sec:conc}. 
Throughout the paper we employed a standard flat $\Lambda$CDM cosmology with $\Omega_M$=0.2865 and $H_0$= 69.32 km s$^{-1}$ Mpc$^{-1}$ \citep{2013ApJS..208...19H}.



\begin{figure}
\centering
\includegraphics[width=\columnwidth]{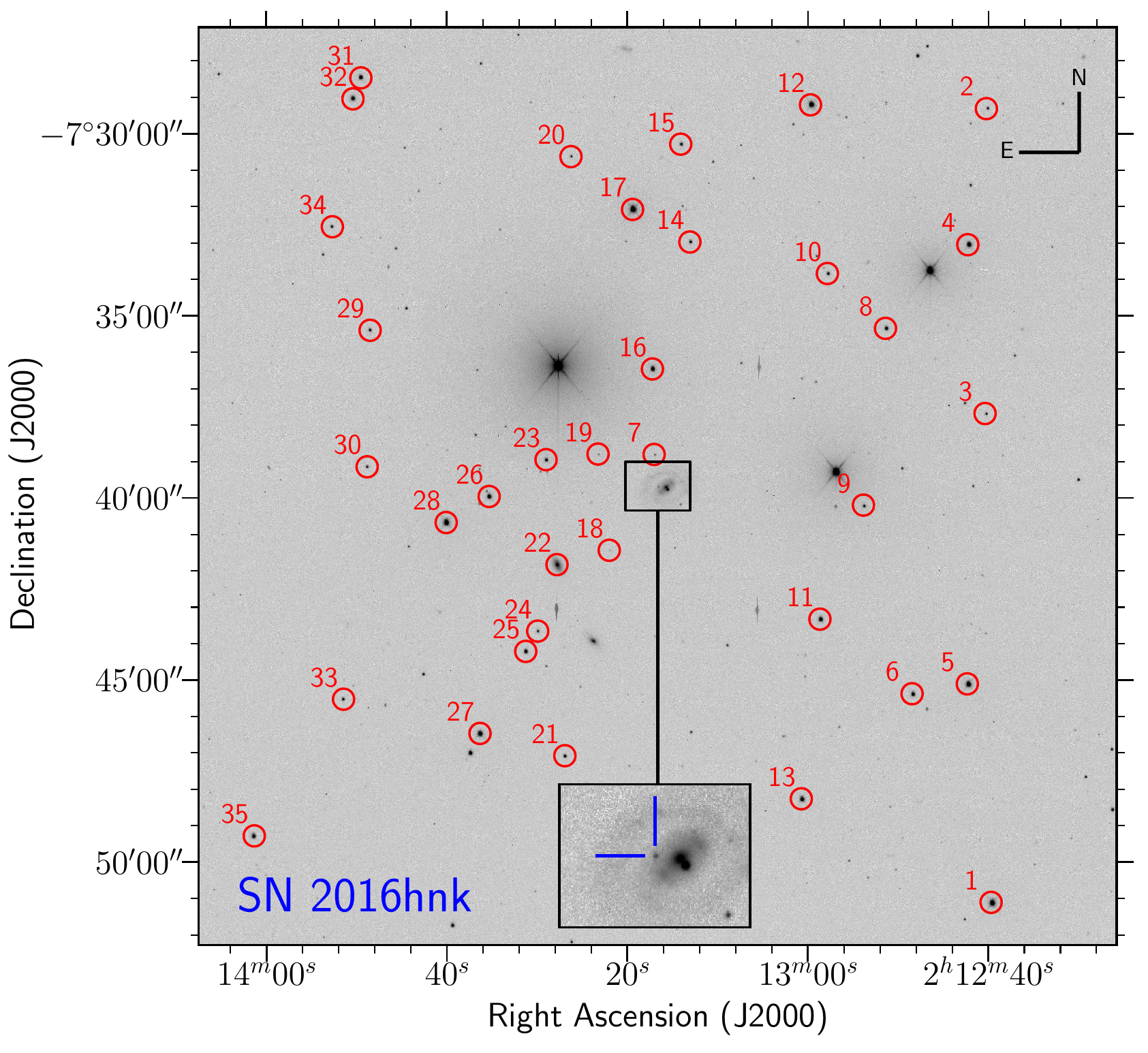}
\caption{Greyscale $B$-band image obtained with the Las Cumbres 1m telescope at Cerro Tololo Inter-American Observatory on November 3rd 2016. Orientation is North-up, East-left. \mbox{SN~2016hnk} is marked with two blue lines at the bottom inner panel, next to its host galaxy \mbox{MCG~-01-06-070}. Red circles represent all 35 standard stars used for photometric calibration in the optical and near-infrared. }
\label{fig:finder}
\end{figure}

\section{\mbox{SN~2016hnk} Observations and reduction} \label{sec:obs}

The discovery of \mbox{SN~2016hnk} (aka ATLAS16dpc) was reported by the Asteroid Terrestrial-impact Last Alert System (ATLAS; \citealt{2018PASP..130f4505T}) using the ACAM1 instrument in the ATLAS Haleakala telescope on 2016 October 27.47 (JD=2457688.97) at $\alpha$ = 02$^{\rm h}$13$^{\rm m}$16\fs63, $\delta$ = $-$07$^{\circ}$39\arcmin40\farcs80 (J2000; see Fig. \ref{fig:finder}) and with a magnitude of 17.91 in the cyan-ATLAS filter.
The object exploded in the type SB galaxy 
\mbox{MCG~-01-06-070} of the Morphological Catalogue of Galaxies \citep{1962TrSht..32....1V}, 11.5 arcsec East of the center, which at the redshift of the galaxy z=0.016 corresponds to  3.71 kpc (see inner panel of Fig. \ref{fig:finder}).
The main parameters of the supernova and host galaxy are listed in Table \ref{table:sn_parameters}.

The last reported non-detection of the object is from 2016 October 19.53 (JD=2457681.03) in the orange-ATLAS filter (>18.66 mag), $\sim$8 days before the first detection. 
Forced photometry of pre-discovery images which have been differenced with the all sky reference frame (see \citealt{2017ApJ...850..149S}), confirmed this as an actual detection, pushing further the last non-detection to the previous image of the region on 2016 October 10.56 (JD=2,457,672.06) in the orange-ATLAS filter, which provides a constraint on the explosion epoch to within 9 days between the last non-detection and first detection.
Throughout the paper, we assume that the explosion epoch is the mid point between the last non-detection and the first detection with an uncertainty of half of this time, on JD=2457676.5 $\pm$ 4.5, however, we base SN epochs on the date of peak brightness (see Section \ref{sec:photanal}).

\mbox{SN~2016hnk} was initially classified by the NOT (Nordic Optical Telescope) Un-biased Transient Survey (NUTS; \citealt{2016TNSCR.872....1C,2016ATel.9703....1C}) on 2016 Nov 01.08 UT as a type I supernova. However, they did note it resembled the spectra of a 1991bg-like SN. 
The spectrum was relatively red, with strong lines dominated by the Ca NIR triplet, \ion{O}{i} $\lambda$7774, and \ion{Si}{ii} $\lambda$6355, and with a \ion{Si}{ii} velocity of $\sim$10,000 km s$^{-1}$. 
A few hours later, on November 2nd at 20:16 UT, the Public ESO Spectroscopic Survey for Transient Objects (PESSTO; \citealt{2015A&A...579A..40S}) confirmed the redshift and provided a definite classification as a type Ia SN spectroscopically similar to SN 1991bg \citep{2016ATel.9704....1D}.
A third report a day later, on November 3rd at 04:24 UT, provided a more detailed classification \citep{2016ATel.9705....1P}.
Although the report was cautious, it pointed out similarities between \mbox{SN~2016hnk} and the peculiar Ca-rich object PTF09dav \citep{2011ApJ...732..118S,2012ApJ...755..161K}, from their peak magnitudes, similar colors, and red spectra with overall similar features.
In particular, two deep absorptions at $\sim$5350 and $\sim$5540 \AA~and another absorption at $\sim$7120 \AA~(in the rest frame) were identified as \ion{Sc}{ii} and \ion{Ti}{ii}, respectively, by \cite{2011ApJ...732..118S}.
However, in Section \ref{sec:specanal} we will show that these features are actually formed by different elements removing any resemblance between both objects, other than the extremely strong calcium features and low luminosity.
Finally, another spectrum was obtained with Hiltner MDM-OSMOS on Oct 29.29, however observations were not reported. 

Given its peculiar nature, it was unusually faint and showed spectral features that clearly distinguish it from other 1991bg-like SNe, \mbox{SN~2016hnk} was recognized as a rare and potentially interesting object. 
A follow up campaign was started immediately the day after the PESSTO Telegram with the New Technology Telescope (NTT) and the Las Cumbres Observatory network, which ran for more than two months.
All the spectra described above and those obtained during the PESSTO and Las Cumbres campaigns, together with extensive photometry from different sources, are presented in this work and described below. 

\begin{figure}
\centering
\includegraphics[trim=0cm 0cm 0cm 0cm,clip=True,width=\columnwidth]{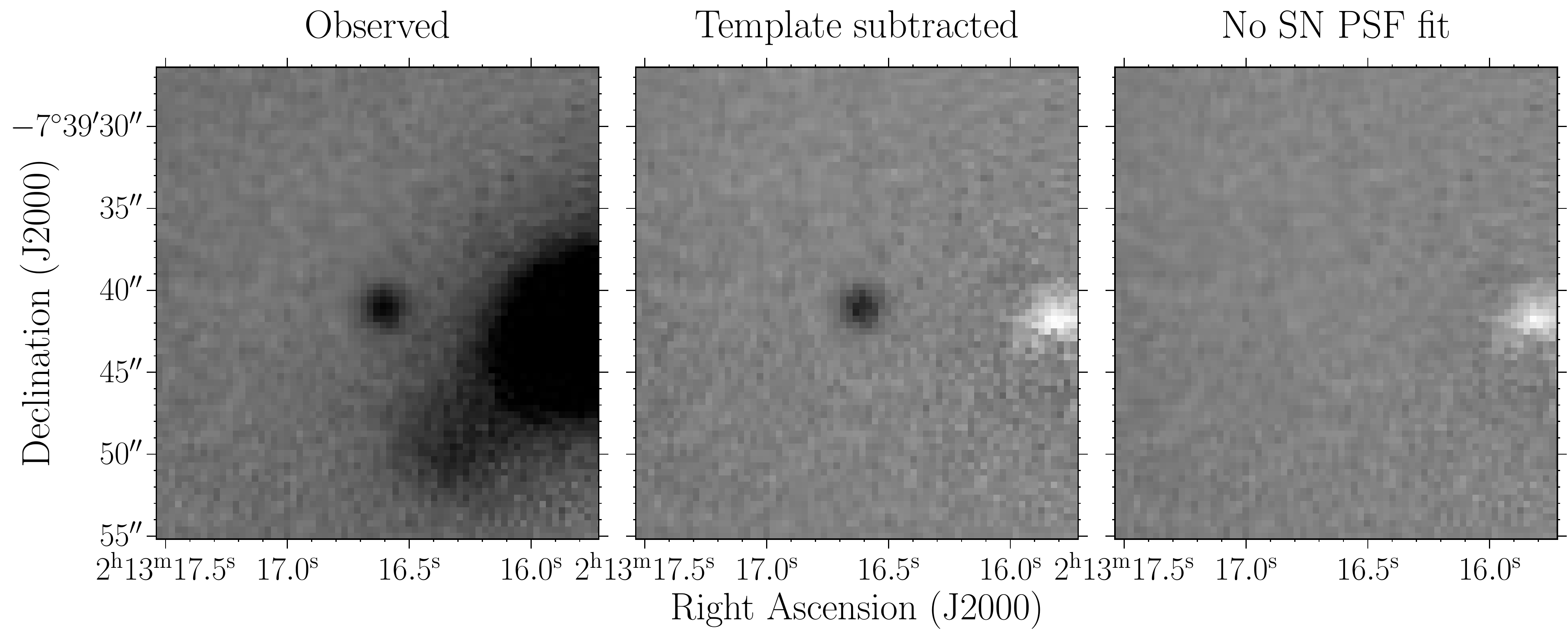}
\caption{$g$-band imaging of \mbox{SN~2016hnk} and its host galaxy, before (left) and after (middle) subtracting the host galaxy template using hotpants integration in \textsc{SNOoPY}. This example shows how the subtracted image is clean enough to perform point-spread-function (PSF) fitting, as shown in the right panel where the SN PSF has been removed. Note the small residual of the galaxy core on the middle and right panels (white blob), however it does not affect SN PSF fitting.}
\label{fig:photometry}
\end{figure}


\subsection{Photometry} \label{sec:phot}

Most of the optical photometric coverage was obtained by the Las Cumbres Observatory 1.0\,m telescope network \citep{2013PASP..125.1031B} through PESSTO allocated time and a number of dedicated SN follow up programmes. 
In addition, supplementary data were obtained from:
the 2.0 m Liverpool Telescope (LT; \citealt{2004SPIE.5489..679S}); 
the twin 0.5\,m ATLAS telescope system \citep{2018PASP..130f4505T}; 
the 0.5\,m DEdicated MONitor of EXotransits and Transients (DEMONEXT; \citealt{2018PASP..130a5001V});the NUTS survey using ALFOSC at the NOT 2.56\,m telescope; 
and pre-discovery data from the All-Sky Automated Survey for SuperNovae (ASAS-SN; \citealt{2014ApJ...788...48S}).
All data have been reduced following standard procedures with dedicated pipelines for each instrument, which consist of bias subtraction, trimming of the over scan region, and flat fielding.

Near-infrared (NIR) follow up broad-band imaging was obtained by both PESSTO and NUTS. 
PESSTO obtained NIR imaging with the 3.58\,m NTT equipped with the Son of ISAAC (SOFI; \citealt{1998Msngr..91....9M}), while NUTS obtained data with NOTCAM on the NOT 2.56\,m telescope.
From these observations we were able to obtain six epochs of $JHK$-band photometry extending from $+$5 to $+$72 d from peak.
NTT images were reduced using the PESSTO pipeline described in \cite{2015A&A...579A..40S}, while the NUTS images were reduced using a slightly modified version of the NOTCam Quicklook v2.5 reduction package\footnote{\url{http://www.not.iac.es/instruments/notcam/guide/observe.html\#reductions}}.
In both cases, the reduction procedures consisted of flat-field, distortion, cross-talk, and illumination corrections, trimming, sky subtraction and stacking of the dithered images
After proper sky-subtraction, obtained through dithering, the individual science exposures were aligned and combined to produce a definitive stacked science image, to which we then applied accurate astrometry. 

Given the location of the SN within the galaxy, we also obtained deep observations of the field in all bands and in all telescopes after more than a year from the discovery of the SN, to be used as a template for galaxy  subtraction.

We performed photometry on the SN and up to 35 reference stars located within the field covered by these different instruments using the SuperNOva PhotometrY (\textsc{SNOoPY}\footnote{\textsc{SNOoPY} is a package for SN photometry using PSF fitting and/or template subtraction developpd by E. Cappellaro. A package description can be found at \url{http://sngroup.oapd.inaf.it/snoopy.html}.}) package.
First, each individual exposure is aligned to the USNO-B catalog \citep{2003AJ....125..984M}, and a point-spread-function (PSF) with a width of the seeing is fitted to all reference stars shown in Figure \ref{fig:finder} to calculate the zero points.
We then subtracted the host galaxy templates using hotPants \citep{2015ascl.soft04004B}, and performed PSF fitting on the SN in the subtracted image (see Figure  \ref{fig:photometry} for an example).
Finally, we calibrated all SN optical photometry against the Sloan Digital Sky Survey Data Release 15 (SDSS; $ugriz$; \citealt{2019ApJS..240...23A}), the AAVSO Photometric All-Sky Survey (APASS; $BV$; \citealt{2015AAS...22533616H}), and NIR photometry to the Two Micron All Sky Survey (2MASS; $JHK$; \citealt{2006AJ....131.1163S}), accounting for corresponding airmass and color corrections in a standard manner.

ATLAS photometry in $c$ and $o$ bands, and ASASS-SN $V$-band photometry are presented in Tables \ref{tab:SN2016hnk_ph_asassn} and \ref{tab:SN2016hnk_ph_atlas}, respectively.
All other photometric observations and SN magnitudes are given in Tables \ref{tab:SN2016hnk_ph} (optical) and \ref{tab:SN2016hnk_ph_nir} (NIR) in the AB system.


\subsection{Spectroscopy} \label{sec:spec}

After the reported peculiarity of the object, we began an intensive optical spectroscopic follow-up campaign using several facilities.
Most of the data (12 epochs, including 1 NIR spectrum) were obtained by the PESSTO collaboration with the NTT using the ESO Faint Object Spectrograph and Camera (EFOSC2; \citealt{1984Msngr..38....9B}) instrument and SOFI (for the NIR spectrum). 
A description of the reduction pipeline can be found in \cite{2015A&A...579A..40S}.

These observations are complemented with data from the following sources:
the LT using the Spectrograph for the Rapid Acquisition of Transients (SPRAT), and reduced and calibrated in wavelength using the pipeline described in \cite{2014SPIE.9147E..8HP};
the NOT using ALFOSC through the NUTS survey. These data were reduced in a standard manner with ALFOSCGUI;
the Yunnan Faint Object Spectrograph and Camera (YFOSC) mounted on the Lijiang 2.4\,m telescope (LJT; \citealt{2014AJ....148....1Z});
the MDM 2.4\,m Hiltner telescope at KPNO using OSMOS \citep{2011PASP..123..187M}. The reductions follow normal procedures \citep{2016SPIE.9906E..2LV,2018PASP..130a5001V};
the 6.5\,m MMT telescope with the Blue Channel instrument \citep{Schmidt89}. Standard procedures to bias-correct, flat-field, and flux calibrate the data using IRAF were followed; 
the 8.1\,m Gemini South Telescope with the FLAMINGOS-2 spectrograph reduced following the prescriptions in \cite{2019PASP..131a4002H};
and the two discovery spectra reported in the TNS and Astronomical Telegrams (ATELs) from the 4.1m Southern Astrophysical Research Telescope (SOAR) and the 11m Southern African Large Telescope (SALT) are also included here. 

\begin{figure*}
\centering
\includegraphics[trim=0.1cm 0cm 0cm 0cm,clip=true,width=0.32\textwidth]{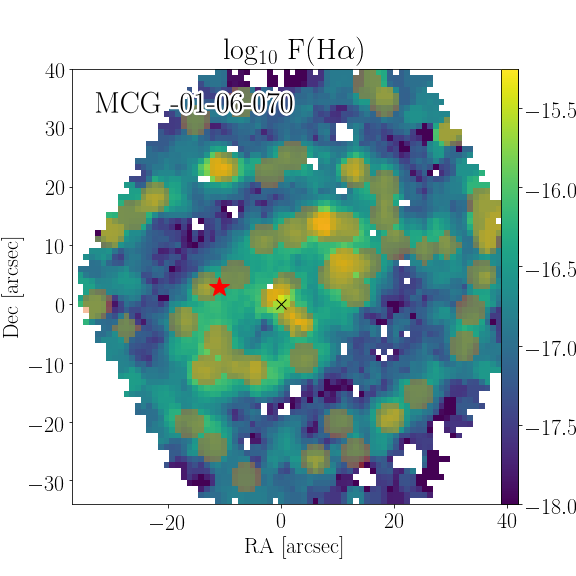}
\includegraphics[trim=0.1cm 0cm 0cm 0cm,clip=true,width=0.32\textwidth]{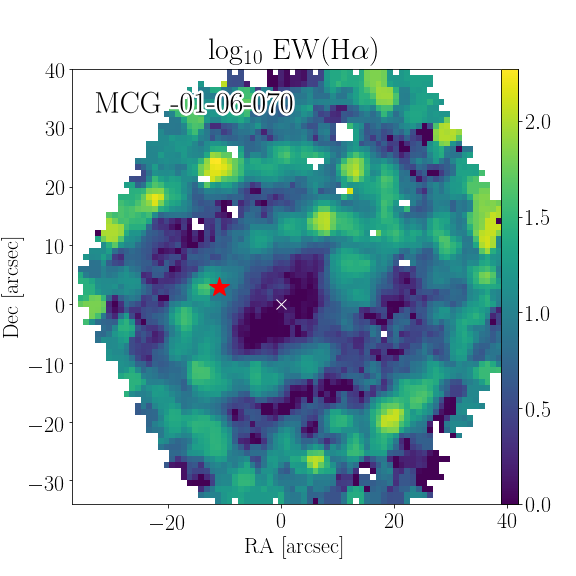}
\includegraphics[trim=0.1cm 0cm 0cm 0cm,clip=true,width=0.32\textwidth]{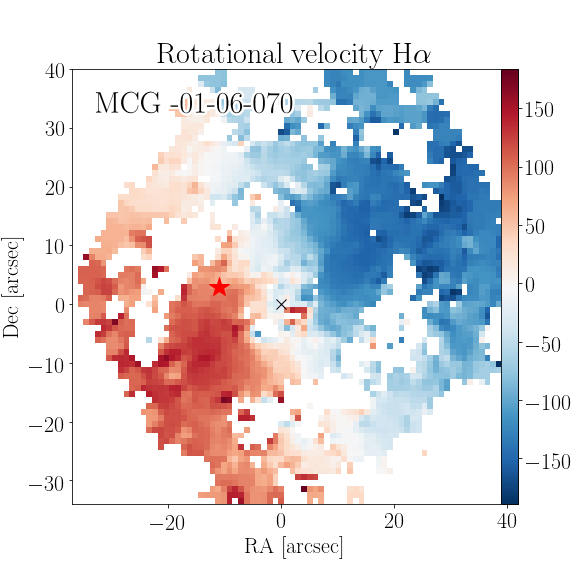}
\caption{Continuum-subtracted H$\alpha$ emission (left), H$\alpha$EW (middle), and H$\alpha$ velocity (right) maps of \mbox{MCG~-01-06-070}. {\sc Hii} regions extracted with {\sc HIIExplorer} are marked in brown polygons in the left panel. The red star and the black/white $\times$ represent respectively the locations of \mbox{SN~2016hnk} and \mbox{MCG~-01-06-070} nucleus.}
\label{fig:host}
\end{figure*}

Before performing any analysis of the spectra, they have been first scaled to match the photometry following a procedure known as {\it color-matching} \citep{2007ApJ...663.1187H}: we first fit a third order spline to both the interpolated observed photometry with gaussian processes at the epoch of each spectrum, and to the same synthetic bands from the observed spectrum; then, the observed spectrum is multiplied by a wavelength-dependent factor that matches the two splines. This procedure ensures that colors from the corrected spectra are the same as those from the photometric measurements, which we assume to be more reliable.

Finally, we obtained six spectra of the nebular phase at four different epochs from +346 to +377 d with X-shooter \citep{2011A&A...536A.105V} under the ESO programme 0100.D-0738(A).
X-shooter provides coverage of the full near-UV (UVB), optical (VIS), and NIR wavelength range (300 to 2,500 nm) in a single exposure in three different channels.
We used slits of 1$\farcs$3 and 1$\farcs$2 for the UVB and VIS arms, respectively. For the NIR arm, we opted for the 0.9"JH slit to optimize the signal-to-noise (S/N) in the $J$ and $H$ bands while blocking the $K$-band region. With this setup, a relatively homogeneous slit transmission of $\sim$85\% is achieved throughout the wavelength range.
Exposure times were fixed to 3060 sec in the UVB channel, 3000 sec in the VIS, and 3000 sec in the NIR.
The six sky-subtracted (but not telluric-removed) observations were combined in a single spectrum to enhance the signal-to-noise ratio.
The full spectral log is presented in Table \ref{tab:spectropcopy}\footnote{
All spectroscopic observations are publicly available at
\url{https://github.com/lgalbany/SN2016hnk}.}.


\section{The host galaxy and local environment of \mbox{SN~2016hnk}} \label{sec:host}\label{sec:ifs}

Integral field spectroscopy of the type SB galaxy \mbox{MCG~-01-06-070} was obtained on 2018 October 11 as part of the PMAS/Ppak Integral field Supernova hosts COmpilation (PISCO; \citealt{2018ApJ...855..107G}).

Observations were performed with the Potsdam Multi Aperture Spectograph (PMAS) in PPak mode  \citep{2005PASP..117..620R} mounted to the 3.5\,m telescope of the Centro Astron\'omico Hispano Alem\'an (CAHA) located at Calar Alto Observatory.
PPak is a bundle of 331 fibers of 2.7\arcsec~diameter ordered in a hexagonal shape with a filling factor of the field-of-view (FoV) of 55\%  \citep{2004AN....325..151V, 2006PASP..118..129K}. 
Three 900~s exposures were performed with the V500 grating of 500 lines mm$^{-1}$, which provides a spectral resolution of $\sim$6~\AA\ and covers the whole optical range from 3750 to 7300~\AA.
The second and third exposures were shifted by a few arcsec with respect to the first pointing to provide spectroscopic coverage of the full hexagonal $\sim$1.3 arcmin$^2$ FoV with 1\arcsec$\times$1\arcsec pixels, which corresponds to $\sim$4000 spectra.
Reduction was done with the pipeline used for the Calar Alto Legacy Integral Field Area (CALIFA) survey Data Release 3 (v2.2; all details can be found in \citealt{2016A&A...594A..36S}).

The analysis was performed in a similar way as that presented in \citealt{2014A&A...572A..38G,2016A&A...591A..48G}.
We used a modified version of {\sc STARLIGHT} (\citealt{2005MNRAS.358..363C,2016MNRAS.458..184L}, priv. comm.) to estimate the fractional contribution of different simple stellar populations (SSP) from the {\it Granada-Miles} base \citep{2015A&A...581A.103G} with different ages and metallicities to the stellar continuum in the spectra, and adding dust effects as a foreground screen with a \cite{1999PASP..111...63F} reddening law and $R_V$ = 3.1.
We then obtained pure gas emission spectra by subtracting the best SSP fit from the observed spectra and estimated the flux of the most prominent emission lines after correcting for dust attenuation from the Balmer decrement (assuming case B recombination, \citealt{2006agna.book.....O}; a \citealt{1999PASP..111...63F} extinction law; and $R_V$=3.1).
Line fluxes were used to estimate a number of galactic parameters, such as the ongoing star-formation rate (SFR; from the H$\alpha$ emission line), the weight of young-to-old populations (from the H$\alpha$ equivalent width; H$\alpha$EW), and oxygen abundances.

In this work we are mainly interested in studying the local properties of the \mbox{SN~2016hnk} environment and the integrated properties of its host.
For the latter, we repeated the above procedures on an integrated spectrum of the host galaxy, summing up the spectra in all spaxels, and accounting for spatial covariance in the error budget.
Wide-field IFS allows SN explosion site parameters to be compared not only to those of the overall host, but to all other stellar populations found within hosts, therefore exploiting the full capabilities of the data.
Following \cite{2016MNRAS.455.4087G} we developed a method to characterize the SN environment comparing its properties to all other nebular clusters in the galaxy.
Using our extinction-corrected H$\alpha$ map we selected star-forming {\sc Hii} regions across the galaxy with a modified version of {\sc HIIexplorer}\footnote{\url{http://www.caha.es/sanchez/HII\_explorer/}} \citep{2012A&A...546A...2S}, a package that detects clumps of higher intensity in a map by aggregating adjacent pixels.
This procedure selected 56 {\sc Hii} clumps with an average radius of $\sim$700 pc.
Once the {\sc Hii} regions were identified, the same analysis described above was performed on the extracted spectra.

\begin{figure}
\centering
\includegraphics[trim=0cm 0cm 0cm 0cm,clip=true,width=\columnwidth]{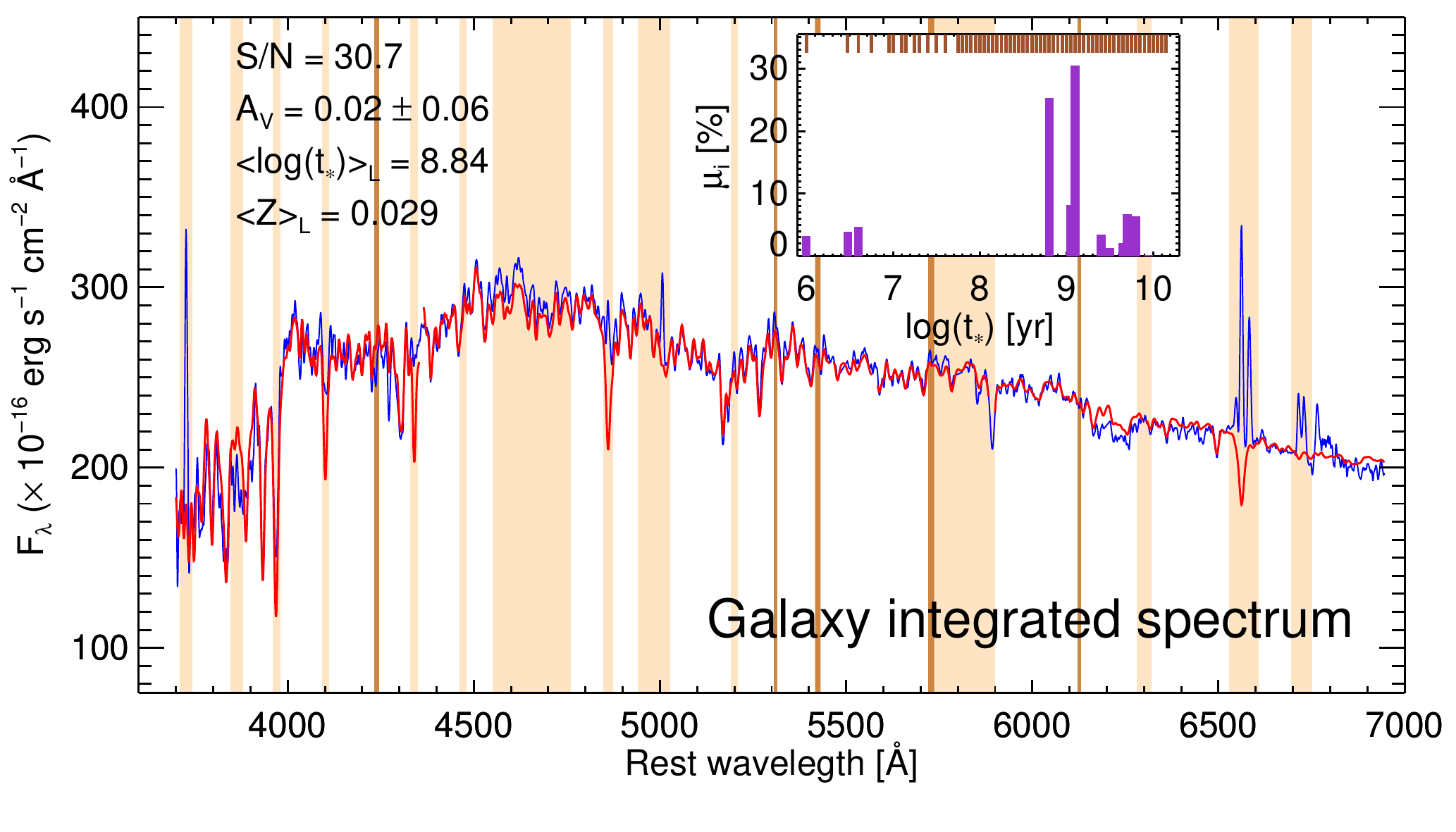}
\includegraphics[width=\columnwidth]{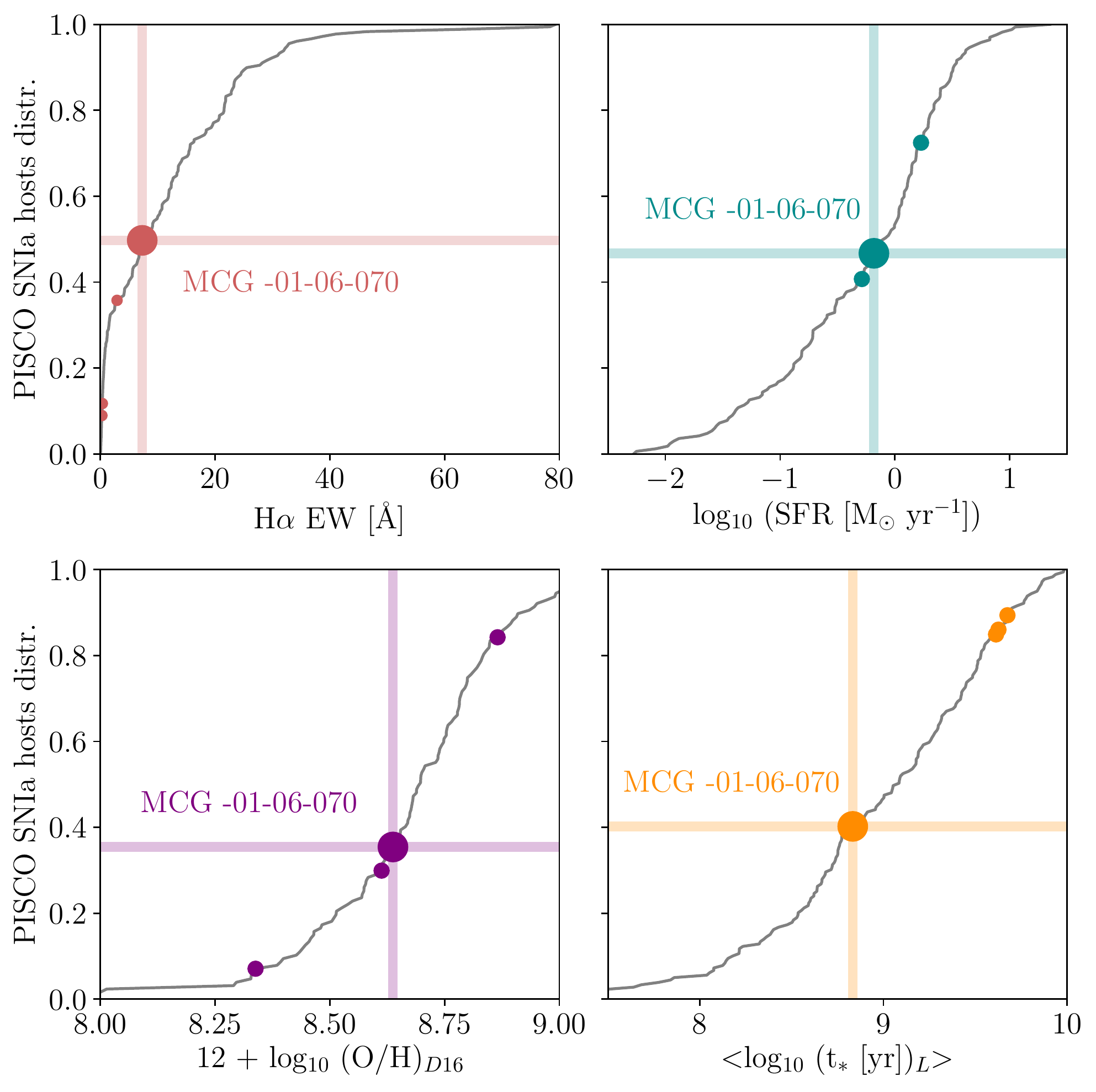}
\caption{Top: \mbox{MCG~-01-06-070} integrated spectrum (blue) and best simple stellar-population fit from STARLIGHT (red). Vertical strips (in beige) represent masked regions in the fit due to known emission lines (such as Balmer lines, oxygen, WR bumps, etc.) and regions with skylines (in brown). The inner panel shows the star-formation history of the spectra (in purple), where upper brown ticks represent the ages of the models used in the fit. Signal-to-noise ratio of the spectrum, optical extinction, average stellar age, and average metallicity are reported in the upper left corners. Lower rows: Distributions of the H$\alpha$ equivalent width, star-formation rate, oxygen abundance, and average stellar age of all 180 SN Ia host galaxies in PISCO \protect\citep{2018ApJ...855..107G}. The large colored dots represent the position of \mbox{MCG~-01-06-070}, and the three small dots represent other 1991bg-like SNIa hosts in PISCO.}
\label{fig:envglobal}
\end{figure}

\begin{figure}
\centering
\includegraphics[trim=0.0cm 0cm 0cm 0cm,clip=true,width=\columnwidth]{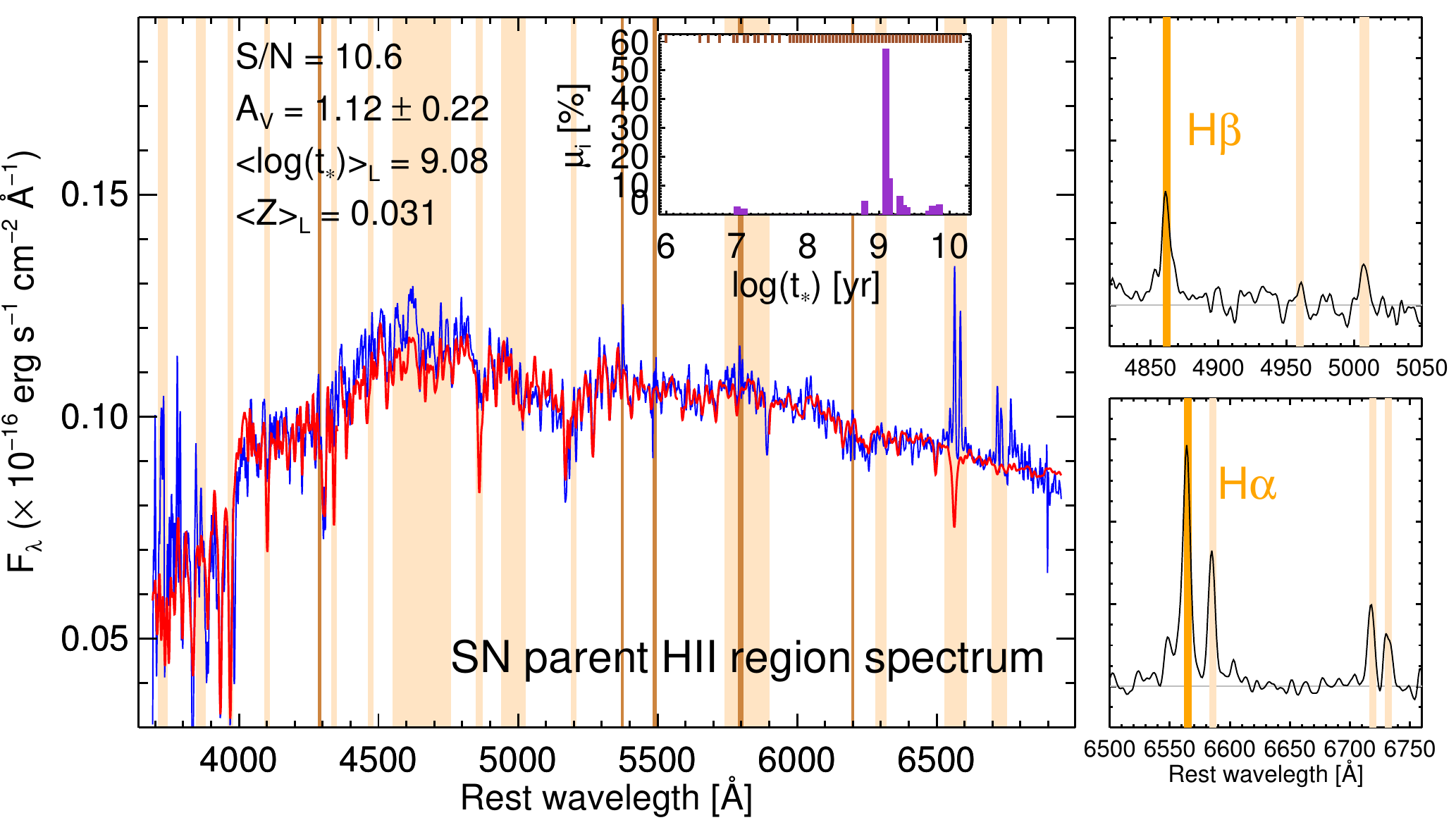}
\includegraphics[width=\columnwidth]{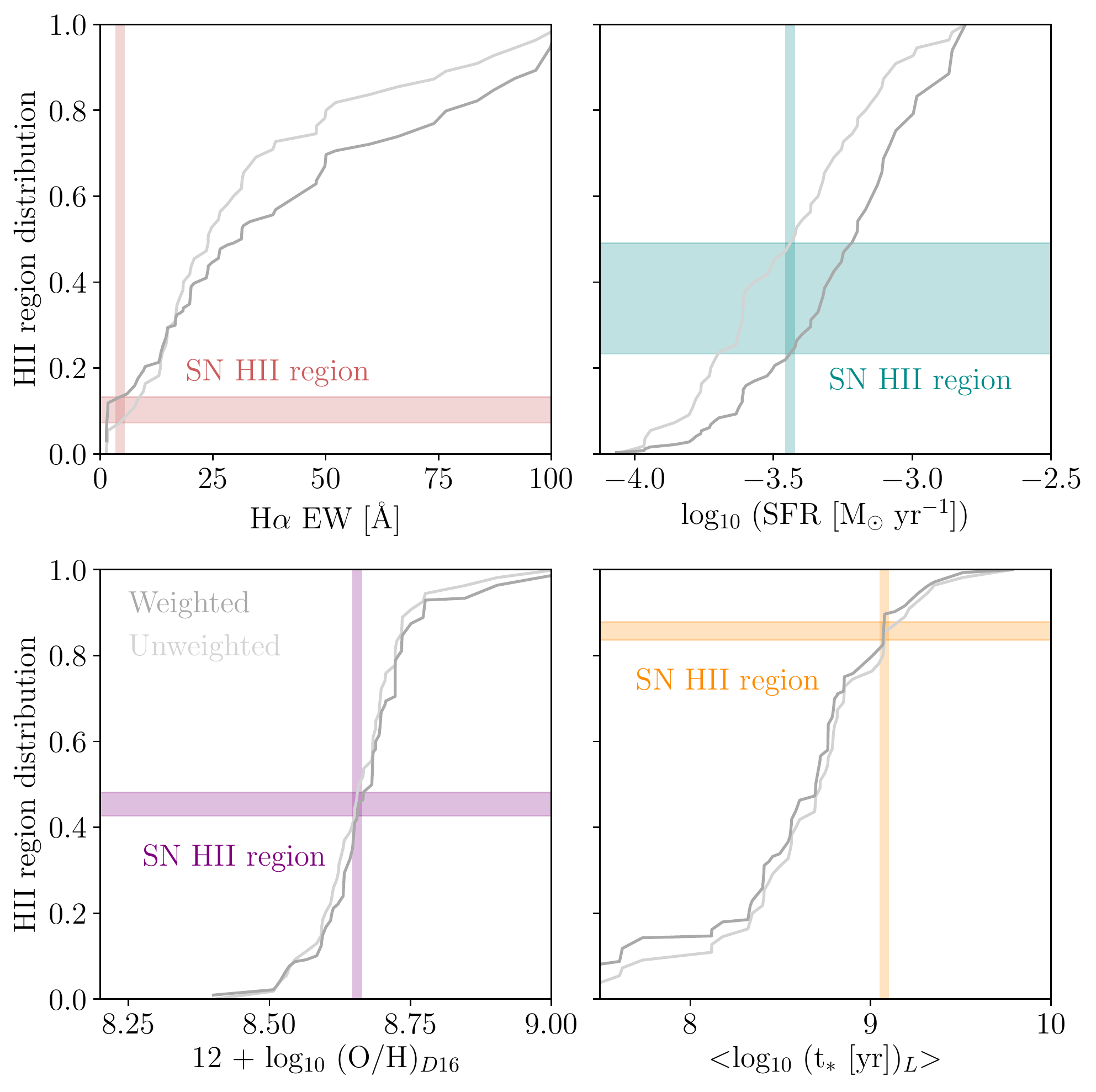}
\caption{Top: Similar to Figure \ref{fig:envglobal}, the spectrum of the \mbox{SN~2016hnk} parent {\sc Hii} region (blue) and best simple stellar-population fit from STARLIGHT (red) are shown. The right panels show the stellar subtracted spectra around the H$\alpha$ and H$\beta$ wavelengths. Lower rows: Distributions of the gas-based parameters of all 56 {\sc Hii} regions in \mbox{MCG~-01-06-070}, weighted by the contribution of the region to the total SFR of the galaxy (dark lines) or unweighted (light lines). In this case, colored horizontal bands represent the band where the \mbox{SN~2016hnk} parent {\sc Hii} region lies within.}
\label{fig:envlocal}
\end{figure}

Figure \ref{fig:host} shows two-dimensional maps of the continuum-subtracted H$\alpha$ emission with the {\sc Hii} region segmentation performed with {\sc\mbox{HIIExplorer}} on top (left), the H$\alpha$ equivalent width (middle), and the rotational velocity field (right). In all panels the red star represents the SN position.
\mbox{SN~2016hnk} occurred between two {\sc Hii} clumps, within the small disc created by the bar present in the host galaxy. 
One of these clumps has both high H$\alpha$ brightness and H$\alpha$EW, pointing to a strong component of young stellar populations and ongoing star formation.
However the SN is closer to the clump showing bright H$\alpha$ emission but lower H$\alpha$EW, which means ongoing star formation in a location dominated by older non-ionizing stellar populations, as it is seen in the star-formation history of Figure \ref{fig:envlocal}.
At the SN location the rotational velocity of the galaxy is 82$\pm$6 km s$^{-1}$, which we subtracted in all SN velocities measured in Section \ref{sec:specanal}.

The resulting integrated spectrum of \mbox{MCG~-01-06-070} is presented in Figure \ref{fig:envglobal} with the best STARLIGHT SSP fit. The recovered star-formation history is also shown in the inner plot.
The stellar mass of the galaxy is 4.84 $\times$ 10$^{10}$ \Msun, and the average stellar age of the galaxy is 690 Myr, although we were able to recover three main peaks of formation: a very old component at around 6 Gyr, an intermediate burst at around 1 Gyr, and a young component of a few Myr.
The best fit also provides an oversolar stellar metallicity of Z=0.03 and a non-significant optical stellar extinction of A$_V$=0.02 $\pm$ 0.06 mag.

The lower panels of Figure \ref{fig:envglobal}, aim to compare the properties of SN~2016hnk and the other three 1991bg-like SNIa host galaxies with those of all other 180 SN Ia host galaxies in PISCO (\citealt{2018ApJ...855..107G}; published sample plus observations up to March 2019 to be presented elsewhere).
We present distributions of the H$\alpha$EW, SFR, oxygen abundance ($12 + \log_{10}$ O/H) in the \cite{2016Ap&SS.361...61D} scale, and the average light-weighted stellar age.
For \mbox{MCG~-01-06-070} we found an H$\alpha$EW of 7.04 $\pm$ 0.18 \AA, a SFR of 0.649 $\pm$ 0.234 \Msun~ yr$^{-1}$, and an oxygen abundance of 8.64 $\pm$ 0.11 dex.
The big colored dot represent the position of \mbox{MCG~-01-06-070} parameters in such distributions, and the vertical and horizontal strips show the actual parameter value of this galaxy and the rank within PISCO. The other three 1991bg-like SN Ia hosts are also represented with smaller collored dots.
Overall, \mbox{MCG~-01-06-070} has no extreme properties and lies within the 35-50\% rank in all measured parameters within the PISCO sample.
However, it has a younger average stellar population age and, equivalently, a higher H$\alpha$EW than the other 1991bg-like SN Ia hosts, which show less evidence of ongoing/recent star formation.

\begin{figure*}
\centering
\includegraphics[trim=0cm 0.3cm 0.2cm 0.2cm,clip=true,width=0.9\textwidth]{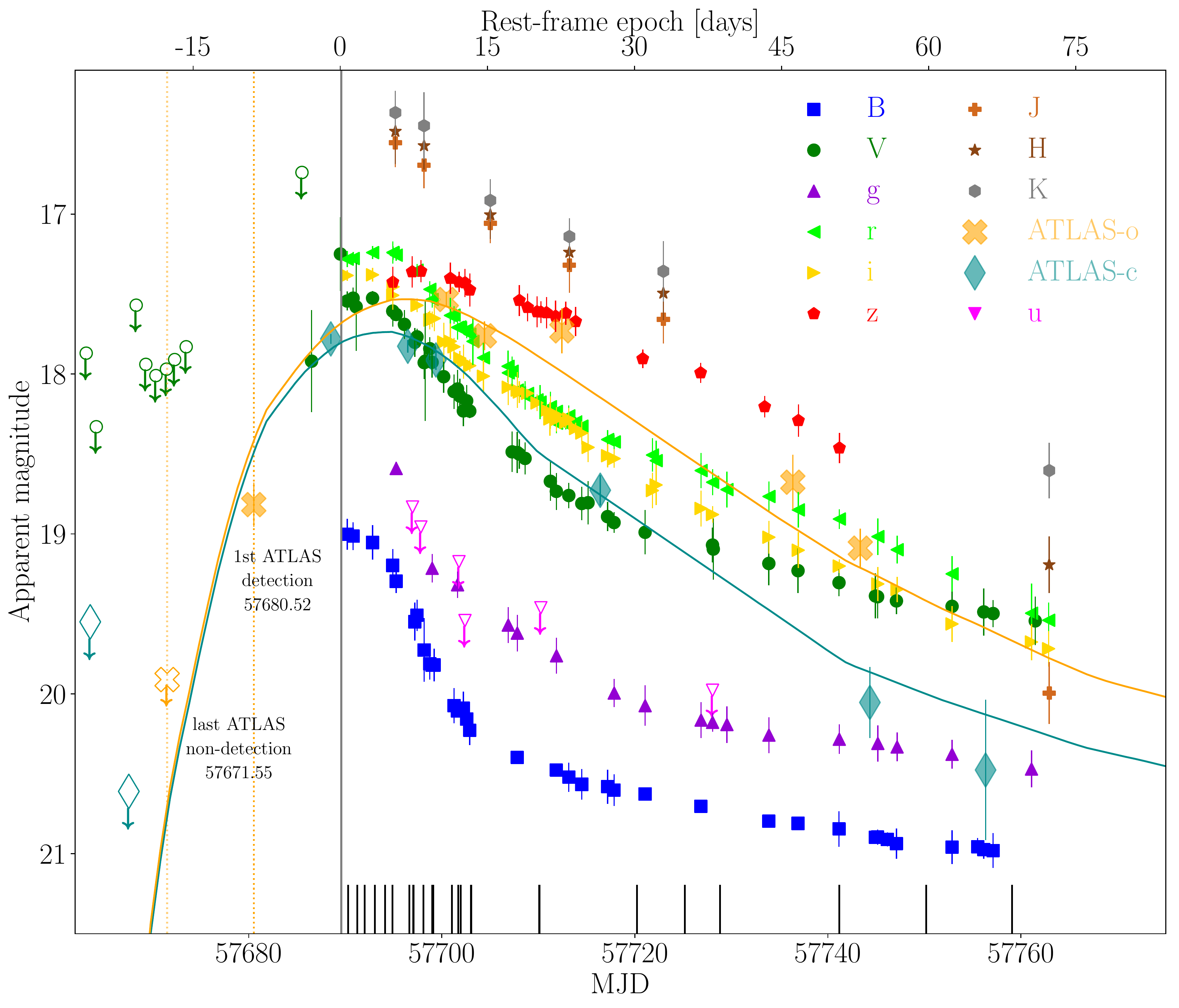}
\caption{Optical and near-infrared light curves. A grey solid vertical line represents the epoch of peak brightness in the $B$-band obtained from a {\sc SNooPy} fit to the $cyan$ and $orange$ ATLAS light-curves, which are shown in orange and cyan solid lines. Ticks at the bottom represent epochs with optical spectra available. Dotted vertical lines mark the last non-detection (yellow) and first detection (orange) epochs. Open symbols with downward arrows correspond to ASAS-SN (circles), ATLAS (crosses/diamonds), and $u$-band (triangles) non-detections.}
\label{fig:phot}
\end{figure*}

Regarding the local environment of the SN, in Figure \ref{fig:envlocal} we present the spectrum of the closest {\sc Hii} region from the segmentation described above.
We also show two panels at the top right centered at Balmer H$\alpha$ and H$\beta$ wavelengths, where the best stellar continuum fit has been subtracted from the observations leaving the pure emission spectra.
In those, the \ion{O}{iii}, \ion{N}{ii}, and \ion{S}{ii} emission lines (in pale pink) and H$\alpha$ and H$\beta$ (in orange) have been highlighted.
The extinction-corrected fluxes of these seven lines were used to calculate the H$\alpha$EW, the SFR, and the oxygen abundance, which are shown in the bottom panels together with the distributions of the same parameters measured from the spectra of all 56 {\sc Hii} regions found in our segmentation of the host galaxy. We included the distribution of the parameters as measured from the spectra, and once weighted by the contribution of the HII region to the total SFR \citep{2018MNRAS.473.1359L}.
For the {\sc Hii} region closest to \mbox{SN~2016hnk} we get a H$\alpha$EW of 4.39 $\pm$ 0.04 \AA, a SFR of $-$3.44 $\pm$ 0.34 \Msun~yr$^{-1}$, an oxygen abundance of 8.65 $\pm$ 0.11 dex, and an average stellar age of 1.2 Gyr, with a dominant peak ($\sim$65\%) at the average age and only a $\sim$4\% contribution from young ($<$15 Myr) populations.
Interestingly, in this case we find that these parameter values deviate from the average properties of all {\sc Hii} regions: it is at the lower end of the H$\alpha$EW distribution indicating a residual contribution of young populations compared to other regions of the galaxy, and can also be seen in the stellar age distribution, in agreement with the expected old progenitors for 1991bg-like SNe Ia; conversely it is in an average star forming region (25 to 50\% rank once weighted for the SFR) and has an average oxygen abundance ($\sim$50th percentiles).

The H$\alpha$ to H$\beta$ flux ratio also provides an estimate of the color excess $E(B-V)$ under certain assumptions (case B recombination; \citealt{2006agna.book.....O}).
We get a ratio of 4.36$\pm$0.35, which compared to the theoretical value of 2.86 for a non-extincted ratio and using a \cite{1999PASP..111...63F} extinction law, gives an $E(B-V)$ of 0.363$\pm$0.049 mag.

\section{Observational photometric properties}  \label{sec:photanal}

Final $ocBVugrizJHK$ light curves are presented in Figure \ref{fig:phot}.
\mbox{SN~2016hnk} photometry has rather sparse coverage before peak, since it was only observed in ATLAS $oc$ bands and ASAS-SN $V$-band during the rise. However, from the shape of the light curves in all bands shown in Figure \ref{fig:phot}, we can infer that the follow up campaign started very close to the epoch of peak brightness (t$_{\rm max}$).

\begin{figure}
\centering
\includegraphics[trim=0.1cm 0cm 0cm 0cm,clip=true,width=\columnwidth]{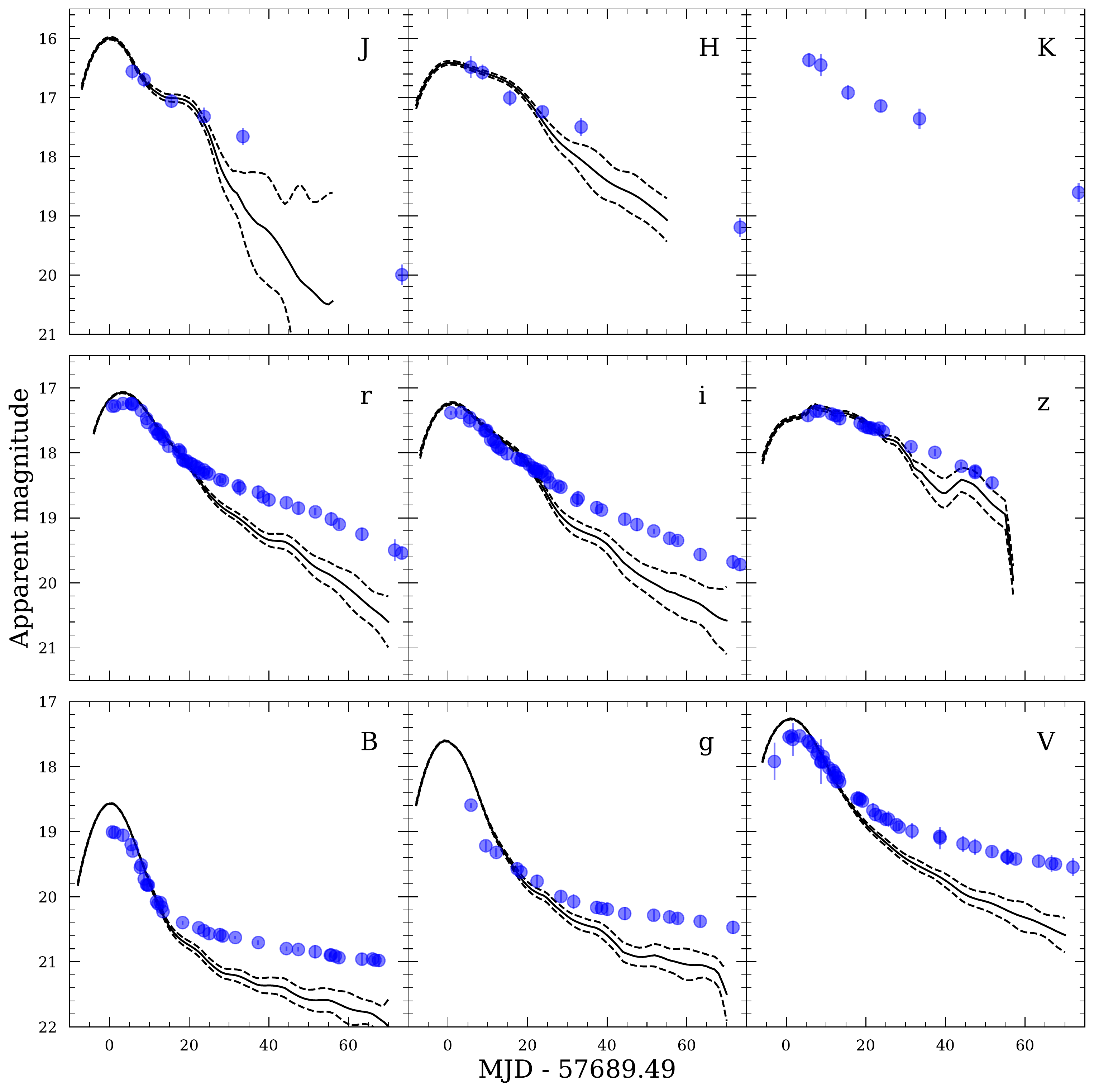}
\caption{Multi-band light curves of \mbox{SN~2016hnk}  (filled circles) compared with a 1991bg-like template (lines). Best-fit \textsc{SNooPy} parameters are $R_V$=2.1 and a host-galaxy color excess of $E(B-V)_{\rm host}$=0.55 mag. Note that beyond $+$20 d the template fit does not succeed in capturing the SN brightness.} 
\label{fig:snoopy}
\end{figure}

To determine t$_{\rm max}$ precisely, we used the SuperNovae in object-oriented Python (\textsc{SNooPy}; \citealt{2011AJ....141...19B,2014ApJ...789...32B}) fitter, which allows us to consider different extinction laws and amounts of reddening in our light curve fits.
Also, \textsc{SNooPy} makes studies involving color curves very flexible, such as the study of color laws, the construction of bolometric light curves, and estimates of the dimensionless $s_{BV}$ parameter, which gives the stretch factor of the $B-V$ color curve with respect to the typical behaviour of a SN Ia having a t$_B^{\rm max}$ $-$ t$_{B-V}^{\rm max}$ of 30 days.
The $s_{BV}$ parameter provides a more physically significant interpretation of the light curve width than similar parameters, as it measures the time at which iron re-combines from \ion{Fe}{iii} to \ion{Fe}{ii} (e.g. \citealt{2014ApJ...789...32B,2018arXiv180506907W}). So, $s_{BV}$ may be providing more information when it comes to fast-decliners and peculiar types, such as \mbox{SN~2016hnk}.
We fit the light curves of the three bands that have pre-maximum observations and the $B$ band, with the 1991bg-like built-in model, and obtain a t$_{\rm max}$ of MJD=56789.5$\pm$3.3 (ATLAS-$co$ fits are included in Figure \ref{fig:phot}). This estimate constrains the SN rise time to 17.4$\pm$5.5 days, hence to the range from 11.9 to 22.9 days, consistent with most SNe Ia and 1991bg-like SNe, but larger than in Ca-rich transients, which have faster rise times of the order 9-15 days \citep{2017arXiv170300528T}. 

A \textsc{SNooPy} light curve fit using the Markov Chain Monte Carlo fitter and the 91bg model was simultaneously performed in all bands and to the whole temporal extent, fixing the t$_{\rm max}$ and also inputing the $E(B-V)$ obtained from the analysis of the most nearby environment as a Gaussian prior.
Our initial attempt provided a reasonable fit up to 15-20 days past maximum (See Figure \ref{fig:snoopy}), but the most remarkable mismatch was found at late times (>+20 d) where \mbox{SN~2016hnk} showed an excess in brightness in all bands.
Best-fit parameters were $R_V$=2.1$\pm$0.4, $E(B-V)$=0.547$\pm$0.064 mag, and $s_{BV}$=0.438$\pm$0.030.

\begin{figure*}
\centering
\includegraphics[trim=0.1cm 0cm 0cm 0cm,clip=true,width=\textwidth]{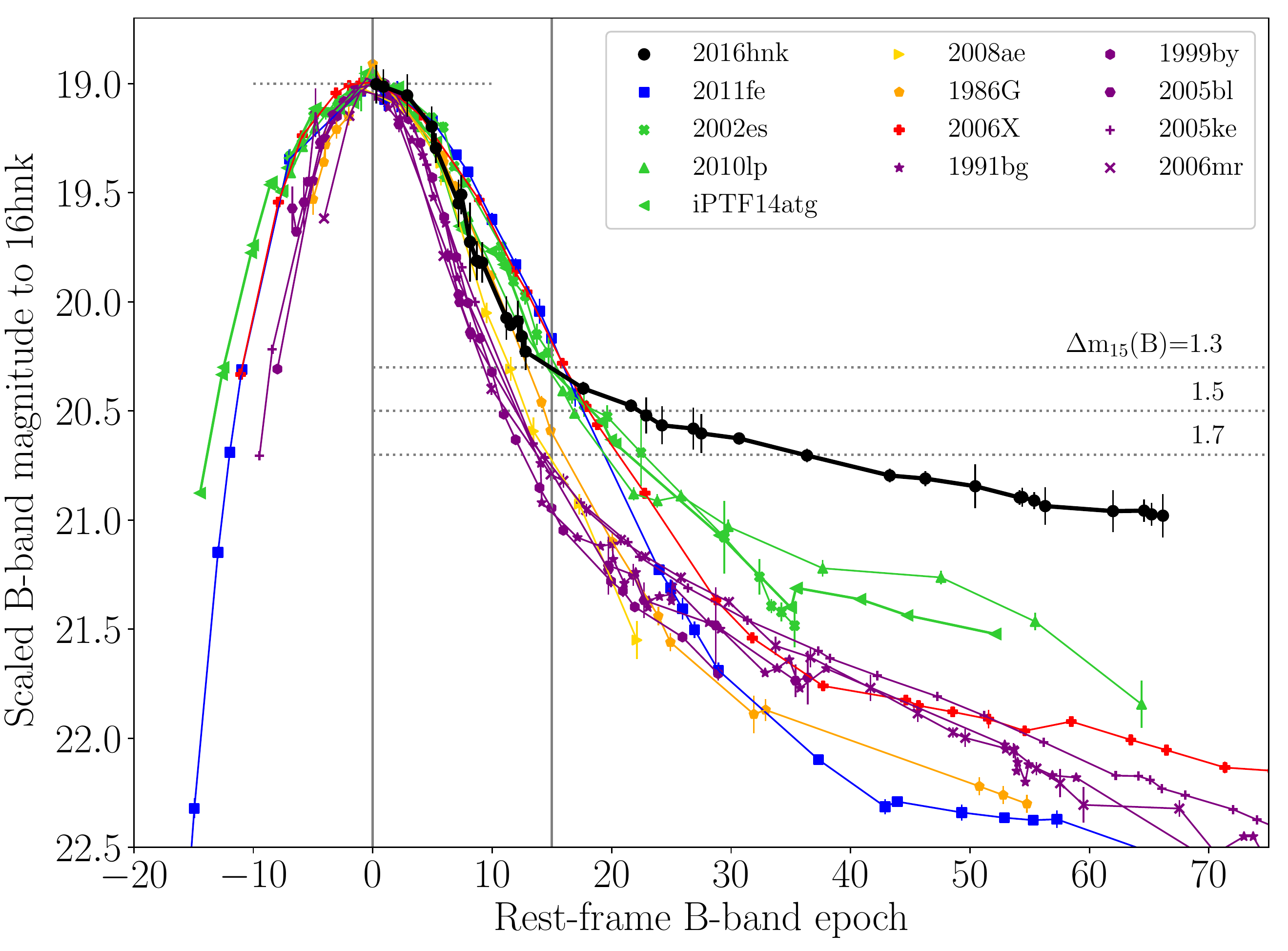}
\caption{$B$-band light curves of \mbox{SN~2016hnk} and a variety of SNe~Ia of different subtypes: normal SN Ia (2011fe), 2002es-like (2002es, 2010lp, iPTF14atg), 2002cx-like (2008ae), transitional (1986G), and 1991bg-like (1991bg, 1999by, 2005bl, 2005ke, 2006mr). All data have been restframed and scaled to match the  \mbox{SN~2016hnk} peak. Vertical lines represent epochs of the peak and +15 days, and horizontal lines mark the crossing point at 15 days of a SN with $\Delta$m$_{15}(B)$=1.3, 1.5, and 1.7 mag, respectively. \mbox{SN~2016hnk} has one of the lowest $\Delta$m$_{15}(B)$ but is the object with the brightest tail in both bands.} 
\label{fig:lccomp}
\end{figure*}

Figure \ref{fig:lccomp} shows \mbox{SN~2016hnk} $B$-band compared to a variety of SNe Ia:
the normal SN Ia 2011fe \citep{2012JAVSO..40..872R};
three 2002es-like SNe Ia, 2002es \citep{2012ApJ...751..142G}, 2010lp \citep{2013ApJ...778L..18K}, iPTF14atg \citep{2015Natur.521..328C}; 
a 2002cx-like SN Ia, 2008ae \citep{2017AJ....154..211K}; 
the transitional SN 1986G \citep{1987PASP...99..592P}; 
the highly extincted SN Ia 2006X \citep{2008ApJ...675..626W,2017AJ....154..211K}; 
and a sample of 1991bg-like SNe Ia including 1991bg \citep{1992AJ....104.1543F,1993AJ....105..301L,1996MNRAS.283....1T,2004AJ....128.3034K}, 1999by \citep{2010ApJS..190..418G,2004ApJ...613.1120G},
2005bl \citep{2008MNRAS.385...75T,2015ApJS..220....9F,2009ApJ...700..331H,2010AJ....139..519C}, 2005ke \citep{2015ApJS..220....9F,2009ApJ...700..331H,2010AJ....139..519C,2017AJ....154..211K}, and  2006mr \citep{2010AJ....139..519C,2010AJ....140.2036S}, the fastest SN Ia ever found.
These 1991bg-like SNe will be used in the following sections as a base for comparison.
All light curves have been restframed and shifted vertically to match the \mbox{SN~2016hnk} at peak brightness. For reference, two vertical lines correspond to epochs at peak and at +15 days, and three horizontal lines cross the +15 d line at the position where light curves with $\Delta$m$_{15}(B)$=1.3 (similar to \mbox{SN~2016hnk}), 1.5 (intermediate), and 1.7 mag (approx. the minimum for 1991bg-like SN Ia) should cross.
\mbox{SN~2016hnk} has a $\Delta$m$_{15}(B)$ most similar to 2002es-like objects, but with two peculiarities: it shows a much narrower light curve before +15 d, and has a much shallower decay at later epoch.
In fact, up to 12-13 days past max, it follows very closely the low-luminosity transitional SN Ia 1986G light curve, being only slightly wider than all other 1991bg-like SN Ia light-curves.
Focusing on the late (>+15 days) light curve, we clearly see that \mbox{SN~2016hnk} has the brightest tail of all objects in comparison.
This figure clearly shows that $\Delta$m$_{15}(B)$ is not capable of capturing all the diversity in SNe Ia and may mask intrinsic differences in light curve shapes, as it is seen in SNe 2002es and 2016hnk.
We expand on the explanation of this light excess in Section \ref{sec:LE}.

\begin{figure}
\centering
\includegraphics[trim=0.3cm 0cm 0.6cm 0cm,clip=true,width=\columnwidth]{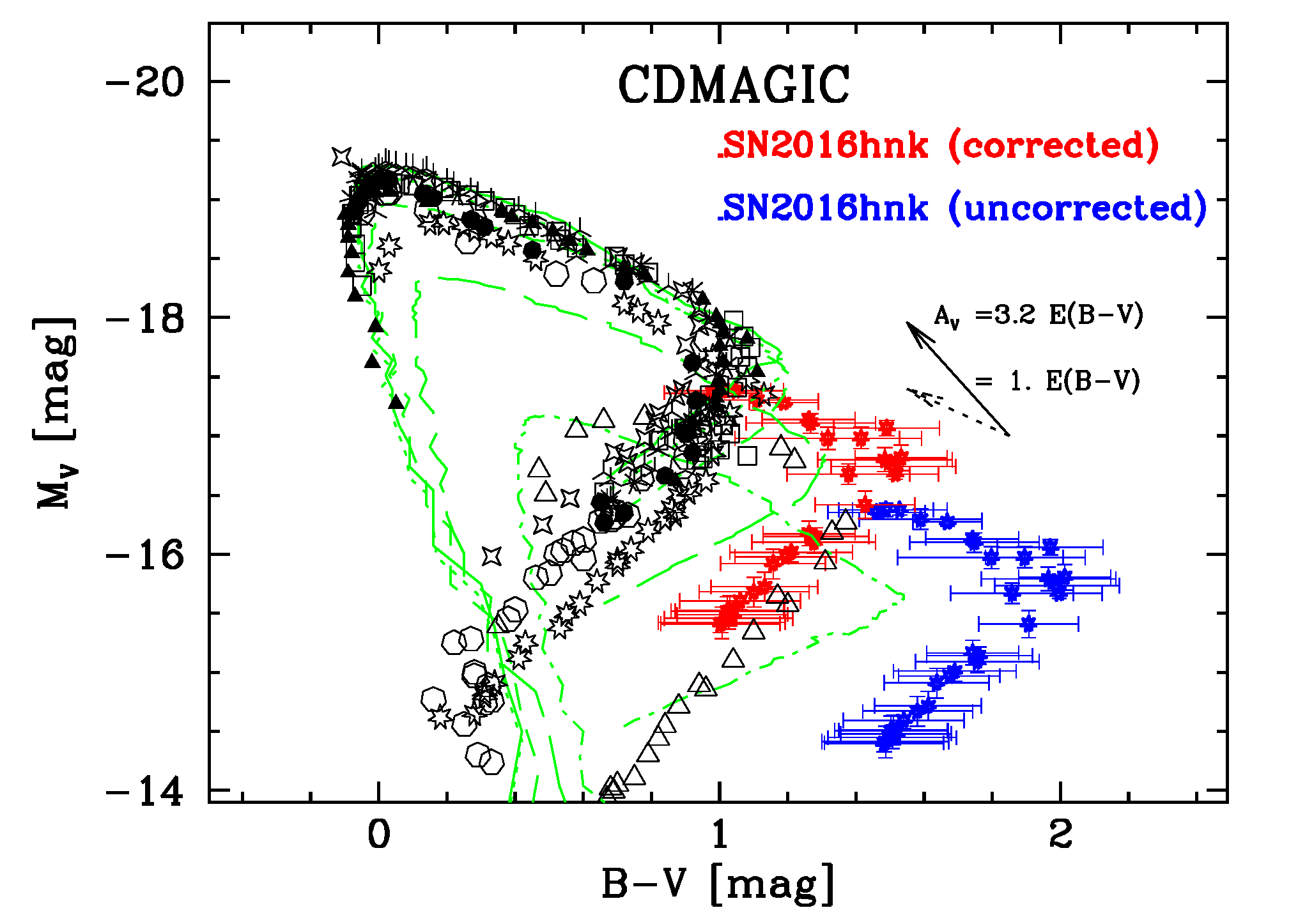}
\caption{Representative panel for the CDMagic method, where a simultaneous fit in the $B$- and $V$-band light curves, the $B$ vs. $(B-V)$ and $V$ vs. $(B-V)$ diagrams, and the $(B-V)$ evolution, is performed. For more details we refer the reader to e.g. Figure 13 in \protect\citealt{2017ApJ...846...58H}. Here we show the absolute $V$-band vs. $(B-V)$ diagram populated with eight normal SNe~Ia from the CSP-I (\protect\citealt{2017AJ....154..211K}; SNe 2004ef, 2005eo, 2005el, 2005iq, 2005ki, 2005M, 2006bh, 2006D; in differnet symbols), the SN Ia 1991bg-like 2005ke (empty triangles), and the uncorrected (blue) and corrected (red) \mbox{SN~2016hnk} datapoints. Models from \protect\citealt{2017ApJ...846...58H} are shown in green for reference, and the two arrows are the reddening vectors representing the direction to which the datapoints are shifted when affected by $R_V$=3.2 and $R_V$=1.0. \mbox{SN~2016hnk} corrected datapoints lie on top of the other subluminous SN Ia 2005ke, and from the shift we estimate an $R_V$=2.1$\pm$0.6.}
\label{fig:cdmagic}
\end{figure}

As an independent check to increase the robustness of our findings, we used the Color-Decline-Magnitude intercept Calibration (CDMagic; \citealt{2017ApJ...846...58H}) to estimate the host-galaxy reddening and preferred dust extinction law. 
First, following \cite{2001ApJ...558..359G}, we measured the stretch factor $s$ that multiplies the temporal axis of the light curve providing the best match of the \cite{2007ApJ...663.1187H} SN Ia template to the observations.
In our case, we restricted the fit to the first 15 days, to avoid the dependence of the light curve on secondary parameters and the light excess detected for SN~2016hnk at later times, and obtained an $s$=0.59$\pm$0.04.
The modified decline rate $\Delta$m$_{15,s}(B)$ is then defined as the magnitude difference between the peak and the reference epoch, $15~{\rm days}\times s$ (measured in the template), divided by the $s$ factor $\Big[\Delta$m$_{15,s}$=$\frac{\Delta m_{(15\times s)}}{s} \Big]$, which in our case is $\frac{\Delta m_{9}}{s}$=$\frac{1.06\pm0.13}{0.59\pm0.04}$=1.803 $\pm$ 0.200 mag.
The measured value of $\Delta$m$_{15,s}(B)$ determines the theoretical model from \cite{2017ApJ...846...58H} that is used as a reference to correct the $B$- and $V$-band light curves, the $B$ vs. $(B-V)$ and $V$ vs. $(B-V)$ diagrams, and the $(B-V)$ evolution of SN~2016hnk.
In the CDMagic fit we fixed the $t^{max}_B$ to the value previously obtained with \textsc{SNooPy} and the distance modulus measured from the assumed cosmology ($\mu$=34.195 mag), and only left as free parameters $E(B-V)$ and $R_V$.
While the color excess is estimated from the shift between the uncorrected (as observed and Milky Way dust corrected) and the corrected points that best match the template, the $R_V$ is measured from the direction of this shift in the color-magnitude diagrams.
In Figure~\ref{fig:cdmagic} we show, as an example, the absolute $V$ and vs. $B-V$ diagram, including a number of SNe~Ia from the Carnegie Supernova Project I (CSP-I; \citealt{2017AJ....154..211K}) and the 1991bg-like SN Ia 2005ke (empty triangles) for reference, models from \cite{2017ApJ...846...58H} with different stretch factor (green lines), and \mbox{SN~2016hnk} data before (blue) and after corrections (red).
We obtained an $E(B-V)$ of $0.45 \pm 0.08$ mag, consistent with previous methods, and an $R_V=2.1 \pm 0.6$, which we will be using hereafter, however, we note that our results are not significantly dependent on the particular choice of $R_V$. 

The observed apparent $B$-band peak magnitude is 19.003 $\pm$ 0.088 (at epoch +0.7 d).
Using t$_{\rm max}$ as a reference, we estimated $\Delta$m$_{15}(B)$ by interpolating the flux at 15 days with a 3rd order spline and obtained a value of 1.279 $\pm$ 0.071 mag. We then applied the correction provided by \cite{1999AJ....118.1766P} to get the true $\Delta$m$_{15}(B)$$_{\rm true}$ = $\Delta$m$_{15}(B)$$_{\rm obs}$ + 0.1 $\times$ $E(B-V)$, and found 1.324 $\pm$ 0.096 mag.
In order to find the absolute $B$-band peak magnitude, we corrected the observed value for Milky Way reddening of $E(B-V)$ = 0.0224 $\pm$ 0.0008 mag from extinction maps of \cite{2011ApJ...737..103S} and the average Milky Way value for $R_V$ of 3.1, and the SN host galaxy dust extinction with the color excess of $E(B-V)$ = 0.45 mag and an $R_V$ of 2.1 found above.
We obtained an A$_V^{\rm MW}$ = 0.069 mag and A$_V^{\rm host}$ = 0.945 mag, which correspond to A$_B^{\rm MW}$ = 0.092 mag and A$_B^{\rm host}$ = 1.395 mag.
Considering both corrections we get a corrected absolute $B$ band magnitude of $-$16.656 $\pm$ 0.162. This is at the low end of the luminosity distribution for SNe Ia \citep{Ashall16}, and confirms SN\,2016hnk as a subluminous SN Ia. 

\begin{figure}
\centering
\includegraphics[trim=0.1cm 0cm 0cm 0cm,clip=true,width=\columnwidth]{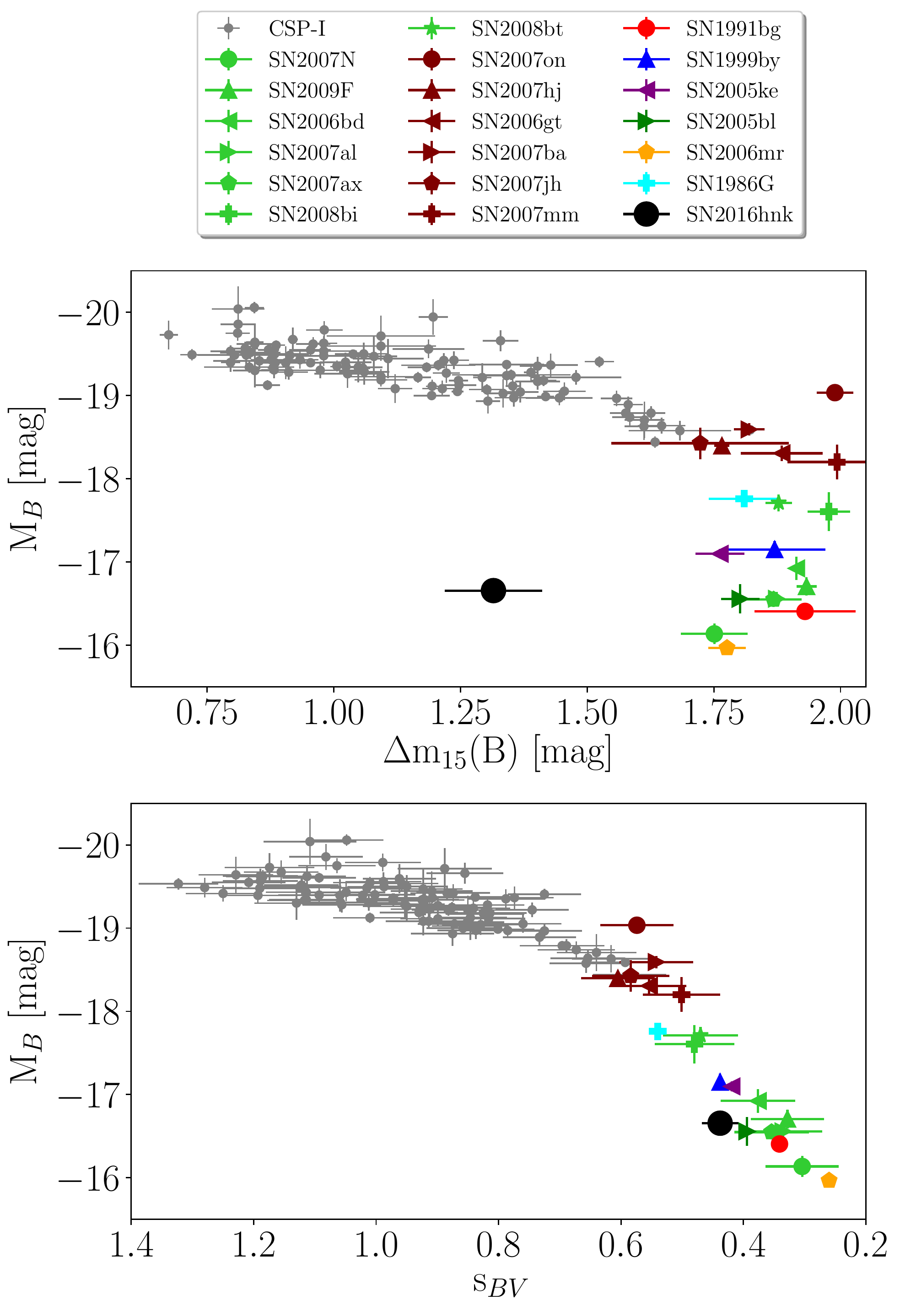}
\caption{The luminosity width relation as a function of $\Dm$\ (top) and $s_{\rm{BV}}$ (bottom) for a selection of CSP-I SNe Ia from \citet{2017AJ....154..211K}, with the addition of SN\,2016hnk (black circle). Note that SN\,2016hnk sits off the $\Dm$\ $-$ $M_{B}$ relationship because of the degeneracy in $\Dm$\ for fast declining SNe Ia.}
\label{fig:LWR}
\end{figure}

This is clearly seen in Figure \ref{fig:LWR}, where we show the absolute magnitude vs. light curve width relation for \mbox{SN~2016hnk} and SNe Ia from the CSP-I \citep{2017AJ....154..211K}.
In the upper panel we used the $\Delta$m$_{15}(B)$ parameter.
While this parameter correlates strongly with the SN peak brightness, it is known to become degenerate for the fainter SNe Ia showing a branch that follows the same relation as brighter SNe Ia, and a second branch that breaks the relation.
In this plot, \mbox{SN~2016hnk} is an outlier, and the reason for this is discussed above and can be seen in Figure \ref{fig:lccomp}. It has a very low $\Delta$m$_{15}(B)$ when compared to other subluminous SNe Ia, which usually have larger values ($>$1.7 mag).
In the lower panel we use the $s_{BV}$ parameter instead.
\cite{2014ApJ...789...32B} showed that this parameter keeps more information and breaks the degeneracy in the peak brightness vs. light curve width relation for the fainter objects. 
In this case, \mbox{SN~2016hnk} perfectly follows the relation.

\begin{figure}
\centering
\includegraphics[trim=0.2cm 0cm 0cm 0cm,clip=true,width=\columnwidth]{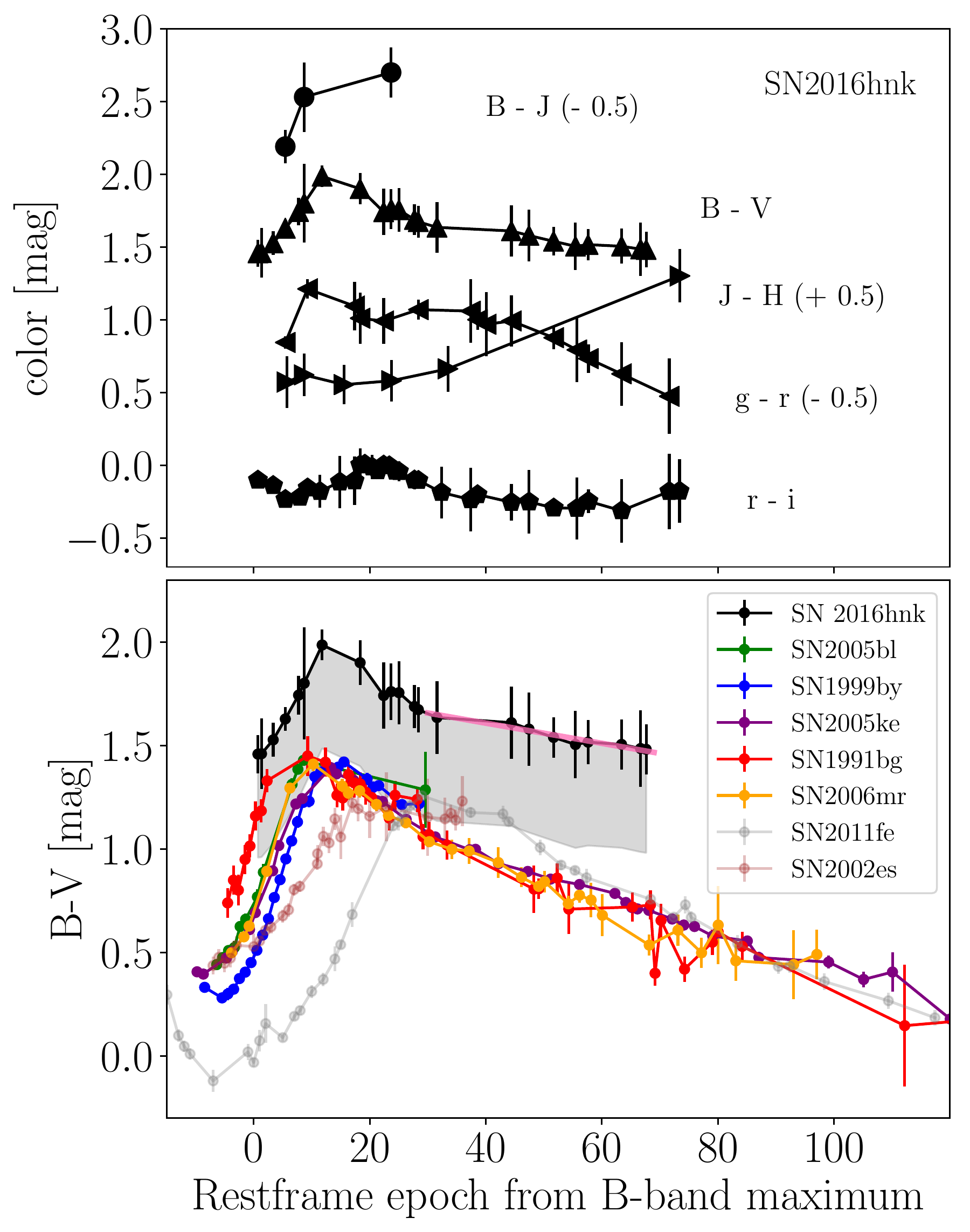}
\caption{\mbox{SN~2016hnk} color curves (top panel), and the $B-V$ color curve of \mbox{SN~2016hnk}, the 1991bg-like comparison sample, and SNe 2002es and 2011fe for reference. The shaded region projects the SN~2016hnk color curve down by 0.5 mag, on top of the bulk of other 1991bg-like SNe Ia. In pink we show the linear fit the to $B-V$ curve from 30 to 70 days past $B$ band maximum.  }
\label{fig:colors}
\end{figure}

In Figure \ref{fig:colors} we present several \mbox{SN~2016hnk} color curves (upper panel), and the $B-V$ color curve compared to the sample of 1991bg-like SNe Ia presented above, SNe 2011fe, and 2002es (bottom panel).
The slope of the $B-V$ color during the Lira law regime (between 35 and 80 days after maximum) provides an independent test on the nature of the reddening process in play \citep{1998AJ....115..234L,1999AJ....118.1766P}. 
Extinction by interstellar material (ISM) produces a shift in the $B-V$ color curve to higher (red) values, without significant changes on the decline rate, whereas sufficiently close dust ($\lesssim$10 pc, either circumstellar material $-$CSM$-$ or ISM) can also affect the decline rate due to an evolving $E(B-V)$ \citep{2013ApJ...772...19F,2018MNRAS.479.3663B}.

It can clearly be seen that the observed $B-V$ curve is well above (redder) those other subluminous SNe at all epochs.
While other 1991bg-like objects have $B-V$ colors at maximum light around 0.4$-$0.8 mag, \mbox{SN~2016hnk} has 1.5 mag. 
Up to +20 days, the color curve is consistently half a magnitude larger when compared to other 1991bg-like SNe Ia. This is in agreement with the $E(B-V)$ values found previously in the CDMagic and \textsc{SNooPy} analyses, and confirms a large amount of reddening.
On the other hand, the post maximum slope is clearly shallower, which is opposite of what is expected for objects affected by large amounts of extinction \citep{2013ApJ...772...19F}.
We fit a straight line between epochs (+30,+70) and found a slope of $-$0.005$\pm$0.001 mag day$^{-1}$, which is actually in the low regime than is expected for objects with insignificant reddening, and is in conflict with the expectations for CSM scattering. 
However, this may also be caused by either difference in the central density of the progenitor at the explosion \citep{2018A&A...611A..58G} or by differences in the transmission filters, which may cause differences in the color curve slope \citep{2017ApJ...846...58H}.

\subsection{Late light curve excess} \label{sec:LE}

Figure \ref{fig:lccomp} shows that SN~2016hnk presents the shallowest brightness decline at later epochs ($\gtrsim$ 12 days) compared with a variety of thermonuclear events. To explain this light excess we performed a number of tests confirming that:
(a) SN~2016hnk is similar to SN~1991bg when similar host-galaxy reddening with low extinction law is artificially applied to SN~1991bg, although it is still redder and no \ion{Na}{i} D lines are detected at any epoch in its spectra; and
(b) when a simple light-echo model is removed from the observed SN~2016hnk light curve, a SN~1991bg-like model fits reasonably well to the observations at late epochs, however no light-echo (spectrum at maximum) features are detected in the late-time spectra. 
Details of those analysis are left for Appendix \ref{sec:lightecho}, here we concentrate on our most plausible explanation.

At late epochs, the slow decline in the SN~2016hnk light curves suggests an additional energy source beyond $^{56}$Co to $^{56}$Fe radioactive decay.
The most likely mechanism for this particular light curve shape is multiple scattering with either a circumstellar or interstellar dust cloud intervening in the line-of-sight, which would in turn cause the total-to-selective extinction ratio to be steeper (lower $R_V$; \citealt{2008ApJ...686L.103G}).
\cite{2011ApJ...735...20A} presented predictions of how scattering by dust affects the SN light curve shape depending on the distance to the interfering cloud and its thickness (opacity). 
Qualitatively, the light curve suffers some broadening before 15 days, and it produces a light excess after 15 days (as observed in e.g. SN~2006X), which is more pronounced with thicker dust clouds (prolonged scattering of SN photons). 
We see similarities between simulated light curves shown in Figure 3 of \cite{2011ApJ...735...20A} for the 10$^{18}$ cm ($\sim$0.1 pc) case, with a broadening and a plateau at latter times, and the \mbox{SN~2016hnk} $B$-band light curve when compared to all other 1991bg-like objects.
Importantly, the net effect of this multiple scattering would be a reduction of the $\Delta$m$_{15}(B)$ parameter (of up to $\sim$0.4 mag) for similar $E(B-V)$ as in \mbox{SN~2016hnk}, as we have also shown in Figure \ref{fig:lccomp}.

\begin{figure}
\centering
\includegraphics[trim=0.1cm 0cm 0cm 0cm,clip=true,width=\columnwidth]{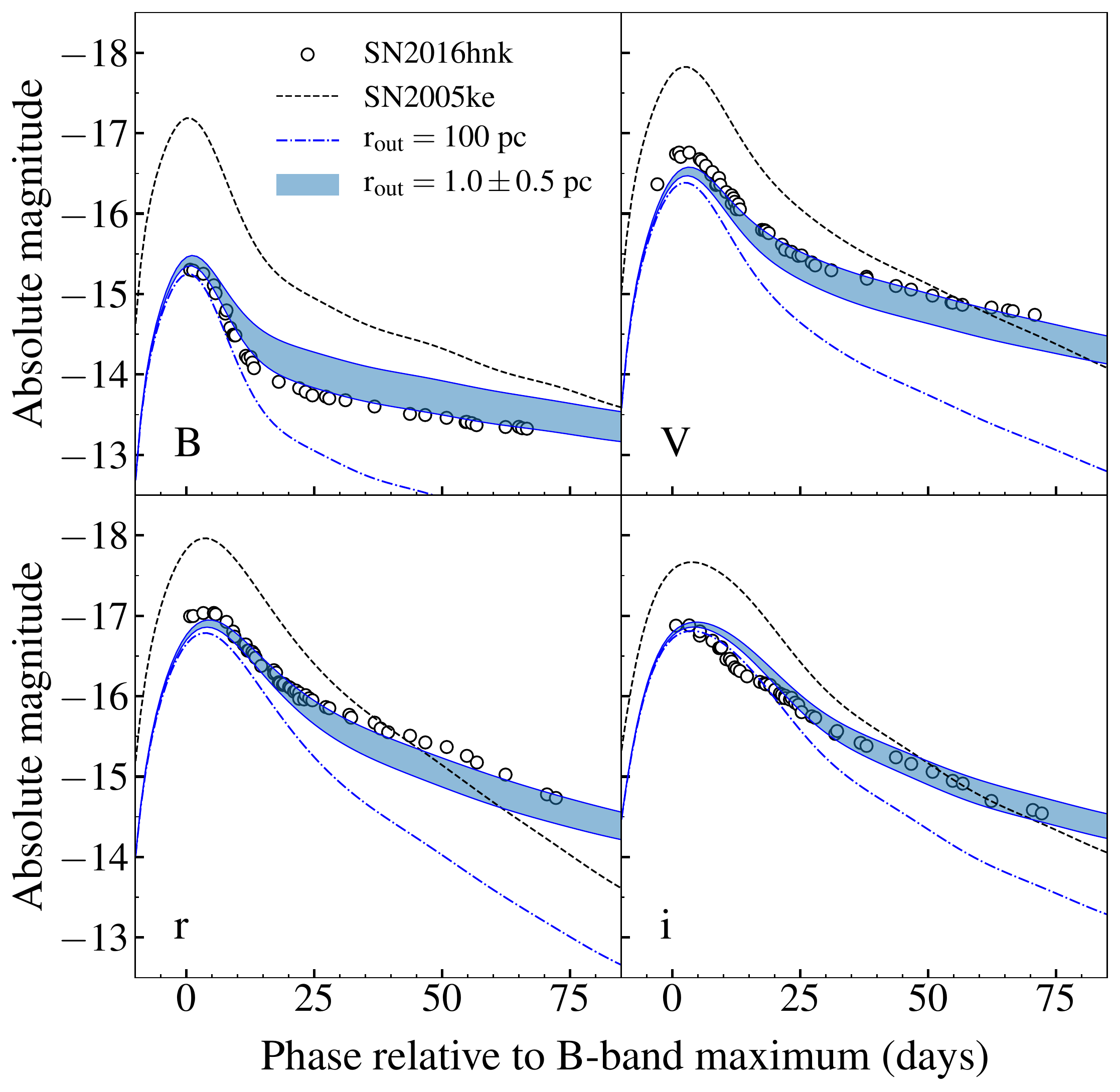}
\caption{$BVri$ observed light curves (in circles), compared with the multi-scattering dust model of \protect \cite{2018MNRAS.473.1918B} using 1991bg-like SN~2005ke as a reference (black dotted line). 
Simulated light curves affected by nearby dust with $E(B-V)$=0.5 mag at a distance of 1.0$\pm$0.5 pc from the SN (blue filled strip) provide very reasonable fits to the observations, while a model where dust is located at larger distance (100 pc; blue dashed line) just dims the light curves by keeping the same shape.} 
\label{fig:bulla}
\end{figure}

We used the Monte Carlo simulations by \cite{2018MNRAS.473.1918B} to estimate the distance to the intervening dust cloud and, at the same time, try to fit multiband light curves of SN~2016hnk with this model.
In Table \ref{tab:scattering} we provide the dust properties of the model for each band, which are based on a LMC-type dust composition from \cite{2001ApJ...548..296W}.
We use the 1991bg-like SN~2005ke \citep{2017AJ....154..211K} light-curves as a reference in our simulation.
In Figure \ref{fig:bulla}, we show the $BVri$ reference light curves and the resulting simulated light curves affected by nearby dust with $E(B-V)$=0.5 mag at a distance of 1.0$\pm$0.5 pc from the SN (blue filled strip), and at larger distance of 100 pc (blue dashed line). 
While dust at larger distances would affect the light curve by dimming the light and keeping the observed light curve shape, nearby dust would broaden it at early times and provide an extra energy source at later times.
SN~2016hnk observations are included in all panels to show the good agreement between data and model.
Based on this simulation, SN~2016hnk would have the same $\Delta$m$_{15}(B)$ than SN~2005ke, if it was not affected by the dust cloud.

\begin{table}\footnotesize
\centering
\caption{Scattering parameters (albedo, average of the cosine of the scattering angle, and absorption cross section divided by dust mass) for optical photons from \cite{2001ApJ...548..296W}
corresponding to interstellar extinction in the Large Magellanic Clouds (LMC).}
\label{tab:scattering}
\begin{tabular}{ccccc} 
\hline\hline
\textbf{Wavelength} & \textbf{Albedo} & \textbf{g=$\langle \cos\theta \rangle$} & \textbf{$\sigma_a$/$m_{dust}$}  & \textbf{Filter} \\
($\mu$m)           &  $\sigma_s/(\sigma_s+\sigma_a)$     &          &   (10$^{3}$ cm$^2$/g)   &                 \\
\hline
0.44 & 0.7159 & 0.6153 & 7.542 & B \\
0.55 & 0.7631 & 0.6059 & 4.666 & V \\
0.62 & 0.7765 & 0.5896 & 3.612 & r \\
0.75 & 0.7749 & 0.5482 & 2.646 & i \\
\hline
\end{tabular}
\end{table}

\begin{figure*}
\centering
\includegraphics[trim=0cm 0cm 0cm 0cm,clip=true,width=0.8\textwidth]{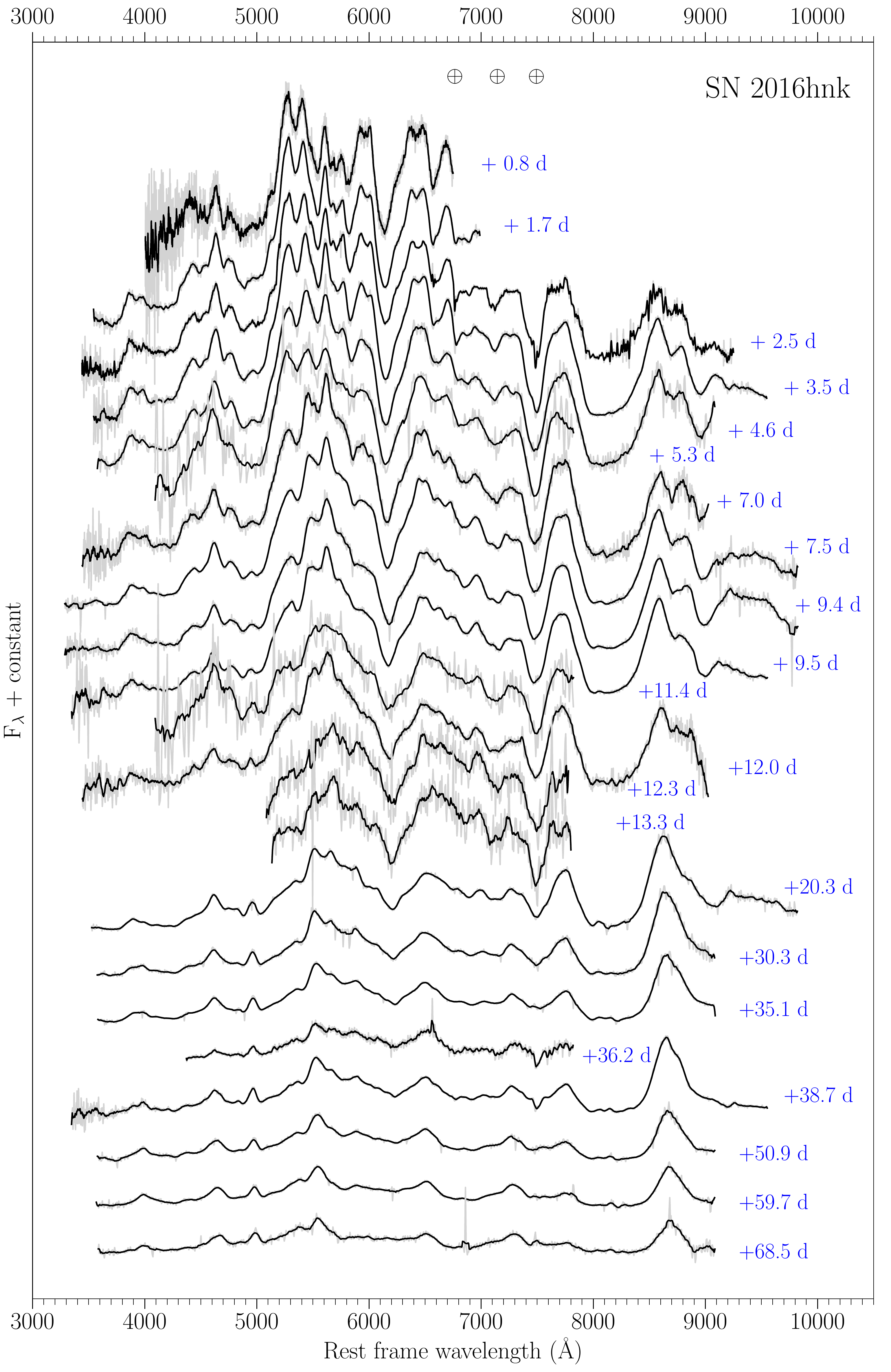}
\caption{Optical spectral sequence for \mbox{SN~2016hnk} including twenty-two observations listed in Table \ref{tab:spectropcopy} (shifted by 1.5e-16 erg s$^{-1}$ cm$^{-2}$ \AA$^{-1}$). In grey we present the color-corrected, Milky Way reddening corrected, and restframed spectra, and in black a smoothed version using a 10 pixel box. Rest-frame epochs for each spectrum are attached in blue.}
\label{fig:sp_opt}
\end{figure*}

\section{Observational spectral properties}  \label{sec:specanal}

In this section, we describe our spectral data in terms of their observed properties, then present the results of a tomographic analysis of \mbox{SN~2016hnk} in Section \ref{sec:abu}, and comparison to explosion models in \S\ref{sec:photNLTE}.

\begin{figure}
\centering
\includegraphics[width=\columnwidth]{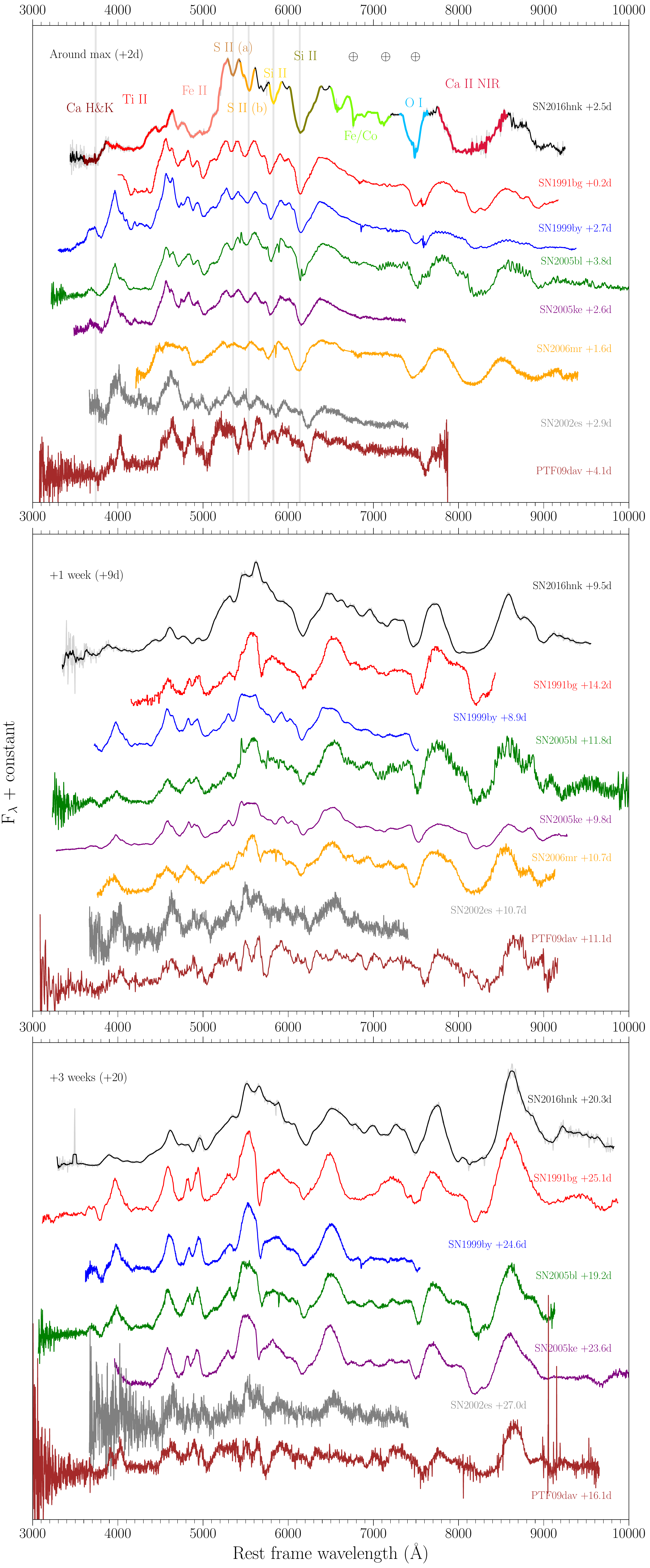}
\caption{Spectra of \mbox{SN~2016hnk} at three different epochs (around maximum, after a week, and around three weeks past maximum light) compared to other subluminous 1991bg-like SNe Ia, PTF09dav (Ca-rich), and SN 2002es (2002es-like SN Ia). The main typical features of 1991bg-like SNe Ia are identified in the upper panel. Vertical lines in the upper panel are located at the minima of a few features, and serve as a comparison of the velocities with respect other objects. All spectra have been Milky Way extinction corrected and shifted to the rest frame.}
\label{fig:comp}
\end{figure}

The sequence of color corrected, Milky Way reddening corrected, restframe optical spectra is shown in Figure \ref{fig:sp_opt}, covering epochs from +0.8 to +68.5 days.
In Figure \ref{fig:comp} we show spectra of \mbox{SN~2016hnk} and the subluminous SN comparison sample at three different epochs: around maximum light, at one week past maximum, and around three weeks past maximum.
We also include spectra of PTF09dav \citep{2011ApJ...732..118S} because \mbox{SN~2016hnk} was claimed to be similar to this Ca-rich SN, and of SN 2002es \citep{2012AJ....143..126B} to enable comparisons to this different SN Ia subclass.
In all three panels we can see the resemblance of \mbox{SN~2016hnk} to other 1991bg-like SNe Ia, especially with respect to the \ion{Ti}{ii} feature at $\sim$4400 \AA, \ion{O}{i} $\lambda$7774, and other intermediate mass elements (IME) characteristic of SNe Ia.
However, there are a number of features that make \mbox{SN~2016hnk} special:
\begin{itemize}
\item First, it is the object with by far the broadest NIR \ion{Ca}{ii} triplet absorption. This absorption remains wide and deep compared to other 1991bg-like objects until at least three weeks past maximum (see bottom panel of Figure \ref{fig:comp}).

\item A second peculiarity are the three features between 6400 and 7300 \AA, especially in the spectrum around maximum light, that fall in a wavelength region that is quite flat in other objects. 
We identified them to be caused by \FeII, \CoI, and \CoII\ lines (see discussion in Section \ref{sec:mod}). Some shallower features are also present in SN 2005bl, and \citet{2009MNRAS.399.1238H} identified them as being mostly caused by \ion{Ti}{ii}, however their Ti abundance is much larger than that which is found through explosion models and they do not reproduce all of the features. 

\item As \cite{2013ApJ...773...53F} previously noted, some extremely cool SNe show a double structure in the profile of the \ion{Si}{ii} $\lambda$5972 absorption with a weak component on the blue side of the main absorption.
This weaker absorption was attributed to the Na I D doublet that grows stronger at lower temperature, or to \ion{Fe}{ii} lines at low velocity.
\mbox{SN~2016hnk} is the only object in the sample that has this bluer absorption detached from the main \ion{Si}{ii} feature.
Related to this, in the +3 week spectra all other 1991bg-like SNe show a deep \ion{Na}{i} D absorption, however \mbox{SN~2016hnk} does not show such an absorption (see also Appendix \ref{sec:lightecho}).

\item All \mbox{SN~2016hnk} spectra are clearly redder than those of the comparison sample. All spectra in Figure \ref{fig:comp} have been corrected for Milky Way reddening and shifted to the rest frame, so that reddening excess has to be caused by host-galaxy reddening or intrinsic color.
As shown in the previous section, we estimated a color excess of 0.4-0.5 mag, which is corroborated here when comparing \mbox{SN~2016hnk} to the spectra of other 1991bg-like SNe Ia.

\item Another interesting thing about \mbox{SN~2016hnk} is the presence of two narrow lines between 5200 and 5600 \AA, which we identify as the typical \ion{S}{ii} "W" feature seen in all SNe Ia, and that were used to claim similarities to the peculiar Ca-rich SN PTF09dav \citep{2016ATel.9705....1P}. 
\cite{2011ApJ...732..118S} were able to reproduce the observed shape of these features in PTF09dav spectra by fitting synthetic spectra simulated with SYNAPPS \citep{2011PASP..123..237T} and inputting \ion{Sc}{ii} instead of \ion{S}{ii}.
We show in Section \ref{sec:mod} that we are able to reliably reproduce these and other features with typical elements found in other 1991bg-like SNe but assuming lower ionization/excitation.
\end{itemize}

\begin{figure*}
\centering
\includegraphics[height=10.25cm]{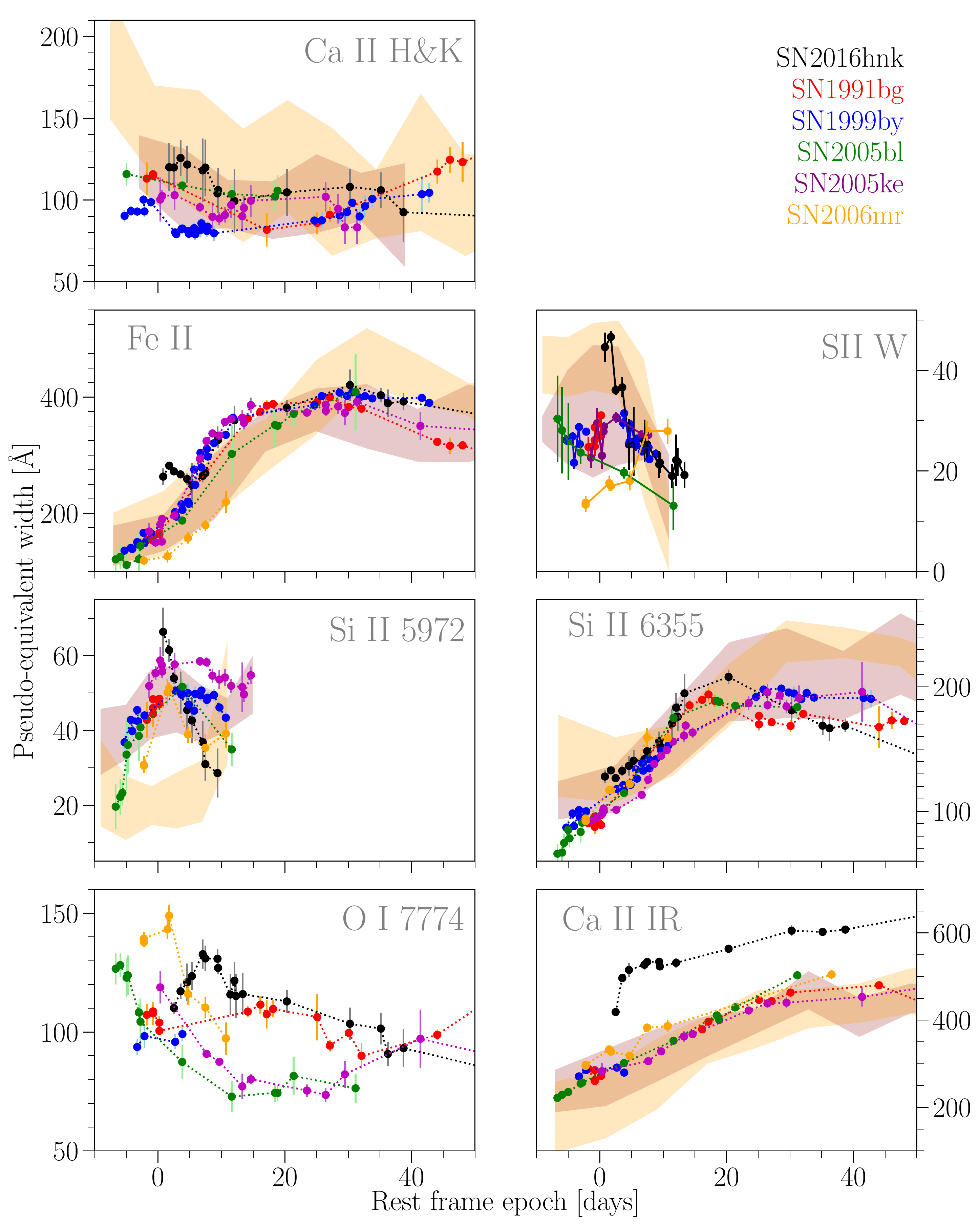} $~~~~~~~~~~$
\includegraphics[height=10.25cm]{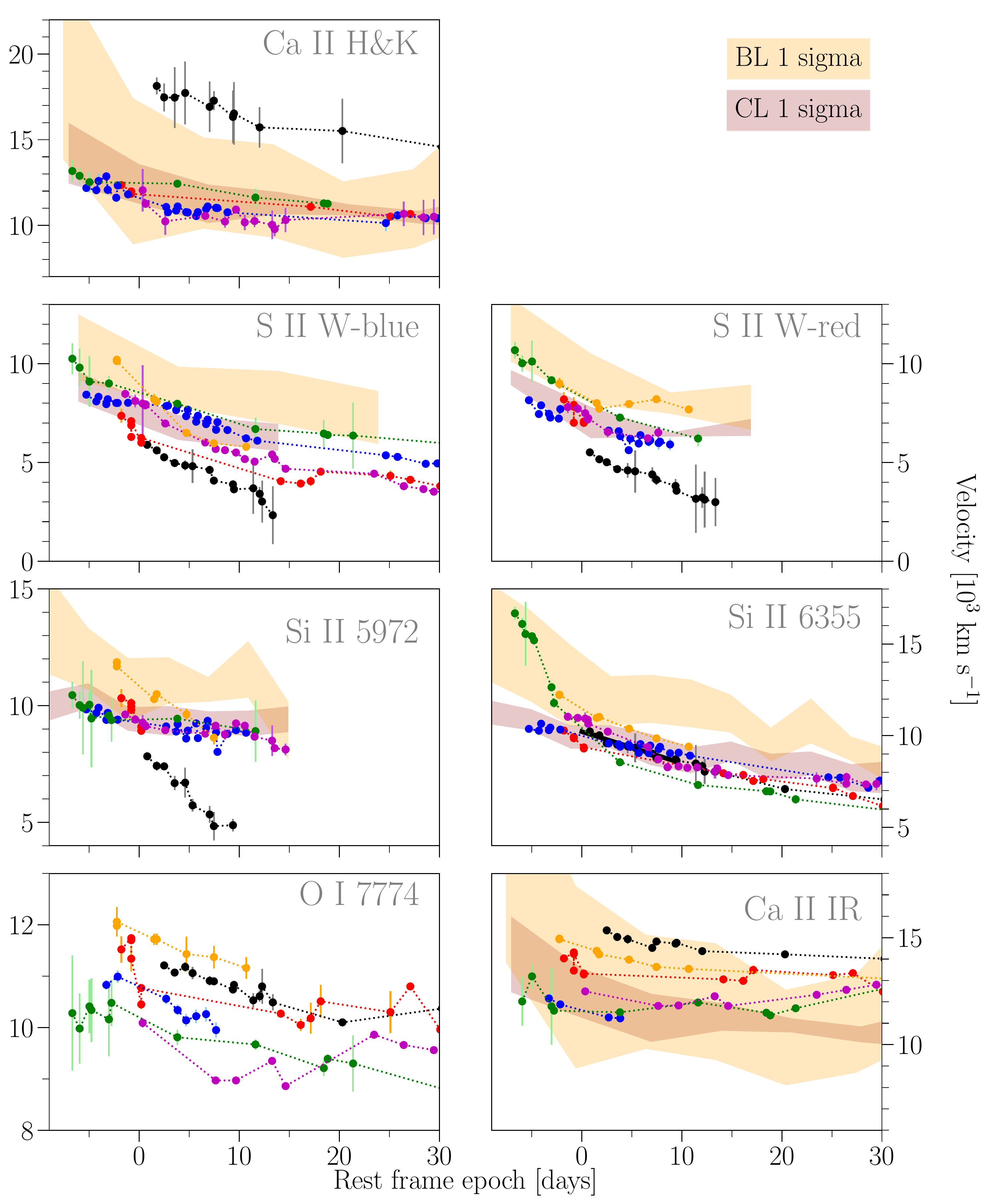}
\caption{Evolution of pseudo-equivalent widths $pEW$ and velocities as a function of time for the most prominent features highlighted in the upper panel of Figure \protect\ref{fig:comp}. 
Shaded regions represent the 1$\sigma$ dispersion of COOL (brown) and BROAD LINE (orange) evolution in the CSP-I sample from \protect\cite{2013ApJ...773...53F}.}
\label{fig:vel}
\end{figure*}

\begin{figure*}
\centering
\includegraphics[width=\textwidth]{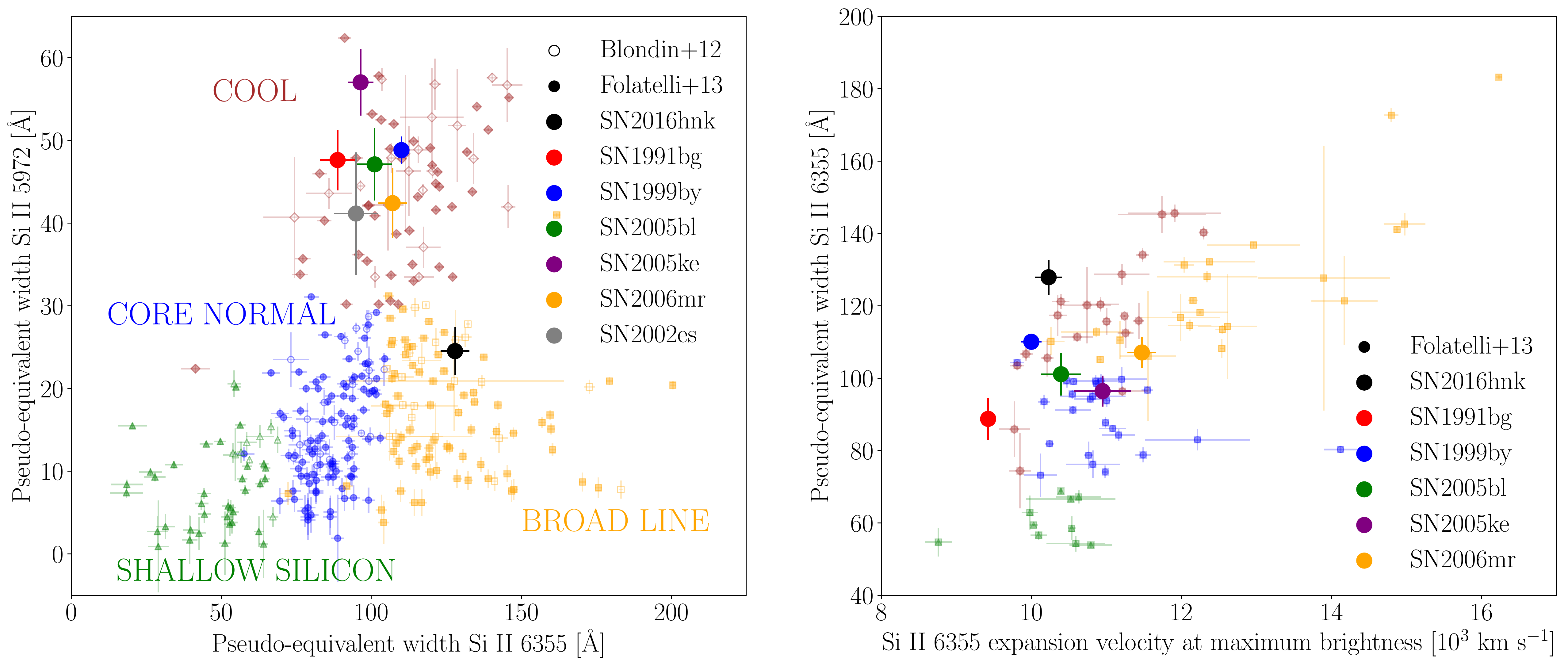}
\caption{\protect\cite{2006PASP..118..560B} and \protect\cite{2009ApJ...699L.139W} diagrams. The left panel has been populated with measurements from the CfA sample by   \protect\cite{2012AJ....143..126B} (open symbols) and from the CSP-I sample by \protect\cite{2013ApJ...773...53F} (filled symbols). In the right panel only the measurements from \protect\cite{2013ApJ...773...53F} are included. 
In both panels we added our own measurements of \mbox{SN~2016hnk} and the sample of subluminous 1991bg-like SNe Ia. SN 2002es has been added only in the left panel because with a velocity of $\sim$5800 km s$^{-1}$ it falls outside the range shown in the right panel.}
\label{fig:diagrams1}
\end{figure*}

\begin{table*}
\centering
\caption{Parameters measured in this work and used in the diagrams presented in Figure \ref{fig:benetti}.}
\label{table:benetti}
\begin{tabular}{lcccccc} 
\hline\hline
\textbf{Parameter} & \textbf{2016hnk} & \textbf{1991bg} & \textbf{1999by} & \textbf{2005bl} & \textbf{2005ke} & \textbf{2006mr} \\
\hline
v$_{\rm grad}$ [km s$^{-1}$ day$^{-1}$] & 155.4 $\pm$ 10.1 & 104.0 $\pm$ 7.0 & 110.0 $\pm$ 10.0 & 100.70 $\pm$ 13.49 & 89.05 $\pm$ 10.83 & 146.45 $\pm$ 15.71 \\
R({\sc Sii}) & 0.45 $\pm$ 0.05 & 0.62 $\pm$ 0.05 & 0.61 $\pm$ 0.06 & 0.68 $\pm$ 0.05 & 0.53 $\pm$ 0.05 & 0.63 $\pm$ 0.1 \\
$\Delta$m$_{15}$ [mag]& 1.324 $\pm$ 0.096 & 1.93 $\pm$ 0.10 & 1.87 $\pm$ 0.10 & 1.802 $\pm$ 0.038 & 1.762 $\pm$ 0.048 & 1.776 $\pm$ 0.037 \\
$\Delta$m$_{15,s}$ [mag]& 1.803 $\pm$ 0.2 & 1.903 $\pm$ 0.1 & 1.93 $\pm$ 0.11 & 1.650 $\pm$ 0.138 & 1.750 $\pm$ 0.048 & 1.92 $\pm$ 0.248 \\
$s_{BV}$ & 0.438 $\pm$ 0.030 & 0.341 $\pm$ 0.014 & 0.438 $\pm$ 0.007 & 0.394 $\pm$ 0.013 & 0.419 $\pm$ 0.003 & 0.260 $\pm$ 0.004 \\
\hline
\end{tabular}
\end{table*}

\begin{figure*}
\centering
\includegraphics[width=\textwidth]{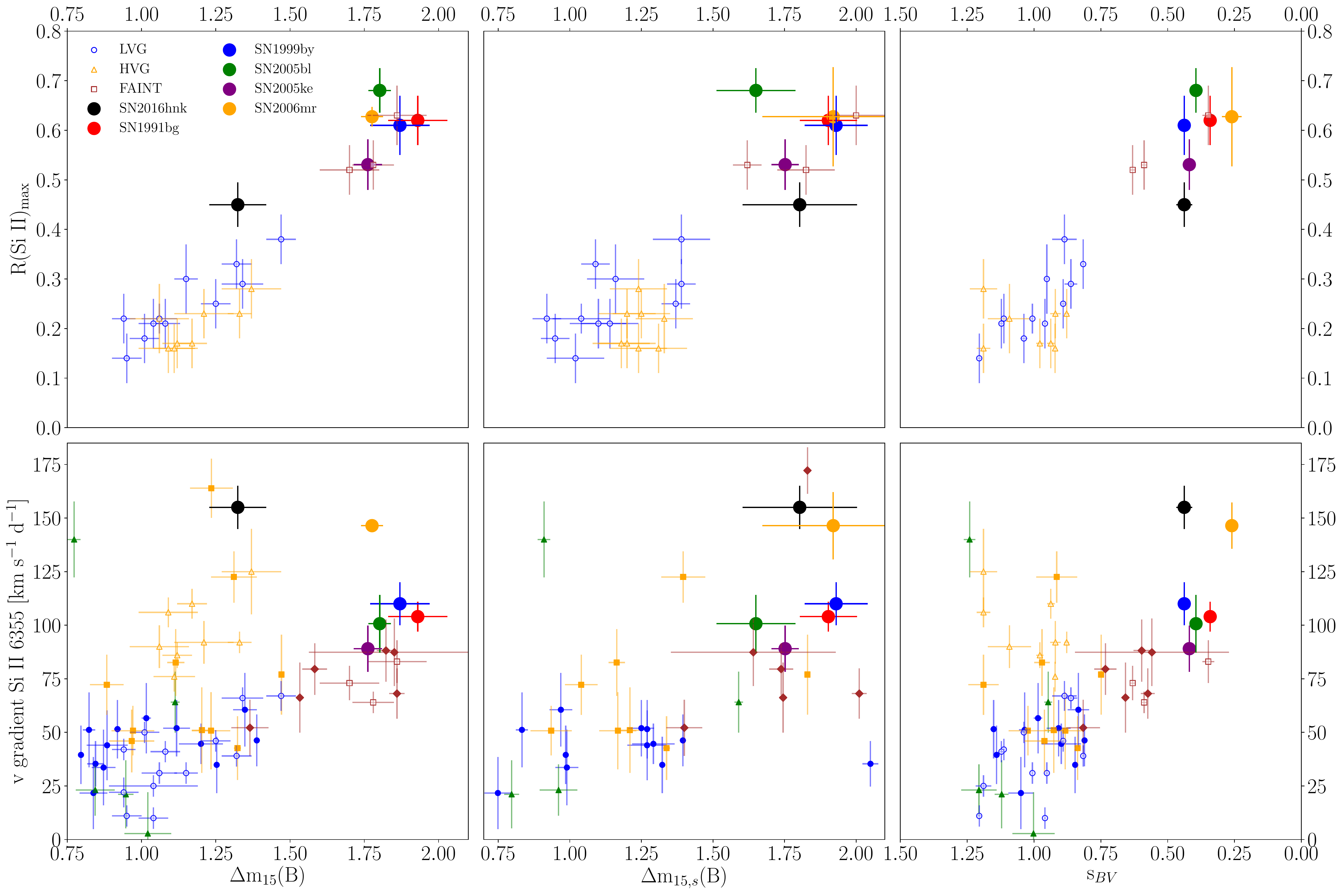}
\caption{\protect\cite{2005ApJ...623.1011B} diagrams of $R(\ion{Si}{ii})$ and \ion{Si}{ii} velocity gradient as a function of $\Delta$m$_{15}$ (left column). We replicated these diagrams using $\Delta$m$_{15,s}$ (center) and $s_{BV}$ (right column) instead as a SN light-curve width parameter.
In all panels, open symbols represent measurements from the original SN sample in \protect\cite{2005ApJ...623.1011B}, while filled symbols are measurements from the CSP-I \protect\citep{2013ApJ...773...53F}.} 
\label{fig:benetti}
\end{figure*}

In the upper panel of Figure, \ref{fig:comp} we highlight a few features in \mbox{SN~2016hnk} that we characterized in the spectral sequence in terms of the evolution of their pseudo-equivalent width ($pEW$) and velocities.
We show our measurements in Figure \ref{fig:vel} and list them in Tables \ref{tab:vel} and \ref{tab:pew}.
The most remarkable characteristics of \mbox{SN~2016hnk} compared to all other subluminous Ia are:
(i) it shows stronger \ion{Ca}{ii} H\&K and NIR features, as well as larger expansion velocities. However, we note that there are indications of a shallow feature on the blue side of the Ca NIR in the spectra of SNe 1991bg and 2005bl, which may simply be deeper in \mbox{SN~2016hnk};
(ii) while SN 20016hnk has larger \ion{S}{ii} $pEW$ than other SNe, the expansion velocities are clearly lower. This can be seen in the upper panel of Figure \ref{fig:comp} where we plotted vertical lines corresponding to the velocities of \mbox{SN~2016hnk} to better illustrate the difference to other objects;
(iii) while the \ion{Si}{ii} velocity and $pEW$ measured from the 6355\AA~line are consistent with other subluminous SNe, the \ion{Si}{ii} $\lambda$5972 velocity is significantly lower and the $pEW$ decreases very rapidly. This may be caused by the detachment of the bluer feature;
(iv) we see indications of two different behaviours regarding the \ion{O}{i} $\lambda$7774 line within the group of subluminous SNe Ia: SNe 2016hnk, 1991bg, and 2006mr show larger $pEW$ and velocities than SNe 1999by, 2005bl, and 2005ke. However, as we show in Section \ref{sec:mod}, this line is caused by a combination of \ion{O}{i}, \ion{Mg}{ii}, and \ion{Si}{ii}, and this excess may be caused by the large concentration of \ion{Mg}{ii} in \mbox{SN~2016hnk}.

The $pEW$ and velocities of the two silicon features have been widely used to identify subgroups within SNe Ia falling in different regions on a number of diagrams.
For instance, \cite{2006PASP..118..560B} proposed mapping SN properties using the $pEW$s of the two \ion{Si}{ii} features at 5972 and 6355 \AA~at the epoch of maximum brightness, and \cite{2009ApJ...699L.139W} proposed a diagram of the velocity and the $pEW$ of the same \ion{Si}{ii} $\lambda$6355 feature.
While the latter uses the same line, the former uses two different lines, taking into account different levels of excitation. 
Less luminous SNe Ia have stronger \SiII\ 5972 features at maximum light. 
This is caused by the recombination from \SiIII\ to \SiII\ happening earlier in less luminous objects, and the 5972\AA~feature having a higher excitation level compared to the 6355\AA~feature, which is saturated at maximum for all SNe Ia \citep{2008MNRAS.389.1087H}. 
Furthermore, if Si lines are at low velocities the 6355\AA\ feature can also be contaminated by \FeII\ $\lambda$6516 (see Section \ref{sec:abu}).   
Another difference between the 5972 and 6355\AA~\SiII\ features is that while the range of $pEW$ for Si 5972\AA\ is due to differences of depth, the range of $pEW$ in the 6355\AA~feature is because of width (e.g. a larger/small Si region in velocity space),
as can be seen in the relation between velocity and $pEW$ of the \cite{2009ApJ...699L.139W} diagram.

We present both diagrams in Figure \ref{fig:diagrams1} with the background filled in with measurements from the CfA SN Ia sample in \cite{2012AJ....143..126B} and the CSP-I sample in \cite{2013ApJ...773...53F}, and with measurements for SN\,2016hnk and the comparison sample shown on top.
As expected, in the \cite{2006PASP..118..560B} diagram (left panel), all 1991bg-like SNe Ia fall in the COOL region, but \mbox{SN~2016hnk} falls in the BROAD LINE region because, although it has consistent \ion{Si}{ii} $\lambda$6355 $pEW$, has a shallower \ion{Si}{ii} $\lambda$5972 feature compared to all other 1991bg-like SNe Ia.

Groups are a bit mixed in the \cite{2009ApJ...699L.139W} diagram (Figure \ref{fig:diagrams1}, right panel), with overlapping regions between COOL and NORMAL and COOL and BROAD LINE SNe Ia.
Accordingly, some 1991bg-like SNe Ia fall on top of the other two groups of CSP-I SNe.
In contrast to the left panel of Figure \ref{fig:diagrams1} (Branch et al. diagram) the location of \mbox{SN~2016hnk} in this diagram is more consistent with the COOL objects, having the largest \ion{Si}{ii} $pEW$ of all objects with similar expansion velocities.
While this diagram is efficient in separating high-velocity SNe (they all fall on the right side of the plot) and overluminous 1991T-like shallow silicon SNe Ia (they fall on the bottom side of the plot), it is not able to clearly separate the core normal and the cool groups, as the Branch et al. diagram does, mostly because it does not capture excitation differences. 

In Figure \ref{fig:benetti} we present two additional historical diagrams from \cite{2005ApJ...623.1011B} (in the left column), which combine spectroscopic parameters with the light curve width parameter $\Delta$m$_{15}(B)$.
In the top panels we show the R(\ion{Si}{ii}) parameter, the ratio of depth of the \ion{Si}{ii} 5972 over 6355 \AA~features, and in the bottom panels we show the \ion{Si}{ii} 6355\AA~velocity gradient as measured in the range from peak up to the epoch when the feature disappears.
We also populated these diagrams with literature measurements from Benetti et al. (open symbols) and the CSP-I (filled symbols).
Corresponding parameters for our sample of 1991bg-like SNe Ia are listed in Table \ref{table:benetti}.
As before, all 1991bg-like SNe Ia fall in the region of the diagram where all other subluminous objects from the literature are located, however \mbox{SN~2016hnk} falls in an unpopulated region closer to NORMAL SNe Ia.
Although its R(\ion{Si}{ii}) would be lower but consistent with COOL SNe, the parameter that is shifting \mbox{SN~2016hnk} far from the COOL group is its low $\Delta$m$_{15}$.

As a comparison we repeat those same two diagrams but using the $\Delta$m$_{15,s}$ and $s_{BV}$ parameters, under the hypothesis that they provide a more physically significant interpretation of the light curve width, as they measure the time at which iron recombines from \ion{Fe}{iii} to \ion{Fe}{ii} and breaks the degeneracy in the absolute magnitude vs. $\Delta$m$_{15}$ (Phillips) relation. 
We filled the diagrams with measurements of $\Delta$m$_{15,s}$ for the CSP-I from \cite{2017ApJ...846...58H} and the original \cite{2005ApJ...623.1011B} sample measured here.
Similarly, we used $s_{BV}$ for the CSP-I sample from \cite{2014ApJ...789...32B} and the \cite{2005ApJ...623.1011B} sample measured here.
In both cases, while 1991bg-like SNe Ia are kept in the region defined by the COOL objects, \mbox{SN~2016hnk} is also shifted towards the COOL group.
It shows how the $\Delta$m$_{15}$ parameter is ambiguous when the SN evolves too fast (but $\Delta$m$_{15,s}$ and $s_{BV}$ do not have this issue).


\begin{figure}
\centering
\includegraphics[width=\columnwidth]{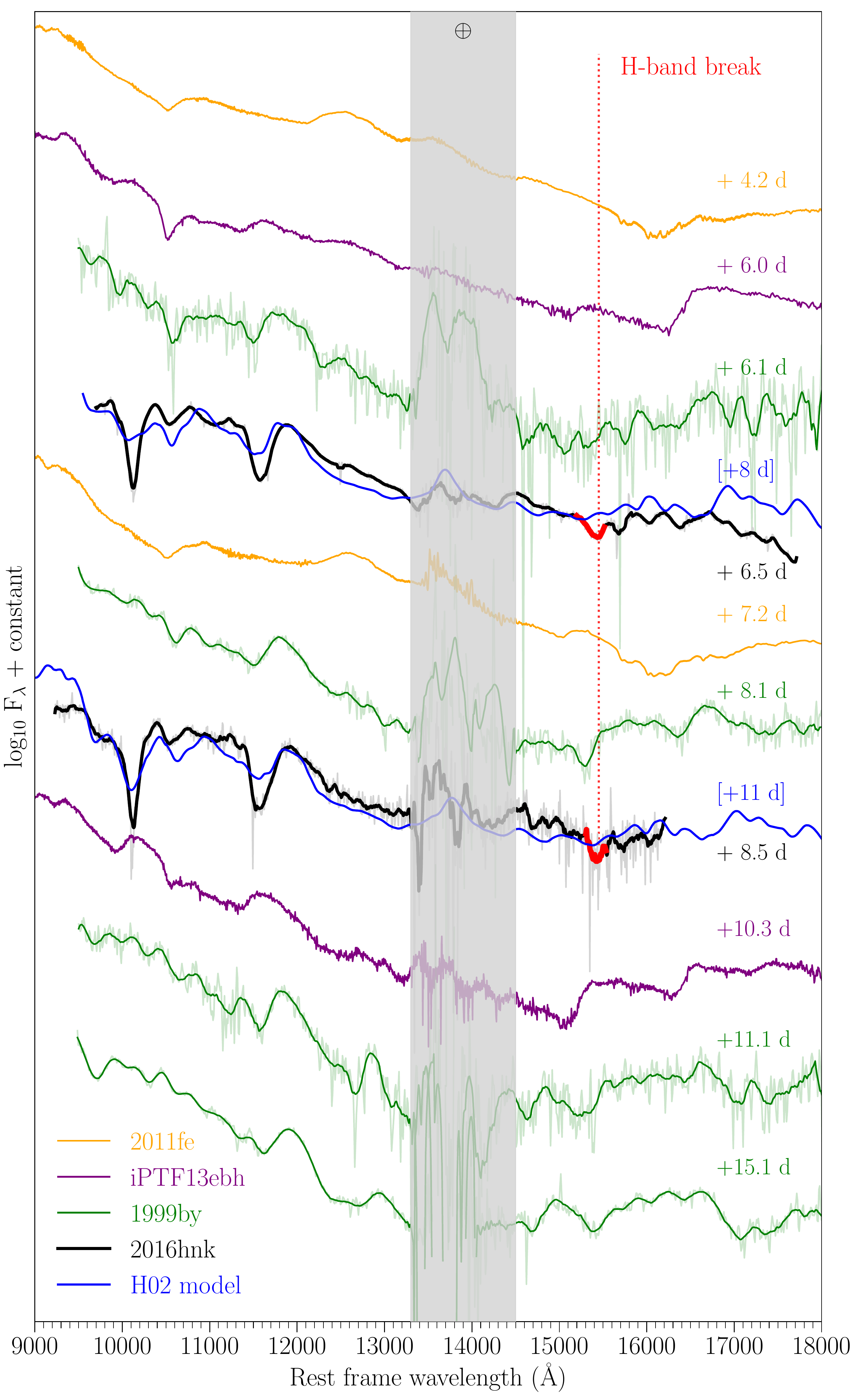}
\caption{Near-infrared spectra of \mbox{SN~2016hnk} at around 6.1 and 8.5 ($\pm$3.1) days and, as a comparison, spectra of SN 2011fe (normal SN Ia), iPTF13ebh (transitional SN Ia), and SN 1999by (SN Ia 1991bg-like).
We also include delayed-detonation models for SN 1999by some 24-27 days after the explosion, which corresponds to about 8-11 days past-maximum light in the $B$ band (see section \ref{sec:photNLTE} for more details). 
The \mbox{SN~2016hnk} spectra are dominated in the $J$ band by blends of singly-ionized iron group elements, \ion{Ca}{ii}, and \ion{Si}{ii}, while the $H$ band appears quite flat with the H-band break marked in red.
The two deep features at around 10,000 and 11,500 \AA~are only seen shallower in SN 1999by spectra.
}
\label{fig:sp_nir}
\end{figure}

\subsection{NIR spectra}

The two NIR spectra available are presented in \mbox{Figure \ref{fig:sp_nir}} together with comparison spectra of a normal SN Ia (SN~2011fe; \citealt{2014MNRAS.439.1959M}), a transitional object (iPTF13ebh; \citealt{2015A&A...578A...9H}), and a SN~1991bg-like (SN~1999by; \citealt{2002ApJ...568..791H}).

\mbox{SN~2016hnk} spectra show a few features between 9,500 and 13,000 \AA, up to the point where a telluric band starts to affect the data.
According to \cite{2003ApJ...591..316M,2009AJ....138..727M}, the two deeper fetaures at $\sim$10,000 and 11,500 \AA~could be caused by \ion{Fe}{ii}  but they are deeper in \mbox{SN~2016hnk} than in other SNe Ia.
Only SN~1999by show similar shallower features at those wavelengths.
However, at a week past maximum SN~1999by did not transition from the Si/S/Ca to the Fe/Ni-dominated regime (see Figure 3 in \citealt{2002ApJ...568..791H}).
SN~2016hnk shows lower velocities and the transition is faster (around 1day), so that the Fe features in the NIR appear earlier.
More detailed line identification is left for section \ref{sec:mod} where the NIR spectra are compared in more detail to SN 1999by models from \cite{2002ApJ...568..791H}.

Although in the middle of the telluric feature, the \ion{Si}{ii} $\lambda$13650 seems to also be present at $\sim$13,500 \AA.
Finally, at $\sim$15,500 \AA, we detect the $H$-band break, a complex formed by \ion{Fe}{ii}/\ion{Co}{ii}/\ion{Ni}{ii}.
The strength of the $H$-band break correlates with $\Delta$m$_{15}(B)$, with brighter SNe Ia having the strongest $H$-band break \citep{2013ApJ...766...72H}, and the outer edge of the velocity of this region is correlated with light curve shape \citep{2019ApJ...875L..14A}.
The outer-edge of this feature has a velocity of v$_{\rm edge}$=5,380 $\pm$ 490 km s$^{-1}$ as measured in the \mbox{SN~2016hnk} spectrum at +6.5 days.
Sub-Chandrasekhar mass explosions are expected to have v$_{\rm edge}$ values of around 7,000 km s$^{-1}$, while M$_{\rm Ch}$ models of sub-luminous SN Ia have lower values, around $\sim$5,000 km s$^{-1}$ \citep{2019arXiv190401633A}.
So, in this picture, \mbox{SN~2016hnk} is compatible with high mass, near M$_{\rm Ch}$ models.

\subsection{Nebular spectrum}

\begin{figure}
\centering
\includegraphics[trim=-2cm 0cm 0cm 0cm,clip=True,width=\columnwidth]{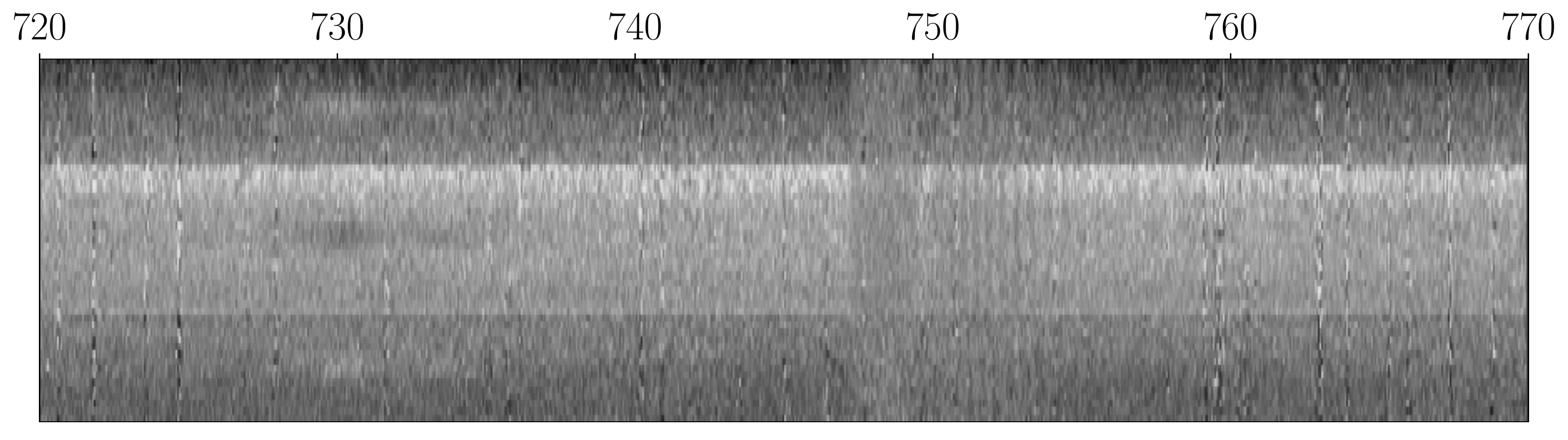}
\includegraphics[width=\columnwidth]{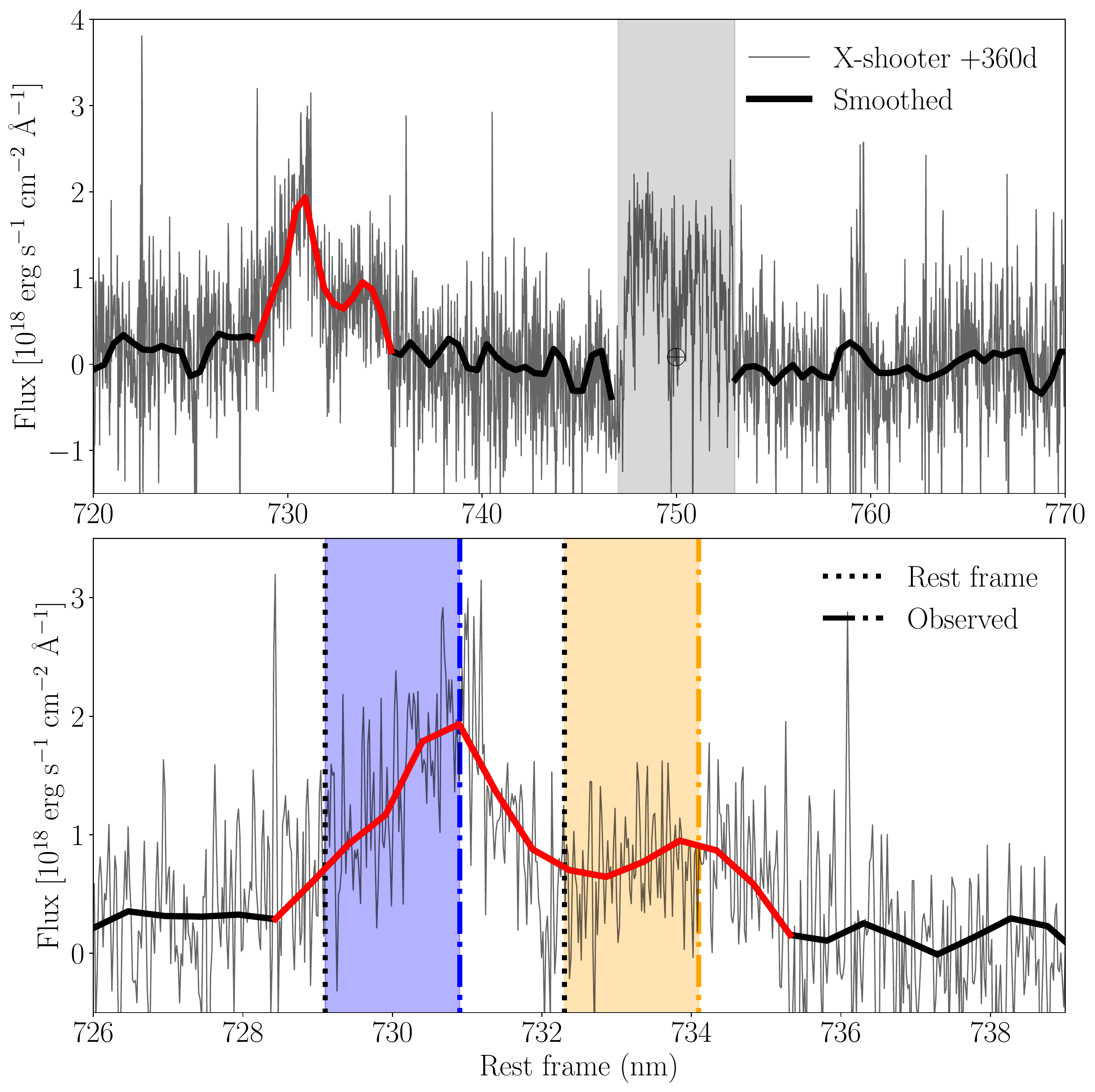}
\caption{Top: 2D frame of the combined X-shooter nebular phase (+360 d) spectrum of \mbox{SN~2016hnk} after sky subtraction, around the region where we detected the only feature. The trace can be clearly seen around 730 nm. The other feature around 750 nm is due to a bad telluric subtraction, since it is present along the spatial direction (vertical).
Middle: The combined 1D X-shooter optical spectrum showing the only two features detected in the whole UV+Opt+NIR wavelength range.
Bottom: Zoom-in to the feature we identified as being due to calcium emission. The blue and orange shaded regions represent the shift of $\sim$700 km s$^{-1}$ between the peak of the emission and the rest frame wavelength.}
\label{fig:xshooter}
\end{figure}

\begin{figure*}
\centering
\includegraphics[width=\textwidth]{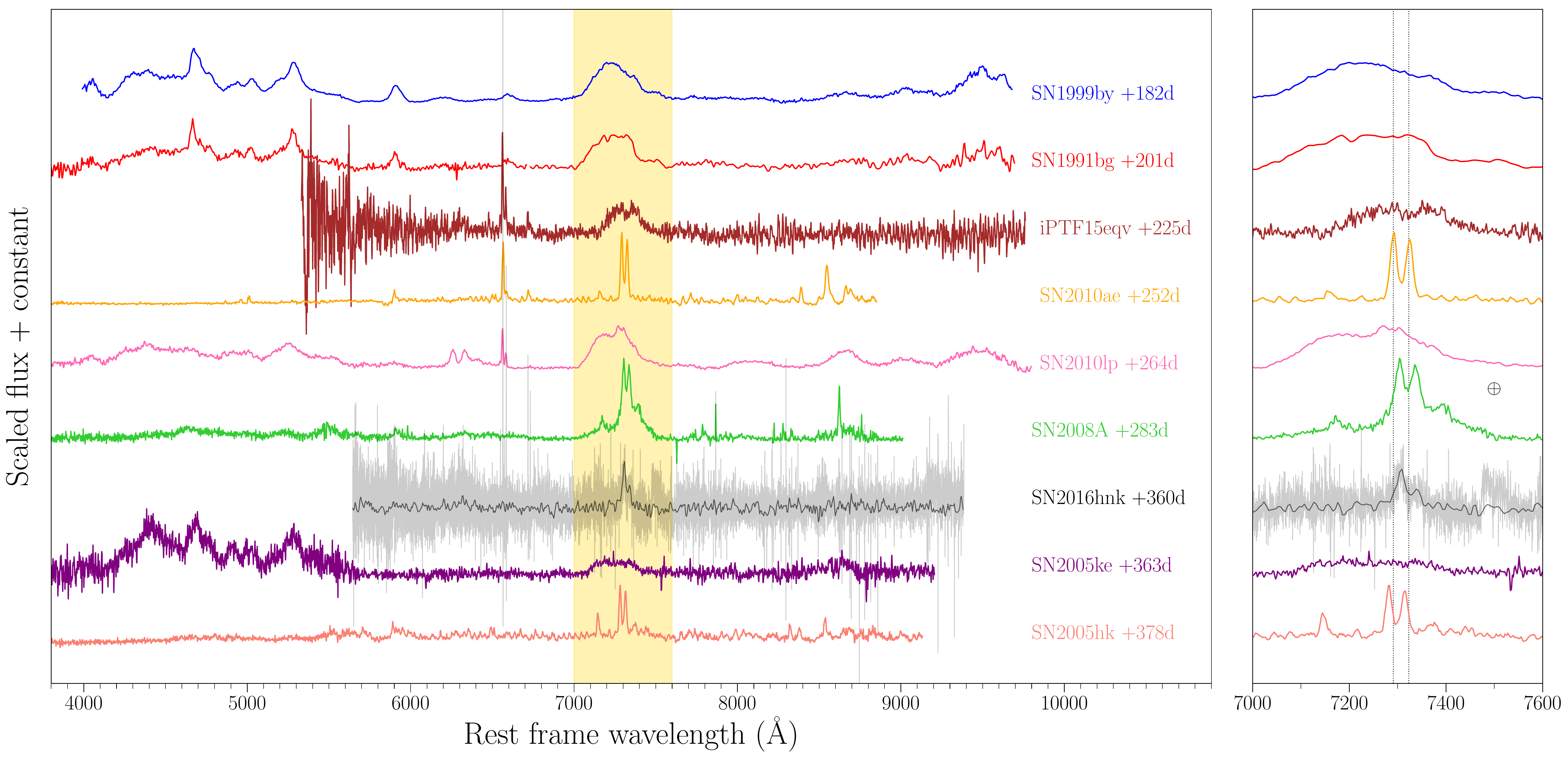}
\caption{Comparison of SN~2016hnk nebular spectra (in black) to other 1991bg-like, 2002es-like, 2002cx-like, and Ca-rich objects. While all these objects show Ca emission at late times, only SN~2002cx-like (and possibly 2002es-like; e.g. SN~2010lp) objects show the narrow emission lines similar to SN~2016hnk. The yellow shaded region is zoomed in on the right panel, where the vertical dotted lines represent the restframe wavelength of the two [\ion{Ca}{ii}]  emission lines. We note that while SN~2005hk lines have negative velocities, SN~2008A lines have positive velocities similarly to \mbox{SN~2016hnk}.}
\label{fig:nebcomp}
\end{figure*}

The resulting stack of the six individual exposures results in a low signal-to-noise spectrum, at an average epoch of +360 days post-explosion.
Carefully scanning all regions from 3000 to 25,000 \AA, we find that the only visible feature is a double emission at $\sim$7300 \AA, as reproduced in Figure \ref{fig:xshooter}, which we associate to the 3$\rightarrow$1 and 2$\rightarrow$1 transitions of the \ion{Ca}{ii} doublet at 7,291.5 and 7,323.9 \AA, respectively.
The peaks of the emission are redshifted by $\sim$740 km s$^{-1}$.
In the upper panel we show a zoom into the 2D spectra where the trace of the SN light is clearly seen at those wavelengths.
A poor telluric correction is responsible for the feature around 7,500 \AA, as the 2D spectrum unequivocally demonstrates.
Although similar narrow [\ion{Ca}{ii}] features have been previously seen in 2002cx-like SN Ia spectra (e.g. SN 2005hk, SN~2008A and SN~2010ae; \citealt{2014ApJ...786..134M,2016MNRAS.461..433F,2014A&A...561A.146S}), and are possibly also present in the 2002es-like SN Ia spectrum of SN~2010lp (\citealt{2013ApJ...775L..43T}; see Figure \ref{fig:nebcomp}), the \mbox{SN~2016hnk} spectrum does not show either the narrow [\ion{Fe}{ii}] $\lambda$7155 and [\ion{Ni}{ii}] $\lambda$7378, nor the broad [\ion{Fe}{ii}] $\lambda$7155 accompanying features, respectively. 
In addition, the lack of other features at bluer wavelengths can be attributed to the low S/N and high reddening that is affecting the SN.
We provide more details in Section \ref{sec:modnebular}.


\section{Spectral modelling} \label{sec:mod}

To get insights on the possible progenitor system and explosion mechanism that explain our observations we turn into modelling of the observed spectra.
In section \ref{sec:abu}, we want to evaluate spectral line identifications and spectral fits using the method of abundance stratification.
This approach is rather independent from the explosion scenario because
the overall envelope structures of most explosions scenarios are similar, and  nuclear physics dominates the specific explosion energy.

In the subsequent sections, we use detailed time-dependent non-LTE models. Based on results of the abundance stratification method and guided by the observed LCs and spectra during the photospheric phase, we identify an explosion model from literature which matches the observations of SN~2016hnk. 
This turns out to be the non-LTE model for SN~1999by from  \cite{2002ApJ...568..791H}.  
Abundance stratification does not allow us to fit NIR spectra, due to the assumption that the photosphere can be approximated as  black body.
Therefore, for verification, we check that the non-LTE model mentioned above can reproduce the 
NIR spectrum of SN~2016hnk, without further tuning of the parameters, we also identity the features produced in the NIR (CI, Mg II and other IME lines). 
The non-LTE models are then used to revisit some of the line identifications of the previous section, and compared to the abundance stratification results. 

In section \ref{sec:modnebular}, we modify our explosion model with the goal to reproduce the narrow [\CaII] feature during the nebular phase. 
We change the initial conditions of the WD and calculate the initial phase of the expansion up to about 10 days \citep{2017ApJ...846...58H}. We also construct detailed non-LTE nebular spectra which include gamma- and positron transport but we neglect radiative and recombination time-scales.
We discuss why energy input and transport by radioactive $^{56}Ni$ decay is important in the central region.

\subsection{Abundance stratification} \label{sec:abu}

The abundance stratification technique utilizes the fact that a SN Ia is in homologous expansion ($v_{\rm ph}(r) \,\propto\,r$) from $\sim$10 s after explosion \citep{2005A&A...432..969R}. With a given density profile, time from explosion ($t_{\rm exp}$), photospheric velocity ($v_{\rm ph}$), bolometric luminosity ($L_{\rm bol}$), and abundance structure, optimally fitting synthetic spectra are produced. This approach has been used for many SNe including subluminous and transitional SNe Ia such as SNe\,2005bl, 1986G, 2007on, and 2011iv \citep[e.g.][]{2009MNRAS.399.1238H,2016MNRAS.463.1891A,2018MNRAS.477..153A}, normal SNe Ia  \citep[e.g.][]{Tanaka11,2014MNRAS.445.4427A,2014MNRAS.439.1959M}, as well as stripped envelope SNe \citep[e.g.][]{2017arXiv170204339A,2018MNRAS.478.4162P}. 
A brief description of the code can be  found in Appendix \ref{app:model}.

\begin{figure}
\centering
\includegraphics[trim=0cm 1.5cm 0cm 2cm,clip=true,width=\columnwidth]{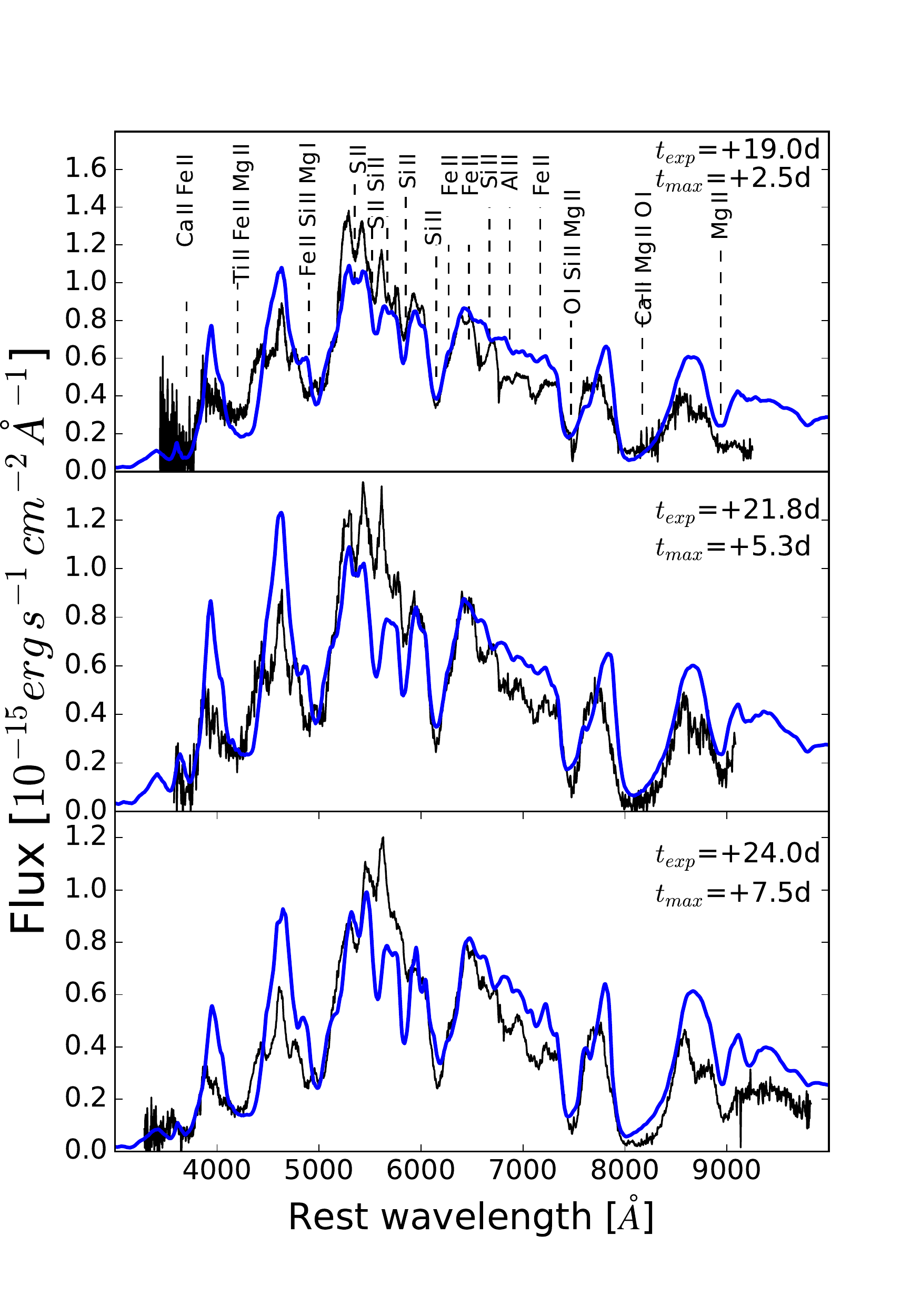}
\caption{A time series of data (black) and spectral models (blue) 
for SN\,2016hnk during the photospheric phase. The data have been corrected for foreground and host galaxy extinction. The observer-frame time from $B$ band maximum and the rest-frame time from explosion is provided for each model. Line identifications from the model are shown in the top plot.} 
\label{fig:photmodel}
\end{figure}

\begin{figure}
\centering
\includegraphics[trim=0cm 0.0cm 0cm 0.0cm,clip=true,width=\columnwidth]{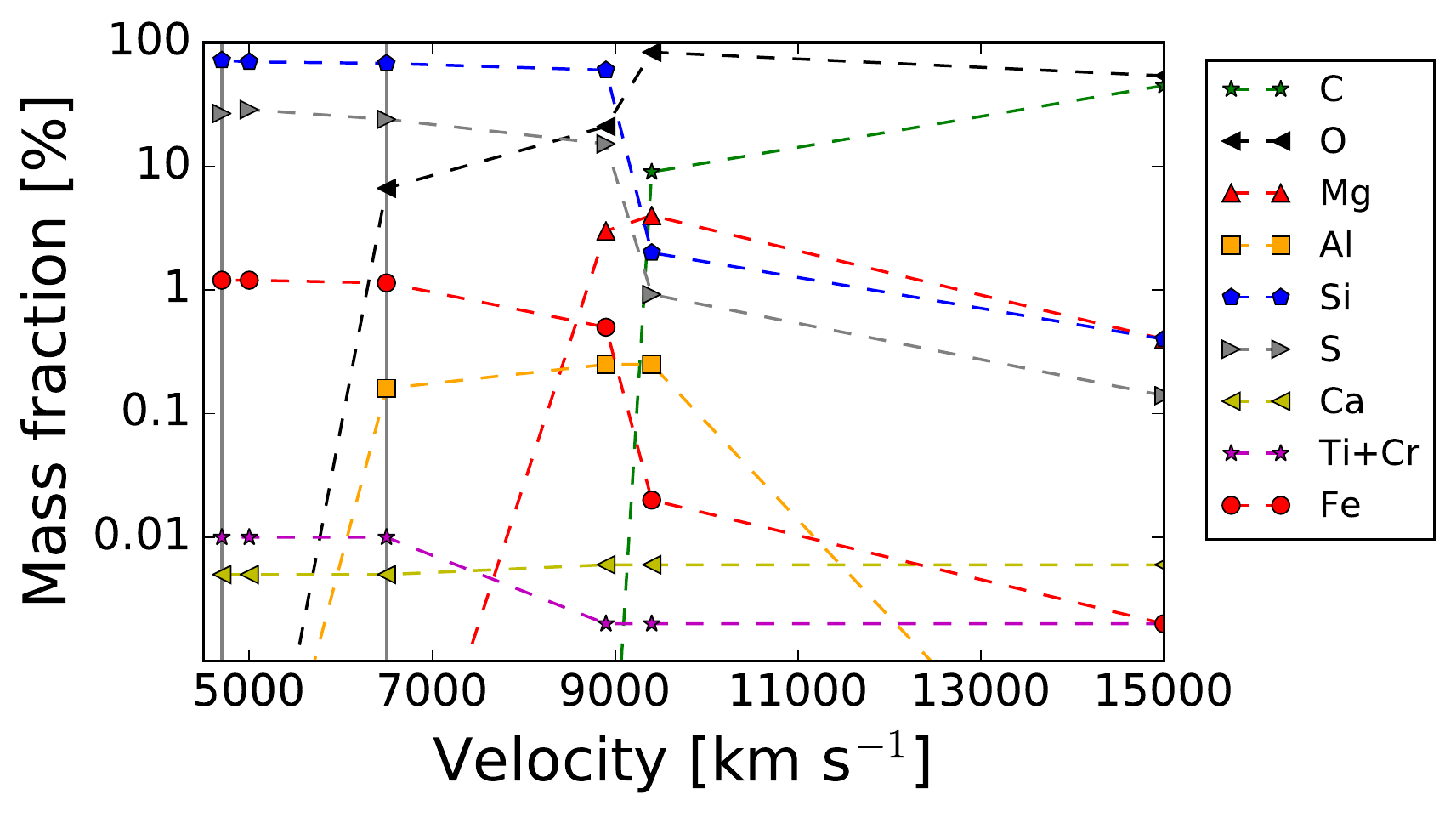}
\caption{The abundance distribution as a function of velocity for SN\,2016hnk. The vertical dashed lines denote the photosphere velocity of the +2.5 and +7.5d models. We note that due to the small time step between our first and last model our results are only sensitive between 4500-10000\kms.}
\label{fig:abustrat}
\end{figure}

\begin{table*}
\centering
\caption{Input parameters for the early-time spectral models.}
\label{table:sn_models}
\begin{tabular}{llllllllllllll} 
\hline
$t_{\rm exp}$&$t_{\rm B_{\rm max}}$&$v_{\rm ph}$&$L_{\rm bol}$&$T_{\rm BB}$&C&O&Mg&Al&Si&S&Ca&Ti+Cr&Fe\\
days&days&\kms&&K&\%&\%&\%&\%&\%&\%&\%&\%&\%\\
\hline
-&-&15000&-&-&45.00&54.00&0.40&0.00&0.40&0.14&6$\times 10^{-2}$&2$\times 10^{-3}$&2$\times 10^{-3}$\\
-&-&9400&-&-&9.00&83.75&4.00&0.25&2.00&0.92&6$\times 10^{-2}$&2$\times 10^{-3}$&2$\times10^{-2}$\\
-&-&8900&-&-&0.00&21.00&3.00&0.25&60.00&15.19&$6\times 10^{-2}$&2$\times 10^{-3}$&0.50\\
19.0 &2.5&6500&8.49& 7900&0.00&6.64&0.00&0.16&68.00&24.00&$5\times 10^{-2}$&1$\times 10^{-2}$&1.14?\\
21.8&5.3&5000&8.48& 8400&0.00&0.00&0.00&0.00&70.00&28.74&$5\times 10^{-2}$&1$\times 10^{-2}$&1.20\\
24.0&7.5&4700&8.44& 7900 &0.00&0.00&0.00&0.00&72.00&26.74&$5\times 10^{-2}$&1$\times 10^{-2}$&1.20\\
\hline
\end{tabular}
\end{table*}

The code requires an input density profile, in the case of \mbox{SN~2016hnk} we used the W7 density profile \citep{1984ApJ...286..644N}. The W7 density profile looks similar to delayed detonation models of sub-luminous SNe~Ia from \citet{2017ApJ...846...58H}. This density profile comes from a fast deflagration explosion of a Ch-mass WD. Spectral models were produced between +2.5 to +7.5 days relative to $B$-band maximum, to obtain the broad structure and chemical composition of the outer layers of the ejecta. We used a rise time of 16.5\,days, which gives the best fit and it is consistent with the value estimated for \mbox{SN~2016hnk} from non-detections in Section \ref{sec:obs}. Figure \ref{fig:photmodel} shows the synthetic spectra and models at three epochs (corresponding to +19.0, +21.8, +24.0 days from explosion). All observed spectra have been first corrected for host galaxy and foreground galactic extinction. The strongest lines contributing to each feature are labeled in the top panel of the plot. The bluest part of the spectra ($<$3900 \AA) is dominated by blends of many Fe group and metal lines, which cause a large amount of line blanketing and reprocesses the flux into the optical. However, the overall red shape of the spectrum is intrinsic and due to \mbox{SN~2016hnk} being less luminous and cooler than a standard object. 

The strongest lines in the $\sim$3900 \AA\ feature are the \CaII\ H\&K lines $\lambda\lambda$3934,3969 with some contribution from \TiII\ $\lambda\lambda$3901,3913. The feature at $\sim$4200 \AA\ is dominated by \TiII\  $\lambda\lambda$4395,4418,4444,4501, \FeII\  $\lambda\lambda$4178,4233,4297,4303,4352,4417,4491, and \MgII\ $\lambda$4481. 
Moving redward the feature at $\sim$5000 \AA\  consists of \FeII\ $\lambda\lambda$4924,5018,5169,5198,5235,5276,5317, \SiII\ $\lambda\lambda$5041,5056, and  \MgI\ $\lambda$5184. The notch at $\sim$5300 \AA\ is produced by \SII\ $\lambda\lambda$5432,5454, and \SiII\ $\lambda\lambda$5466,5467. The absorption at $\sim$5450 \AA\ consists of \SII\ $\lambda\lambda$5606,5639  and \SiII\ $\lambda\lambda$5670,5707, although note that in the data there could be an emission line in this area of the spectra.
The feature at $\sim$5800 \AA\ is produced by \SiII\ $\lambda\lambda$5958,5979, and the small absorption between the two \SiII\ features is made by \FeII\ $\lambda\lambda$6147,6150,6238.
The absorption at $\sim$6200 \AA\ is produced by \SiII\ $\lambda\lambda$6347,6371, with the unusual red side of the feature being dominated by \FeII\ $\lambda\lambda$6456,6516, which gives it a broader asymmetrical shape. 
This is seen in most sub-luminous SNe Ia (see Figure \ref{fig:comp}). However, for 
brighter objects, where the Si region is further out in velocity space, this \FeII\ $\lambda$6516 line is on the emission component  of  the \SiII\ $\lambda$6355 feature, as the velocity of the Fe and Si-rich regions may not scale in by the same factor. In the latter case it may be easy to misidentify this feature as \CII.

Between 6200 and 7700 \AA\ there are a number of small absorption features not typically seen in brighter SNe Ia. Many of these are low excitation \FeII\ lines that are produced at lower temperatures, which is consistent with the low luminosity of \mbox{SN~2016hnk}. The strongest lines in the models between the \SiII\ 6355\AA\ feature and the \OI\ $\lambda$7774 features are \FeII\  $\lambda\lambda$6456,6516,7308,7462,7711, \SiII\ $\lambda\lambda$6818,6830, and \alII\ $\lambda\lambda$7042,7057\footnote{Note that we doubt our the \alII\ line identification here as it requires an abundance that is larger than what is predicted by explosion models.}. The feature at \ab7770 \AA\ is dominated by \OI\ $\lambda\lambda$7772,7774,7775 with strong contributions from \SiII\ $\lambda\lambda$7849,7850, and \MgII\ $\lambda\lambda$7877,7896,7896. The deep feature at 8200 \AA, often attributed to only \CaII, is in fact produced by \CaII\ $\lambda\lambda$8498,8542,8662, \OI\ $\lambda\lambda$8446,8446,8447, and \MgII\ $\lambda\lambda$8213,8235. It is important to note that the blue side of this feature consists of \MgII, \CaII, and \OI, however  \MgII\ is not one of the main sources of opacity in the feature and its contribution is minor.
Finally the absorption at 8900 \AA\ is produced by \MgII\ $\lambda\lambda$9218,9244. Throughout the spectra modeled in Figure \ref{fig:photmodel} the line identification does not change significantly, it is only the strength of each line that varies. 

Table \ref{table:sn_models} shows the abundances as a function of velocity that were used to compute the synthetic spectra. We have also provided the abundances in the shells outside the formation region of the first spectrum. These shells are required to produce the stratified abundances and to match the observations. For example, if we assume the abundances in the 6500 \kms\ shell are the same as the very outer layer, the models would contain high velocity Si absorption. This would be caused by the ejecta at high velocities containing 68\%  Si. 

In the first epoch the modeled $v_{\rm ph}$ is 6500 \kms, the abundances near the photosphere are dominated by Si (\ab68\%), S (\ab24\%), and C+O (7\%), with Ca, Ti, Cr, V, and Fe making up the remaining 1\%. The next epoch has a  $v_{\rm ph}$ of 5000 \kms, the Si and S abundances have increased slightly (\ab70\% and \ab29\%, respectively). This model has no C or O at the photopshere, with  Ca, Ti, Cr, V, and Fe making up the rest of the mass. By the third epoch the $v_{\rm ph}$ is 4700 \kms\ and the Si abundance is \ab72\%, S is \ab27\% and the remaining mass consists of Ca, Ti, Cr, V, and Fe. For a summary of errors using this method see \citet{2008MNRAS.386.1897M}. 

The derived abundance distribution obtained through spectral modeling is consistent with a low luminosity delayed-detonation SN Ia. Our models are sensitive out to velocities of \ab15000 \kms. Intermediate mass elements dominate the spectra at the epochs observed, however the presence of significant  O is required above 9000 \kms. It can be assumed that the bulk of the \Nifs\ is located below our innermost shell. There could be a small abundance of \Nifs\ above the innermost photosphere, but it is not large enough to be optically thick. There is less effective burning in this SN explosion compared to a more luminous event. 

In these models at +24 d past explosion there are 0.8\Msun\ above the photosphere.
As there is a large amount of material below the photosphere, it is not possible for us to constrain the total mass of the explosion. However the velocity of the $H$-band break (see above) and the nebular phase spectrum will help, see Section \ref{sec:photNLTE}. 

A comparison of Figure \ref{fig:abustrat} to the abundance distributions of other transitional and sub-luminous SNe Ia show similar results. 
However, for SN 2016hnk we find no evidence of \Nifs\ in our photospheric models, whereas in the sub-luminous SN~2005bl \cite{2009MNRAS.399.1238H} found \Nifs\ out to $\sim$7000\kms.
A comparison to the abundance distribution to the transitional SN~1986G \citep{2016MNRAS.463.1891A} shows similar results to SN\,2016hnk, with the only discrepancy being that SN~1986G has \Nifs\ located out to 5500\kms. As discussed below, we note that our abundance distribution is similar to the delayed detonation model of SN\,1999by of \citet{2002ApJ...568..791H}.

\subsection{Explosion models photospheric phase} \label{sec:photNLTE}

We calculate the final phase of the thermonuclear runaway, the explosion, LCs and spectra with our code HYDRA \cite{2017ApJ...846...58H}, and references therein. Here, we use delayed detonation models \cite{1989MNRAS.239..785K} in which the explosion starts as a slow laminar deflagration. The free parameters are the main sequence mass $M_{MS}$ and metallicty $Z$ of the progenitor, the central density of the explosion $\rho_c$. The flame starts as laminar deflagration  and transitions to a detonation at a density $\rho_{tr}$ which, in spherical geometry is equivalent to the amount of burning during the deflagration phase. 
To setup the explosion, we calculate the final phase of the accretion towards the thermonuclear runaway with our code but neglect the second derivatives to allow for large time steps similar to \cite{1982ApJ...253..798N}.
Here, the constant accretion rate is adjusted so that the thermonuclear runaway occurs at a given central density $\rho_c$ \citep{2017ApJ...846...58H}. 
To first order, the explosion is triggered when compressional heating due to increasing mass by accretion from the companion star cannot be balanced by heat conduction and neutrino cooling (including the convective URCA process).  Close to the AIC, the detailed parameters such as the accretion rate depends sensitively on the equation of state and neutrino cooling, and the history of the binary evolution (see final discussion and conclusions).

As mentioned in Sect. 6.1, \mbox{SN~2016hnk} has many observational characteristics similar to the subluminous SN 1999by. To produce sub-luminous SNe~Ia in the framework of DDT models, the pre-expansion of the WD needs to be sufficiently large during the deflagration phase, that the density has dropped enough so that little $^{56}Ni$ is produced during the subsequent detonation phase, and we need a partial suppression of Rayleigh-Taylor (RT) mixing \citep{2002ApJ...568..791H,2017ApJ...846...58H,2018ApJ...858...13H}. The model for SN1999by had $\rho_c=2\times10^9 g/cm^3$, $\rho_{tr}=8 \times 10^6 g/cm^3$, and $MS = 5M\odot$, $Z=Z_\odot$.  Our tomography approach for the early spectra resulted in an abundance structure that closely resembled the one of the delayed-detonation model for SN1999by which shows that the NSE elements are within $\approx 4500~km/sec$, followed
by a Si/S-rich layer and a transition to layers of explosive carbon burning (Mg/O) at $\approx 9000 km/sec$.

\begin{figure}
\centering
\includegraphics[trim=0cm 0.5cm 0cm 0cm,clip=true,width=\columnwidth]{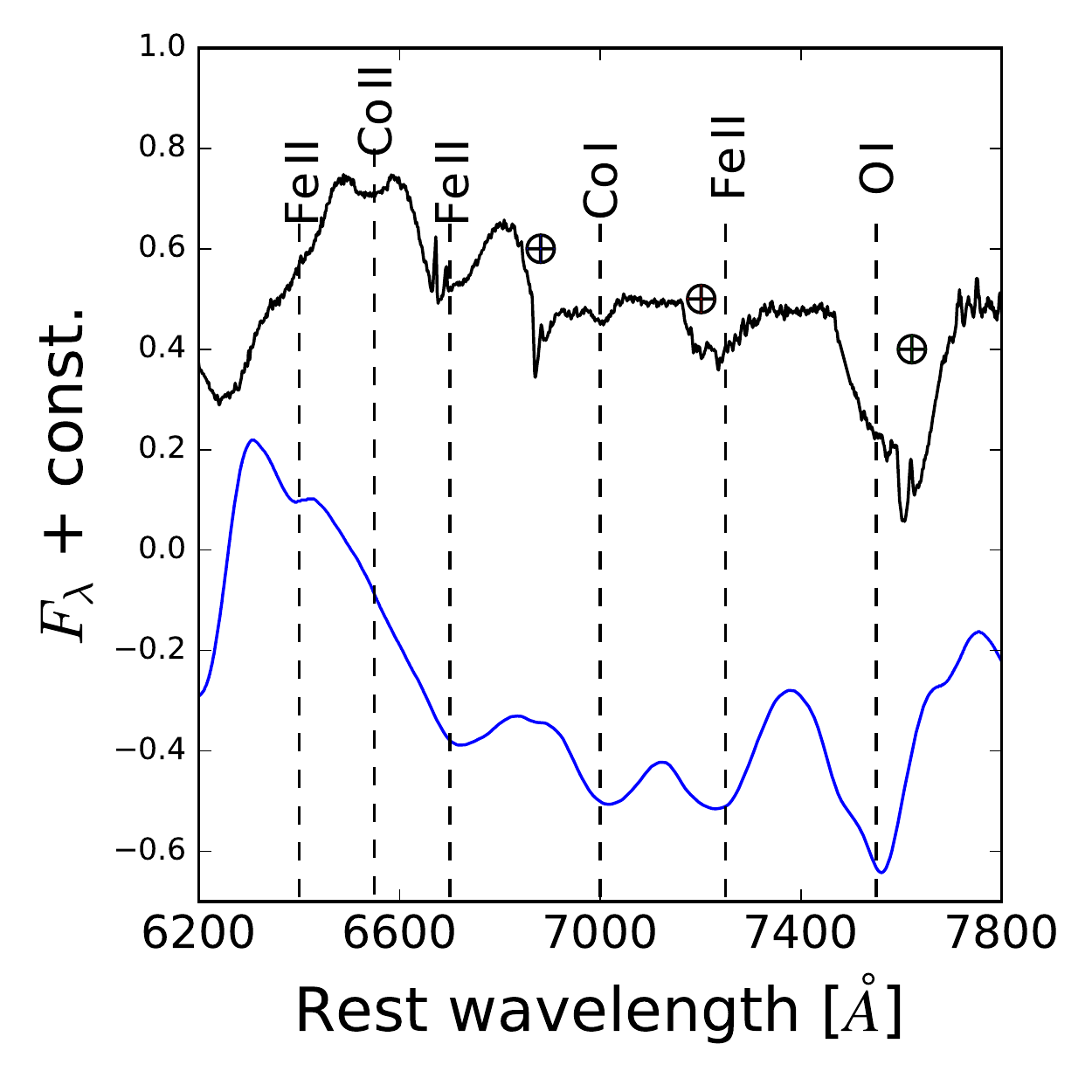}
\caption{Comparison of SN\,2016hnk (black) +2.5\,d past maximum light and the non-LTE model of SN\,1999by at +7\,days past maximum light (blue) from \citet{2002ApJ...568..791H}. The telluric features are marked in the observed spectra. }    
\label{fig:diagrams3}
\end{figure}

NIR observations provide a useful constraint on the explosion because its sensitivity to whether the photosphere is formed in a mostly C/O rich mixture, IME elements, or NSE elements. As such, it provides a useful probe for the time of the explosion, and the explosion model.
In Figure \ref{fig:sp_nir} a NIR spectrum of \mbox{SN~2016hnk} is compared to the non-LTE delayed-detonation model of SN 1999by at 26 days past explosion, corresponding to about 11 days past maximum.
Overall, the observed and theoretical spectra agree. 
The theoretical spectrum is dominated by blends of singly ionized Si, Ca, Fe, and Co. The strongest features are \ion{Fe}{ii} ($\approx 1.02 \mu m$), \ion{Si}{ii} ($\approx 1.20 \mu m$), \ion{Fe}{ii} ($\approx 1.06 \mu m$), \ion{Ca}{ii}, \ion{Fe}{ii} and \ion{Co}{ii} ($1.16 \mu m$), \ion{Co}{ii} and \ion{Fe}{ii} ($1.15 $ and $1.38  \mu m$), as well as a multiplet of \ion{Fe}{ii}/\ion{Co}{ii}/\ion{Ni}{ii} which is weak but appears in the  $H$-band region.
Note that the feature at $0.98 \mu m$ is a \ion{Fe}{ii} dominated blend of \ion{Ca}{ii}/\ion{Fe}{ii}/\ion{Co}{ii}. In the model it is more defined or in the observation the suppression of flux is stronger.
Because it is formed in a steeply declining wing, it is ambiguous whether the metallicity, $Z$, of \mbox{SN~2016hnk} is significantly larger than $Z_\odot$ in the model leading to a suppression of flux and a widening of the adjoining feature to form a shoulder, or if $Z$ is significantly smaller than $Z_\odot$ resulting in a narrower feature. 
This strongly places \mbox{SN~2016hnk} into the realm of the sub-luminous 1991bg-like SNe Ia and emphasizes the NIR as a chronometer for the time since maximum because of  the rapid spectral changes. The time of this model is consistent with our lightcurve-based estimates for maximum light, as well as the abundance stratification models.   

The features between 6000--7500 \AA\ in the photospheric phase spectra are not normally seen in bright SNe Ia.  
With the abundance stratification models
we produce many of these lines but at different flux levels. 
Therefore, to see if our line identifications are correct, 
we compare SN\,2016hnk to the full non-LTE models of SN\,1999by from \citet{2002ApJ...568..791H}. 

To conclude the photospheric section, the element identification should be revisited in light of the explosion based non-LTE models.
Figure \ref{fig:diagrams3} shows the +2.5$\pm$3.3\,d spectra of SN\,2016hnk and the non-LTE model of SN\,1999by at +7\,d in the wavelength region 6200-7800 \AA. Note that we have limited our analysis to the published spectra from \citet{2002ApJ...568..791H}. We note that in the model at +2\,d these lines have not yet formed, but given the error on the time of maximum from the observations and the quick evolution of SN\,2016hnk, we speculate that these lines begin to emerge in the models of SN\,1999by between +2 and +7\,d, which is consistent with the observations and NIR models. 
The model and data have similar spectral features, shown with the vertical dashed lines. The line identification is mainly consistent with the abundance stratification models, where  \FeII\ lines dominate the spectra in the  6000--7500 \AA\ region. 

Some of the other features in the 6200-7800 \AA\ region are \FeII\ $\lambda$6664 and \CoI\ $\lambda$7084. 
The one feature which cannot be explained by the models is the small absorption at 6550 \AA.
However, the \citet{2002ApJ...568..791H} models assume no mixing.   With a small amount of mixing 
it could be expected that this feature is caused by \CoII\ $\lambda\lambda$6571,6576, as mixing will enhance the Co abundance in these layers. In the wavelength region 6000--7500 \AA\ we favour the line identifications from the non-LTE models, as these solve for atomic level populations explicitly and are produced through full radiation-hydrodynamical simulations and not a parameterized to fit the data.


\subsection{Explosion models for the nebular phase} \label{sec:modnebular}

Due to the similarity of SN\,2016hnk to SN\,1999by, we have adapted the  previous $M_{\rm Ch}$ mass spectral simulations of  SN\,1999by \citep{2002ApJ...568..791H,2017ApJ...846...58H} to produce models for SN\,2016hnk. 
In SN\,2016hnk an interesting observation in the nebular spectrum is the appearance of a single, narrow dominating spectral feature at $\approx$7200 \AA~that has a Doppler width of $\approx 1000$ \kms. 
This is very different from late-time spectra of normal bright SNe Ia that are dominated by iron-group elements.
Because the delayed-detonation model for SN\,1999by does not produce narrow features on less than a  $2000-3000$ \kms\ scale, the inner region of the original model needs to be modified.  

We note that, in principle, multidimensional effects such as Rayleigh-Taylor (RT) instabilities produce plumes with  an individual velocity spread on the scale of $\approx 1000$ \kms. However they produce multiple plumes which would make the total velocity spread too large to explain the observation of SN\,2016hnk. Moreover, various observations and analyses strongly suggest that the RT instabilities are present but need to be partially suppressed \citep{h06,fesen15,2017ApJ...846...58H,2018ApJ...861..119D,2018ApJ...858...13H}.
Therefore, in this work spherical delayed-detonation models are used, and we focus on nuclear physics based effects in combination with progenitor parameters to explain the observations of SN\,2016hnk. 
The hydrodynamical simulations are in the comoving (mass) frame to avoid numerical advection of abundances, which may affect both nuclear burning and small scale structures.
The simulations utilize a nuclear network of 315 isotopes during the early phase of the explosion based on the nuclear reaction network using the library REACLIB \citep{2010ApJS..189..240C} and weak rates from \cite{2014PhRvC..89d5806M}, \cite{2003RvMP...75..819L}, and \cite{1985ApJ...293....1F}. We use detailed, time-dependent non-LTE models for atomic level populations, including $\gamma$-ray and positron transport and radiation-hydrodynamics (RH; \citealt{1995ApJ...443...89H,2003hydra,2014ApJ...795...84P,2017ApJ...846...58H}). 
After an initial deflagration phase, a detonation is triggered when the density at the flame front drops to a transition density, $\rho_{\rm tr}$, which is a parameterization of the amount of deflagration burning. 

The model presented originates from a progenitor with a main sequence mass (M$_{\rm MS}$) of $7~M_\odot$ with solar metallicity.
The models of SN\,2016hnk require a higher central density ($\rho_{c}=6 \times 10^9$ g cm$^{-3}$) than that of SN\,1999by, which implies a larger binding energy, and to compensate for the longer phase to lift the WD, we used a slightly higher $\rho_{\rm tr}= 10^7$ g cm$^{-3}$. The rate of subluminous SNe~Ia diminish at high redshifts indicating a long evolutionary time scale \citep{2011ApJ...727..107G}. Thus, we assume an increased $^{22}$Ne/$^{20}$Ne ratio ($\approx$ factor of 2) in the center of the WD due to gravitational settling over billions  of years \citep{2002ApJ...580.1077D}. Note that most of the Ne is inherited from the formation of the progenitor star in the  molecular cloud. In effect, our model starts at a slightly lower $Y_e$ compared to models without gravitational settling. The front starts with a laminar
speed. To increase the electron capture, we assume a slow deflagration front ($C1=0.1 $ of Eq. 1 in \citealt{2000ApJ...528..854D}).

\begin{figure}   
\includegraphics[angle=360,width=\columnwidth]{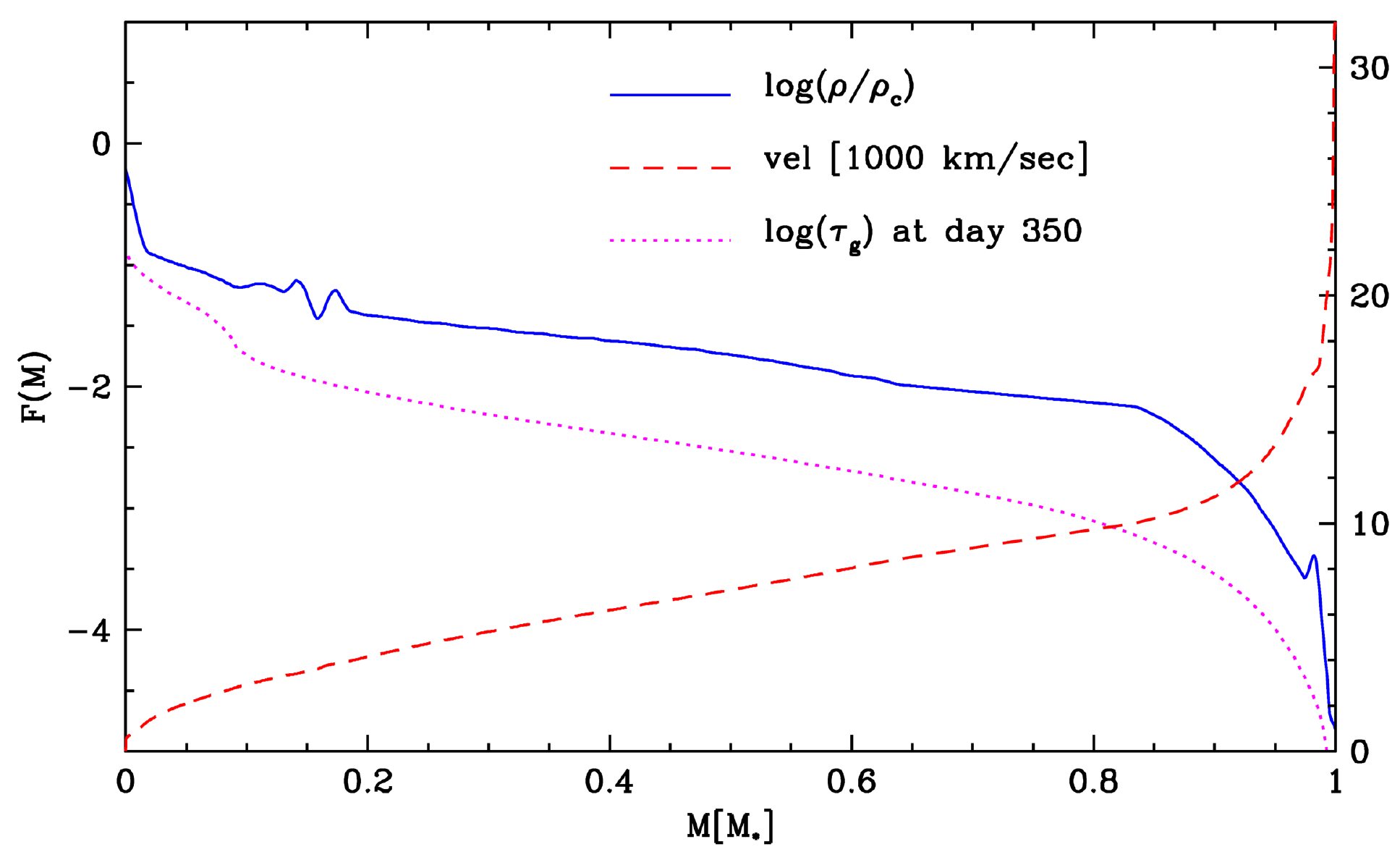}
\caption{Structure during the phase of homologous expansion as a function of the mass coordinate in WD-masses for our subluminous model with high central density. We give the normalized density as $\log(\rho/\rho_c)$ (left scale), velocity (right scale) and the mean optical depth 
$\log(\tau _\gamma)$ (left scale) in $\gamma$-rays at day 350.}
\label{models}
\end{figure}  

\begin{figure}   
\includegraphics[trim=0cm 0cm 0cm 0cm,clip=true,angle=360,width=\columnwidth]{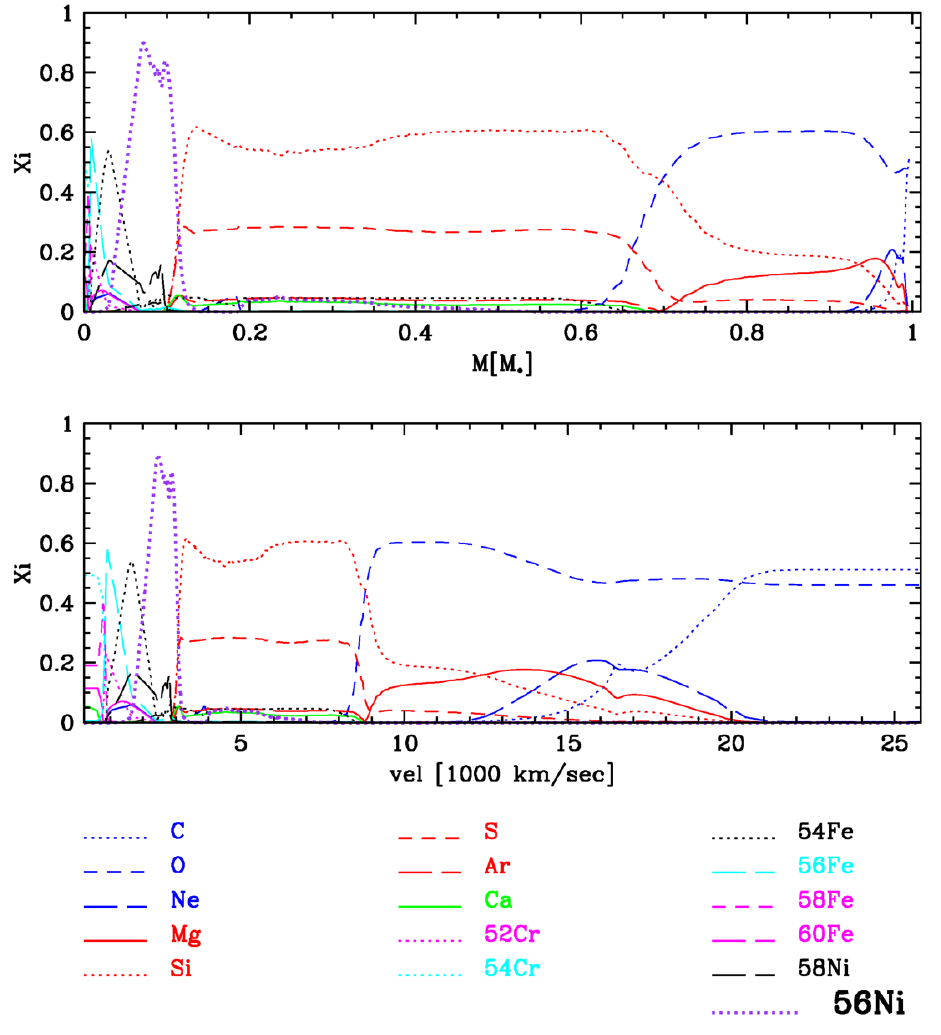}
 \caption{Same as Figure \ref{models} but showing the overall abundance structure at large times as a function of mass coordinate (top) and  expansion velocity (bottom).
It is very similar to the subluminous model for SN 1999by and consistent with the chemical structure based on our abundance tomography given in Table 3. In addition, we show the initial \mbox{$^{56}$Ni $\rightarrow$ $^{56}$Co $\rightarrow$ $^{56}$Fe} at the time of the explosion (purple dotted line).}
\label{model}
\end{figure}  

The density and velocity structure of the model during the homologous expansion, after $\approx$10 days, is shown in Figures \ref{models} and \ref{model}. The density structure is very similar to the model used for SN~1999by in most respects. As in \cite{2002ApJ...568..791H} models and also in W7 \citep{1984ApJ...286..644N}, a local maximum in the outer density structure, $M \approx 0.98 M_{\rm WD}$, marks the transition from burned to unburned material. The feature at $M \approx 0.15 M_{\rm WD}$ marks the imprint of the refraction wave during the deflagration to detonation transition, which is inherent to spherical models. The increased density in the central region of the explosion\footnote{Note this is different from the central density of the exploding WD.} is not present in the \cite{2002ApJ...568..791H} models, as it is caused by radiation hydrodynamical effects. 
After the explosion, the envelope enters the phase of `almost' free expansion on time scales of seconds to minutes. Adiabatic cooling results in a rapid drop of temperature in the envelope and an overall free expansion. 
This explains the similarity in the outer layers between the new model and the old SN\,1999by model.
However, the energy production by radioactive $^{56}$Ni decay corresponds to a change of the specific energy equivalent to a velocity of $\approx$ 3000 km s$^{-1}$. This effect changes the density structure of low-velocity layers because the `dynamics' is dominated by pressure equilibrium. In models with lower $\rho_c$, the central region becomes partially optically thin for $\gamma$-rays and low energy photons. 
Thus, the entire inner region is affected by the heating. However, in our high-density model with $\rho_c = 6 \times 10^{9}$ g cm$^{-3}$, the average $\gamma$-ray optical depth $\tau_g$ stays larger than 2$-$3, even at twice the decay time of $^{56}$Ni, leading to a compression of layers with $\lesssim$1,500 km s $^{-1}$ (Figure \ref{models}). Note that these layers are well inside the $^{56}$Ni layers as discussed below. However, this increase will have severe effects on the $\gamma$-ray heating, because the central  $\tau_g$ is an order of magnitude larger than the rest of the envelope even at +350 d. In part, this is responsible for the late-time spectral formation discussed below.

\begin{table}\small
\centering
\caption{Asymptotic isotopic yields at large times for the delay-detonation model for \mbox{SN~2016hnk} discussed in Section \ref{sec:photNLTE}. The mass is given in units of M$_{\rm Ch}$. }
\label{table:sn_isotopes}
\begin{tabular}{llllll} 
\hline
    $^{12}$C    &  8.109$\times 10 ^{-3}$   &   $^{16}$O    &  1.891$\times 10 ^{-1}$   &   $^{20}$Ne    &  7.443$\times 10 ^{-3}$     \\
    $^{22}$Ne    &  1.351$\times 10 ^{-4}$  &  $^{23}$Na    &  7.526$\times 10 ^{-5}$   &   $^{24}$Mg    &  3.371$\times 10 ^{-2}$ \\
         $^{25}$Mg    &  9.375$\times 10 ^{-5}$ & $^{26}$Mg    &  1.273$\times 10 ^{-4}$   &   $^{27}$Al   &  1.503$\times 10 ^{-3}$  \\
    $^{28}$Si    &  3.943$\times 10 ^{-1}$  & $^{29}$Si    &  1.492$\times 10 ^{-3}$  &    $^{30}$Si    &  3.393$\times 10 ^{-3}$    \\ 
      $^{31}$P    &  7.924$\times 10 ^{-4}$& $^{32}$S    &  1.669$\times 10 ^{-1}$    &   $^{33}$S    &  5.248$\times 10 ^{-4}$    \\  
     $^{34}$S    &  3.415$\times 10 ^{-3}$ & $^{35}$Cl    &  2.136$\times 10 ^{-4}$   &   $^{37}$Cl    &  3.536$\times 10 ^{-5}$ \\
     $^{36}$Ar    &  2.413$\times 10 ^{-2}$  & $^{38}$Ar    &  1.477$\times 10 ^{-3}$ &      $^{39}$K    &  1.217$\times 10 ^{-4}$ \\   
  $^{40}$Ca    &  1.597$\times 10 ^{-2}$ & $^{42}$Ca    &  3.616$\times 10 ^{-5}$     & $^{48}$Ca    &  8.148$\times 10 ^{-4}$   \\
     $^{46}$Ti    &  2.200$\times 10 ^{-4}$ &$^{48}$Ti    &  2.358$\times 10 ^{-4}$   &   $^{49}$Ti    &  1.098$\times 10 ^{-4}$ \\ 
   $^{47}$V    &  2.826$\times 10 ^{-3}$&  $^{48}$V    &  2.076$\times 10 ^{-4}$      & $^{51}$V    &  1.183$\times 10 ^{-2}$   \\
       $^{50}$Cr    &  3.032$\times 10 ^{-3}$ &$^{52}$Cr   &  3.272$\times 10 ^{-3}$    &  $^{53}$Cr    &  2.318$\times 10 ^{-3}$   \\
       $^{55}$Mn    &  7.401$\times 10 ^{-4}$  &$^{54}$Fe    &  1.572$\times 10 ^{-3}$   &   $^{56}$Fe    &  1.045$\times 10 ^{-1}$  \\   
        $^{57}$Fe    &  1.710$\times 10 ^{-3}$ & $^{58-64}$Ni   &  1.170$\times 10 ^{-2}$\\
\hline
\end{tabular}
\end{table}

\subsubsection{Ca line in the late-time spectrum}

The main effect on $Y_e$ is the density of burning. A central density of $6 \times 10^9$ g cm$^{-3}$ is required to reduce $Y_e$ to values that produce centrally located Ca as observed.
With regard to the electron capture, the settling of Ne and slow deflagration phase has the same effect of an increased $\rho_c$ by $\approx 10\%$. The transition to a detonation is triggered at $\rho_{\rm tr} = 10^7$ g cm$^{-3}$ to compensate for the larger binding energy of the initial WD. The WD has a final explosion energy $ E{_{\rm kin}}=8.83 \times 10^{50}$ erg and produces $M({\rm ^{56}Ni})= 0.108 M_\odot$. The overall density and chemical structure is given in Figure \ref{model}, and the element production is shown in Table \ref{table:sn_isotopes}. The overall chemical structure resembles the \cite{2002ApJ...568..791H} model for SN~1999by and, more importantly, the results of the abundance tomography analysis (Table \ref{table:sn_models}): a) the unburned C-O layers reach down to about 15,000 km s$^{-1}$; b) products of explosive carbon burning (O-Ne-Mg) are seen down to $\approx$9500 km s$^{-1}$; c) followed by incomplete burning (Si-S) down to about 3500 km s$^{-1}$.

\begin{figure}   
\includegraphics[angle=360,width=\columnwidth]{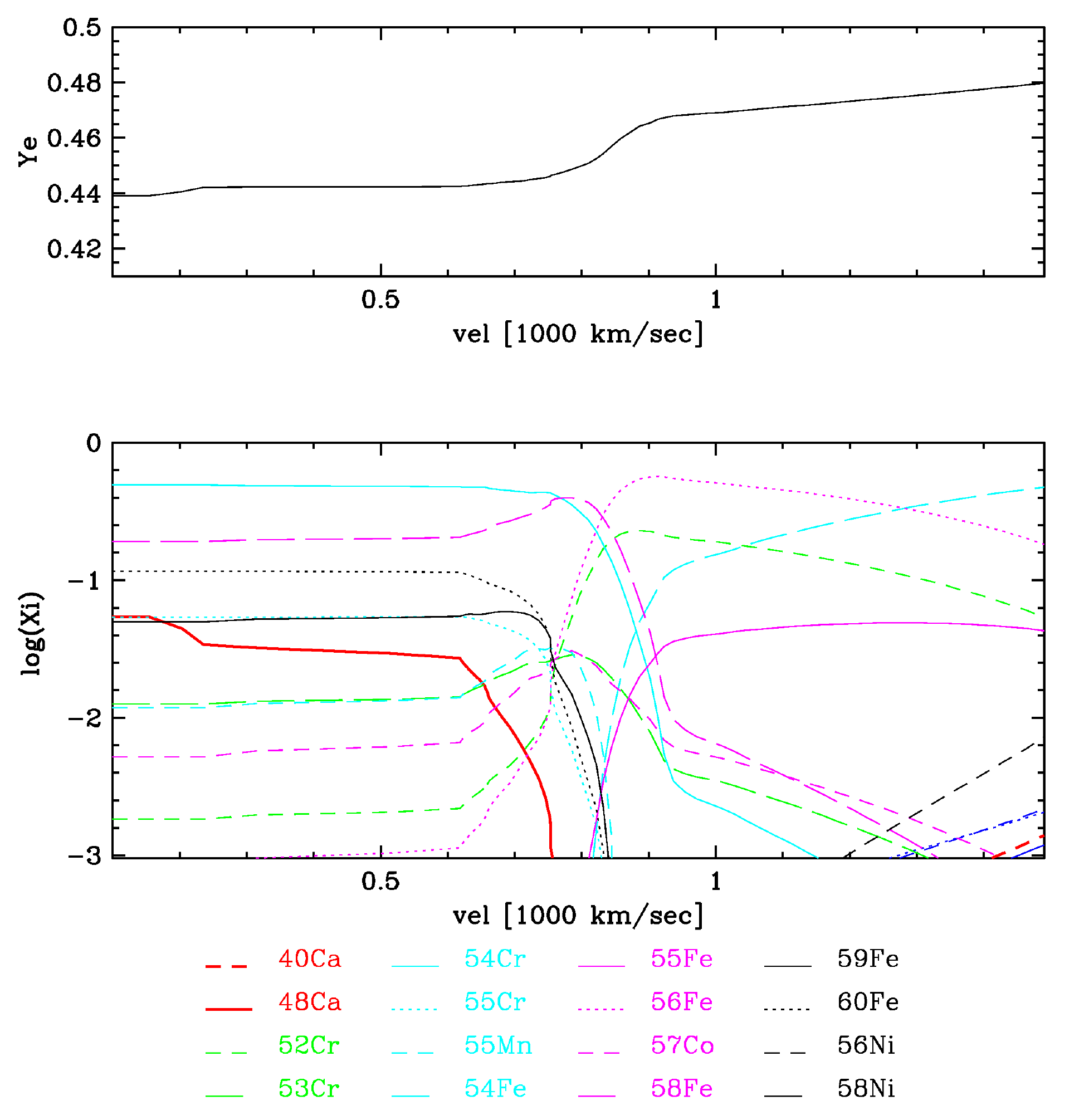}
 \caption{Distribution of most abundant isotopes in log($X_i$) and $Y_e$ as a function of velocity for the central region are dominated by neutron-rich isotopes at several minutes after the explosion. Subsequently, radioactive isotopes decay  $^{55}$Cr (3.5 m) $\rightarrow$ $^{55}$Mn, $^{55}$Fe(2.7 y) $ \rightarrow ^{55}$Mn, $^{59}$Fe (44.5 d) $\rightarrow ^{59}$Co, $^{56}$Ni (6 d) $\rightarrow$ $^{56}$Co (77 d) $\rightarrow ^{56}$Fe, $^{57}$Co (271 d) $\rightarrow$ $^{57}$Fe. }
\label{model2}
\end{figure} 

\begin{figure}   
\includegraphics[width=\columnwidth]{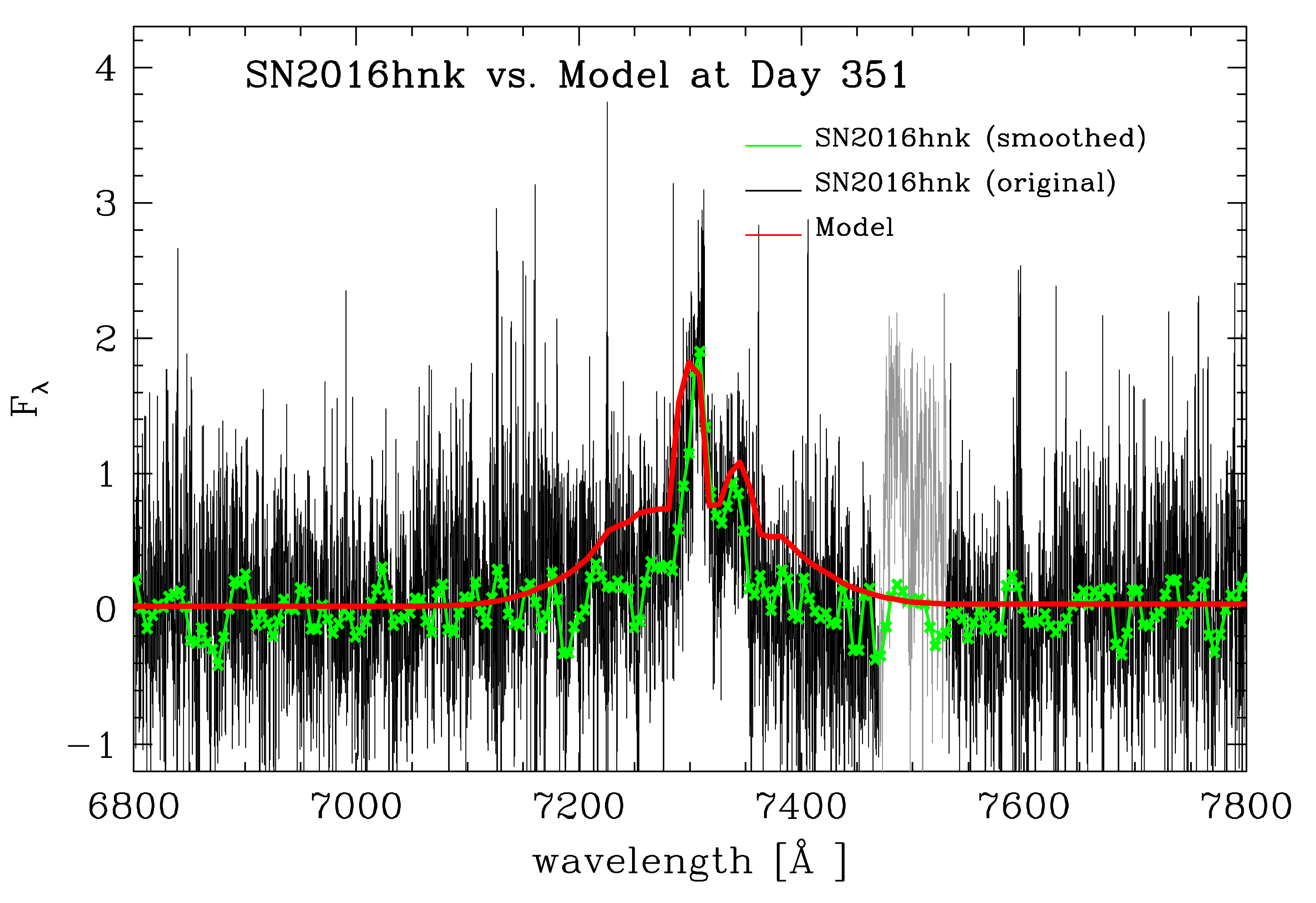}
 \caption{Comparison of the nebular spectrum of \mbox{SN~2016hnk} with our high-density delayed-detonation model in the rest frame assuming a redshift of $z =0.01610$. The flux is in $10^{-18}$ erg s$^{-1}$ cm$^{-2}$ on the same scale as Figure \ref{fig:xshooter}. The telluric region in the original spectrum is plotted in gray, around 7,500\AA. The theoretical spectrum is dominated by a narrow, forbidden [\ion{Ca}{ii}] doublet at  7291 and 7324\AA~above a quasi-continuum at about the 3\% level formed by many transitions of mostly heavier elements including Fe, Co, Ni, Cr, etc.}
\label{ltmspec}
\end{figure}  

In the inner region, electron capture shifts the nuclear statistic equilibrium from $^{56}$Ni to neutron rich isotopes, as the $Y_e$ decreases towards the central region, see Figure \ref{model2}. The decrease is from a $Y_e$ of 0.497, which produces  $^{56-58}$Ni, $^{55}$Co, $^{54}$Fe, to a $Y_e$ of $0.43-0.44$, which produces $^{58-60}$Fe, $^{54-56}$Mn, $^{52-56}$Cr and most importantly $^{47-48}$Ca.   

In Figure \ref{ltmspec}, the comparison between the theoretical and observed late-time spectrum is given. The observation shows narrow emission from the forbidden [\ion{Ca}{ii}] doublet at  $\lambda\lambda$7291,7324 whose lower level is the ground-state of \ion{Ca}{ii}. 
This narrow [\ion{Ca}{ii}] doublet dominates with respect to the other features because most of the energy emitted in Ca is confined to a very narrow region of central Ca powered by radioactive decay. 
The strength of these other features is only up to the 5\% of the Ca doublet and are lost below the noise level.
The central energy deposition is large because that region is still optically thick to $\gamma$-rays, whereas $\gamma$-rays  from the outer layers mostly escape (Figure \ref{models}). The underlying continuum is formed by many overlapping allowed and forbidden transitions of mostly Cr, Mn, and  Fe group-elements, which distribute the energy over a wide wavelength range from the UV to the mid-IR and produce many lines commonly seen in SNe~Ia, but on much lower levels of about $\sim$3\% relative to the strength of the dominant narrow Ca feature. 
Note that the density at the inner region is still high at $\approx 3 \times 10^7$ particles cm$^{-3}$. Thus, forbidden lines dominate for Ca for transitions to the ground state but not for the more complex ions. 
In the model, the central feature is formed by $^{48}$Ca with expansion velocities $v_{\rm exp}$ up to about 700 km s$^{-1}$ whereas a shoulder is the result of $^{40}$Ca of the region expanding at $v \approx$ 3000-9000 km s$^{-1}$. 
Within the scatter in the observed spectrum (black, Figure \ref{ltmspec}), the synthetic spectra agree with the observations.

The comparison with the smoothed spectrum (green, Figure \ref{ltmspec}) shows some discrepancies, namely the hint of the $^{40}$Ca shoulder in the models as just discussed and, in particular, a Doppler shift of the central Ca feature by about 700 km s$^{-1}$. Possible factors that could contribute to this offset and that can be discriminated by future observations of high-resolution spectra include a peculiar motion of the SN progenitor within the galaxy or a passive drag of pre-existing turbulent fields during the smoldering phase \citep{2002ApJ...568..779H,2004A&A...419..623Y,2007ApJ...656..333Z,2011ApJ...740....8Z} during the 2-3 seconds of a typical deflagration phase. The former would show variations but would be correlated with the rotation of the host galaxy, while the latter would show variations between objects.
We note that the  smoothed spectrum shows structure on the 10\% level that is not predicted, but is also not at a significant level. The strong jump at $7500$ \AA\ is telluric.


\section{Discussion and conclusions} \label{sec:conc}

We presented photometric and spectroscopic observations of \mbox{SN~2016hnk} and IFS of its type SB host galaxy \mbox{MCG~-01-06-070}. 
For the first time, these data are combined with a unique late-time spectrum at $\approx$360\,d and with IFS data providing a unique link to the progenitor and environment of sub-luminous SNe~Ia and, possibly, their nature.
The IFS places \mbox{SN~2016hnk} close to a region with significant H$\alpha$ emission and with high dust content near the nucleus of \mbox{MCG~-01-06-070}. For the ISM, we consistently found $E(B-V)\sim0.4$ mag with three different methods and a low $R_V=2.1 \pm 0.6 $ typical of highly extinguished SNe~Ia \citep{2014ApJ...789...32B}. 

The \mbox{SN~2016hnk} light curve was most similar to 2002es-like SNe Ia up to +15 days past maximum, 
however, optical and NIR spectra of SN\,2016hnk during the photospheric phase show that the SN closely resembles the prototypes for subluminous SNe~Ia, namely SNe 1991bg and 1999by.
The $\Delta$m$_{15,s}$(B) of 1.80$\pm$0.20\,mag and $s_{\rm{BV}}$ of 0.438$\pm$0.030 confirms the similarity between SN\,2016hnk and subluminous SNe~Ia. 
Based on detailed analyses of the optical and NIR light curves, we detect a flux excess after $\approx$15 d caused
by scattering processes with an interstellar medium at around 1.0 pc from the SN, suggesting that the subluminous SN\,2016hnk is embedded in a dense environment.
We note that the extinction by dust is significantly larger in our object than in SNe\,1991bg, 1999by and 2005ke \citep{1999A&A...345..211L,2002ApJ...568..791H,2012A&A...545A...7P,2014ApJ...795...84P}.
The under-representation of sub-luminous SNe~Ia in high-redshift samples is usually attributed to long progenitor evolution, but may also be magnified by subluminous SNe Ia occurring in dusty regions or at low galactocentric distances \citep{2012ApJ...755..125G}, or simply be attributed to selection effects \citep{2011ApJ...727..107G}.

The early-time spectra of SN\,2016hnk have unique absorption features between 6000-7000 \AA, which we identified as low excitation \FeII, \CoII\, and \CoI\ lines. Furthermore, it was shown that the line on the redward side of the \SiII\ 6355\AA\ feature is not produced by \CII\ but rather by \FeII\ $\lambda$6516. 
In the instance of sub-luminous SNe Ia the ionization state of carbon at the epochs studied would be neutral and carbon would not be present in the optical spectrum, furthermore it could be expected to be further out in velocity space and not visible at these epochs.
This \FeII\ line has also been seen by a few groups for normal bright SNe Ia (see \citealt{1995ApJ...443...89H}) and with the PHOENIX code (E. Baron private communication).  

We performed a detailed comparison of the optical spectra to spectral abundance stratification/tomography. The results are consistent with a $M_{\rm Ch}$ mass WD with characteristics found in a previous analysis of SN\,1999by based on delayed-detonation models \citep{2002ApJ...568..791H}. These characteristics include: a) low maximum-light luminosity and ionization state;
b) low photospheric velocities;
c) low-velocity iron-group elements;
and d) an extended region (in velocity space) of products of explosive C-burning and incomplete O- and Si-burning. 
The model spectrum places our early-time NIR spectrum at +11 days after maximum light, which is $\approx$2 days later than inferred from maximum-light curve fits. 
Given the fast evolution in the NIR and the large uncertainty in $t_{\rm max}$, we show that NIR spectra can be used efficiently to refine the time of maximum light and place the first optical spectrum at about +2-3 days after maximum.
Both the tomography and the previous analysis of SN~1999by provide consistent line identifications for spectra dominated by neutral and single-ionized Fe, \ion{S/Si}{ii}, and \ion{O}{i}.

Moreover, and unlike other SN Ia subtypes, there are indications of  asphericity for subluminous SNe Ia.
The early continuum polarization of SN~1999by shows a good agreement with the same models from \cite{2002ApJ...568..791H} used in this work when an asphericity of about $\sim$20\% is included \citep{2001ApJ...556..302H}.
However, the resulting anisotropy in the luminosity is about $\sim$5\% and hardly effects the spectra \citep{2012A&A...545A...7P}.

The late-time nebular spectrum obtained by X-shooter shows only one prominent, narrow feature with a width of $\approx$ 700 km s$^{-1}$. This feature has not been predicted by the delayed-detonation model for SN\,1999by despite the fact that most of the optical and NIR features at early phases are rather similar between SN\,2016hnk and SN\,1999by, as discussed above.
This narrow late-time feature can be attributed to the [\ion{Ca}{ii}] doublet at $\lambda\lambda$7291,7324. 
It can be understood as a result of a nucleosynthesis effect, namely electron capture. 
For high central densities of a WD close to the limit for an accretion induced collapse, with $Y_e$ less than 0.44, results in a shift of the nuclear statistical equilibrium towards $^{48}$Ca. 
In our model, a separate Ca-layer is produced at expansion velocities less than 750 km s$^{-1}$. As discussed in Section \ref{sec:photNLTE}, the exact density depends on the initial conditions. 
For example, the gravitational settling of heavy elements (i.e. $^{22}$Ne and $^{20}$Ne) over long time-scales may increase the central density by about 10 \% to achieve lower $Y_e$. 
There is a high ratio of $^{48}$Ca/$^{40}$Ca in solar composition \citep{t1}, and the result here may be suggestive of some subluminous SNe~Ia as a main production sites for  $^{48}$Ca.  
However, we have late-time, high-resolution observations only for SN\,2016hnk and for none of the other subluminous SNe~Ia, and we lack high S/N observations of the underlying iron-group lines that are predicted on the few percent level compared to the [\ion{Ca}{ii}] feature. 
We note that sub-$M_{\rm Ch}$ models seem to fall well short of producing $^{48}$Ca \citep{Brachwitz00}, despite having a compressional detonation wave, which increases the peak density in a WD \citep{2018arXiV181005910H}. 

We note our interpretation of the narrow nebular [\ion{Ca}{ii}] feature may not be unique. However, RT instabilities are unlikely to cause this feature. As discussed in  Section \ref{sec:photNLTE}, RT instabilities would produce many plumes rather than a single narrow feature and, even if one plume were to dominate from the region of incomplete burning, we would expect a mixture of Ca/Fe and, potentially, S/Si. 
Conversely, a narrow feature similar to the observations could be produced within the ejecta in a narrow jet-like Ca-rich structure. 
In this scenario, the single strong feature does not consist of iron-group elements and must be viewed almost equatorially. If viewed pole on we would not see a narrow feature, but rather a high-velocity one. 

Another explanation could be in the framework of interacting binaries with material stripped from the donor star. This can occur within both the $M_{\rm Ch}$ and sub-$M_{\rm Ch}$ scenarios. In principle, the low-velocity Ca may be attributed to material stripped from a companion star. However, in this case, we may expect very strong forbidden [\ion{O}{i}] lines at 6300, 6364 (and 6393) \AA~regardless of whether the accreted material is H, He, or C-O rich. To rule out or confirm this possibility we would require a time series of spectra of more objects at those late epochs. 

Finally, we want to address the possible link to progenitor systems. 
Potential progenitor systems may either consist of systems of two WDs (called double degenerate), a single WD with a donor which may be main sequence, red giant, or He star (called a single degenerate), or explosions within an asymptotic red giant branch star \citep{1995PASP..107.1019B,2010ApJ...724..502H,2014MNRAS.445.3239P,2015MNRAS.447.2696D,2018SSRv..214...67N,yoon2007}. 

However, the link of SN2016hnk is highly uncertain.
As discussed, SN2016hnk is in a galaxy of low star formation rate. This may point towards a progenitor scenario with a long evolutionary time scale. Though, there is still some ongoing star formation. 
This problem cannot be solved by direct analysis of the ejecta due to the lack of 'classical' observations such as high-precision light curves, as well as flux and polarization spectra well before maximum light \citep{patat12,hoeflich2013,gall2018}.  The lack of early time LCs and optical and IR spectra does not allow us to probe the outermost layers to measure $Z$ in the unburned layers,  or the amount of stable iron-group elements.

This leaves the the high central density of the WD and the nearby material in the environment as a hint towards the progenitor.  Within our models, we need a low accretion rate during the final phase of accretion to achieve a high central density, which results in a low $Y_e$. The nearby material may indicate a high mass loss rate prior to the explosion, e.g. during a common envelope phase. For many of the common progenitor systems, a drop in the final accretion rate may be expected because binary systems tend to become wider due to mass-loss.  
However, the low accretion rate depends on the treatment of the convective URCA process \citep{1972ApL....11...53P,1990ApJ...355..602B,2006ApJ...643.1190S,2015MNRAS.447.2696D,2017ApJ...836L...9P}, and a faster accretion rate may automatically lead to an explosion close to the AIC.


\begin{acknowledgements}
Endavant les atxes.
LG would like to thank the organizers of the {\it Transients in New Surveys: the Undiscovered Country} workshop held in July 2018 in Leiden where this work started as a result of a huge collaborative effort, and the Lorentz Center staff for providing such a great work environment.
CA and EYH acknowledge the support provided by the National Science Foundation under grant No. AST-1613472.
PH acknowledges support by the National Science Foundation (NSF) grant 1715133, and thank thank F.K. Thielemann and G. Martinez-Pinedo for helpful discussions and advice on the current state of electron-capture rates.
MS is supported  in part by a generous grant (13261) from VILLUM FONDEN and a project grant from the Independent Research Fund Denmark. 
EB acknowledges partial support from NASA grant NNX16AB25G, and thanks the Aarhus University Research Fund (AUFF) for a guest researcher grant.
SB is partially supported by China postdoctoral science foundation grant No.2018T110006.
MB acknowledges support from the Swedish Research Council (Vetenskapsr\aa det), the Swedish National Space Board and the research environment grant "Gravitational Radiation and Electromagnetic Astrophysical Transients (GREAT)".
CPG and MS acknowledge support from EU/FP7-ERC grant number 615929.
JH acknowledges financial support from the Finnish Cultural Foundation.
DAH, CM, and GH were supported by NSF grant AST-1313484.
Support for TdJ have been provided by US NSF grant AST-1211916, the TABASGO Foundation, Gary and Cynthia Bengier.
SWJ acknowledges support from NSF award AST-1615455.
KM acknowledges support from STFC (ST/M005348/1) and from H2020 through an ERC Starting Grant (758638).
Research by DJS is supported by NSF grants AST-1821987 and 1821967.
XW is supported by the National Natural Science Foundation of China (NSFC grants 11325313 and 11633002)
JZ is supported by the National Natural Science Foundation of China (NSFC, grants 11773067, 11403096), the Youth Innovation Promotion Association of the CAS (grants 2018081), and the Western Light Youth Project of CAS.

Based in part on data obtained with the LCO network under programmes CLN2016A-005 (PI: Prieto), NAOC2017AB-001 (PI: Dong), CON2016A-008 (PI: Stanek), and ARI2017AB-001 (PI: Bersier).
This work has made use of data from the Asteroid Terrestrial-impact Last Alert System (ATLAS) project. ATLAS is primarily funded to search for near earth asteroids through NASA grants NN12AR55G, 80NSSC18K0284, and 80NSSC18K1575; byproducts of the NEO search include images and catalogs from the survey area.  The ATLAS science products have been made possible through the contributions of the University of Hawaii Institute for Astronomy, the Queen's University Belfast, and the Space Telescope Science Institute.
ASAS-SN is supported by the Gordon and Betty Moore Foundation through grant GBMF5490 to the Ohio State University and NSF grant AST-1515927. Development of ASAS-SN has been supported by NSF grant AST-0908816, the Mt. Cuba Astronomical Foundation, the Center for Cosmology and AstroParticle Physics at the Ohio State University, the Chinese Academy of Sciences South America Center for Astronomy (CASSACA), the Villum Foundation, and George Skestos. We thank the Las Cumbres Observatory (LCO) and its staff for its continuing support of the ASAS-SN project.
NUTS is supported in part by the The Instrument Center for Danish Astrophysics (IDA). 
This research uses data obtained through the Telescope Access Program (TAP), which has been funded by "the Strategic Priority Research Program- The Emergence of Cosmological Structures" of the Chinese Academy of Sciences (Grant No.11 XDB09000000), the Special Fund for Astronomy from the Ministry of Finance, and the National Astronomical
Observatories of China.
Funding for the LJT has been provided by the Chinese Academy of Sciences (CAS) and the People's Government of Yunnan Province. 
Based on observations obtained at the Gemini Observatory under program GS-2016B-Q-22 (PI: Sand). Gemini is operated by the Association of Universities for Research in Astronomy, Inc., under a cooperative agreement with the NSF on behalf of the Gemini partnership: the NSF (United States), the National Research Council (Canada), CONICYT (Chile), Ministerio de Ciencia, Tecnolog\'ia e Innovaci\'on Productiva (Argentina), and Minist\'erio da Ci\^encia, Tecnologia e Inova\c{c}\~ao (Brazil). The data were processed using the Gemini IRAF package. We thank the queue service observers and technical support staff at Gemini Observatory for their assistance.
\end{acknowledgements}

\bibliographystyle{aa}
\bibliography{16hnk}

\newpage 
\begin{appendix}

\section{Light curve excess tests} \label{sec:lightecho}

We performed a number of tests to try to explain the late light curve excess observed in SN~2016hnk. In the main text we provided the most plausible explanation, multi-scattering by dust clouds at 1.0$\pm$0.5 pc from the SN, while here we give more details of other tests.

\subsection{Reddening SN~1991bg spectrum and \ion{Na}{i} D absorption}

\begin{figure}
\includegraphics[trim=0.5cm 0.5cm 0cm 0.8cm,clip=true,width=\columnwidth]{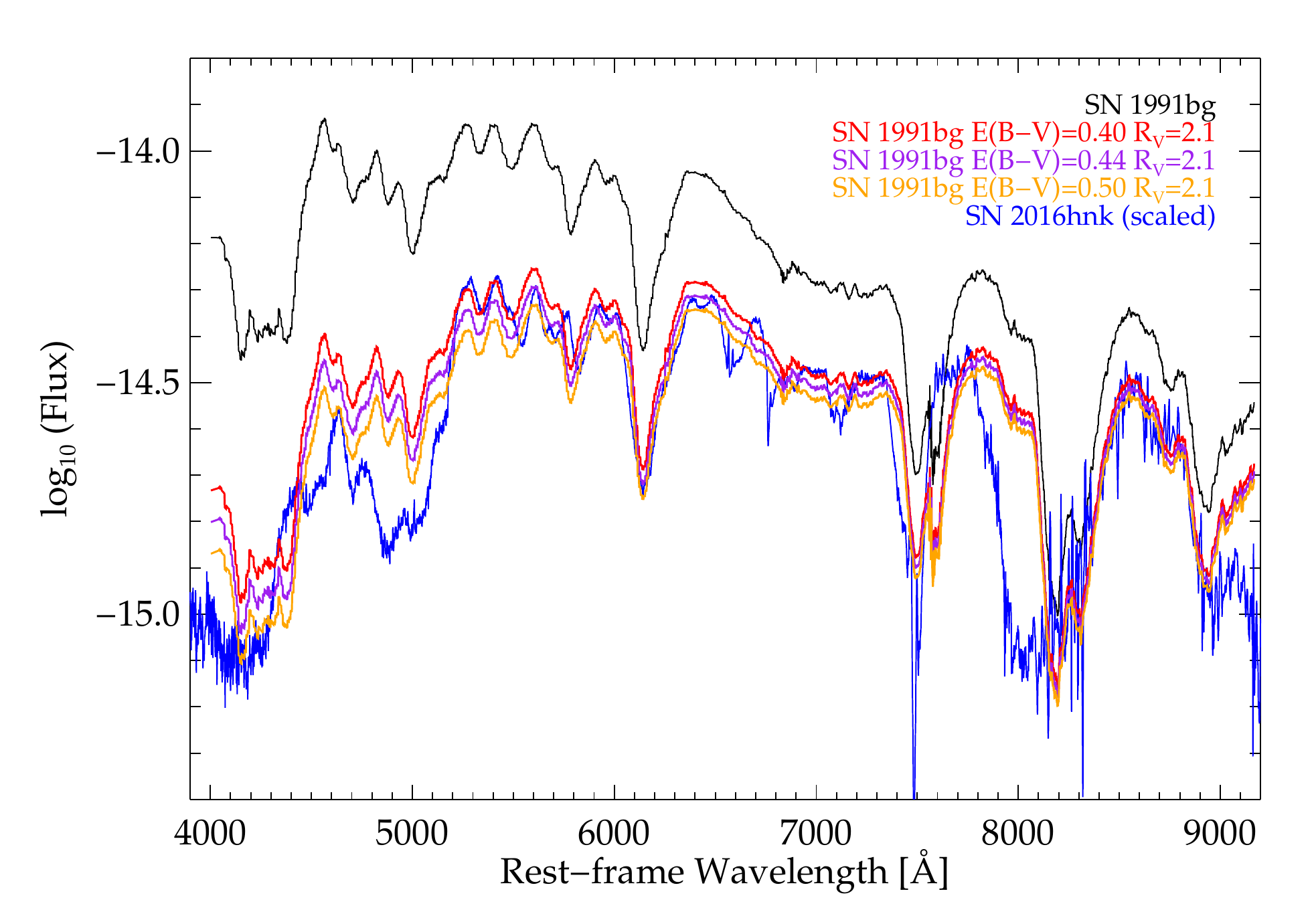}
\caption{SN~1991bg (black) and SN~2016hnk (blue) spectra at maximum, together with three artificially reddened SN~1991bg spectra using an $R_V$=2.1 and $E(B-V)$=0.40 (red), 0.45 (purple), and 0.50 (orange) mag. All spectra have been first corrected for Milky Way extinction and rest-framed.}
\label{fig:91bgcomp}
\end{figure}

From Figures \ref{fig:lccomp} and \ref{fig:comp}, it is obvious that SN~2016hnk light and color curves show some peculiarities. The first question to address is: is SN~2016hnk actually reddened, or is its red color intrinsic? For that, we took SN~1991bg spectrum at maximum light and, after correcting for Milky Way extinction and shifting to the restframe, we artificially applied the same reddening we found for SN~2016hnk ($E(B-V)$=0.4$-$0.5 mag with an $R_V$=2.1).
Figure \ref{fig:91bgcomp} shows the SN~1991bg spectrum at maximum (in black), together with the corresponding SN~2016hnk spectrum also at maximum (in blue; also corrected for Milky Way reddening and shifted to the restframe), and three SN~1991bg reddened spectra with $E(B-V)$=0.40, 0.45, and 0.50 mag (in red, purple, and orange, respectively).
Besides the differences already pointed out in Section \ref{sec:spec} (broader NIR Calcium absorption, and features in the range 6400-7300 \AA), we can see an almost perfect match between SN~2016hnk and all reddened SN~1991bg spectra redwards of 5200 \AA.
However, the blue part of the spectrum is where most of the discrepancy comes from: both the aggregate \ion{Fe}{ii}, \ion{Si}{ii}, and \ion{Mg}{ii} features at $\sim$5000 \AA, and the \ion{TI}{ii} at $\sim$4200\AA, are deeper in SN~2016hnk than in the reddened SN~1991bg spectra, which makes SN~2016hnk color curves even redder than expected by simple dust extinction. 
So, we conclude that there should be an intrinsic effect, in addition to reddening responsible for the large red colors of SN~2016hnk.

\begin{figure}
\includegraphics[trim=0.5cm 0cm 0.5cm 0.8cm,clip=true,width=\columnwidth]{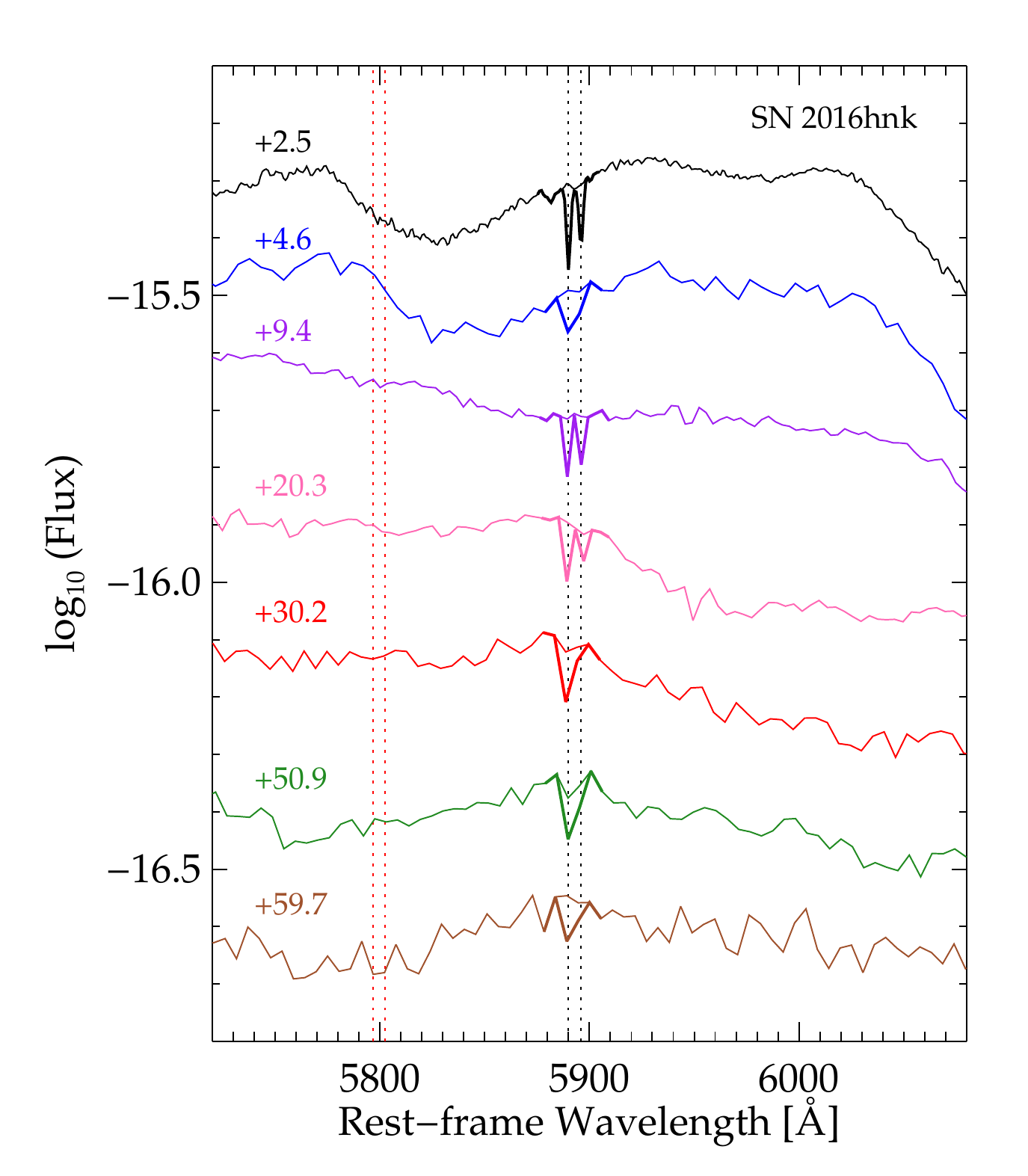}
\caption{Zoom into SN~2016hnk spectra around the \ion{Na}{i} D doublet (restframe 5890 and 5896\AA; in vertical black dotted lines) absorption, at seven epochs from +2.5 to +59.7 days after maximum. Thick lines represent \ion{Na}{i} D absorption with $pEW$=1.3\AA~at the spectral resolution of each spectrum. Red vertical lines mark the position where the \ion{Na}{i} D doublet from the Milky Way should be found. In all cases, we do not detect the \ion{Na}{i} D lines in SN~2016hnk spectra.}
\label{fig:nad}
\end{figure}

Another peculiarity of SN~2016hnk is the lack of \ion{Na}{i} D absorption at any epoch, even when we obtain large $E(B-V)$ (>0.4 mag) with different methods.
Although the relation between extinction and \ion{Na}{i} D $pEW$ has been put in question when measured in low-resolution spectra \citep{2011MNRAS.415L..81P}, it is usually accepted that the presence of \ion{Na}{i} D $pEW$ absorption is an indication of extinction by ISM \citep{2013ApJ...779...38P}.
The relations from \cite{1990A&A...237...79B} and \cite{2012MNRAS.426.1465P}, predict both \ion{Na}{i} D $pEW$ of at least 1.3\AA, from the $E(B-V)\sim0.4$ mag measured in SN~2016hnk.
We artificially input \ion{Na}{i} D absorption with $pEW$ of 1.3\AA~in several SN~2016hnk spectra at different epochs, as shown in Figure \ref{fig:nad}, taking into account the spectral resolution of each spectrum.
Even for the latter spectrum at +59.7 d that has the lowest signal-to-noise ratio, we can significantly discard any indication of \ion{Na}{i} D absorption.
Therefore, in the case of SN~2016hnk, the large extinction does not come together with  \ion{Na}{i} D absorption.

\begin{figure}
\includegraphics[trim=0.1cm 0cm 0cm 0cm,clip=true,width=\columnwidth]{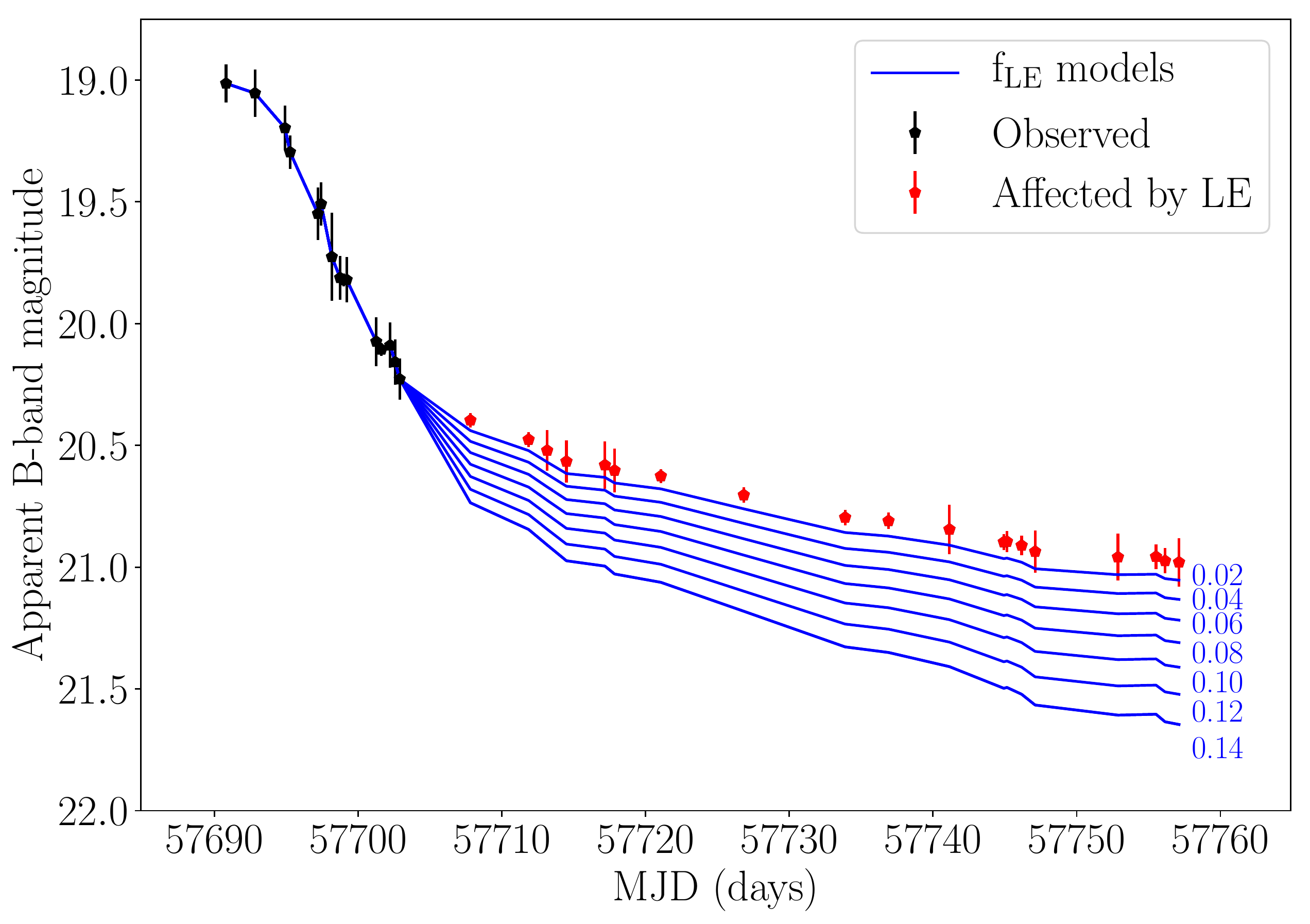}
\caption{Effect of subtracting different fractional contribution of the light-echo to the \mbox{SN~2016hnk} $B$ band observed light curve.}
\label{fig:LE}
\end{figure}

\subsection{Simple light echo model}

\begin{figure}
\includegraphics[trim=0.1cm 0cm 0cm 0cm,clip=true,width=\columnwidth]{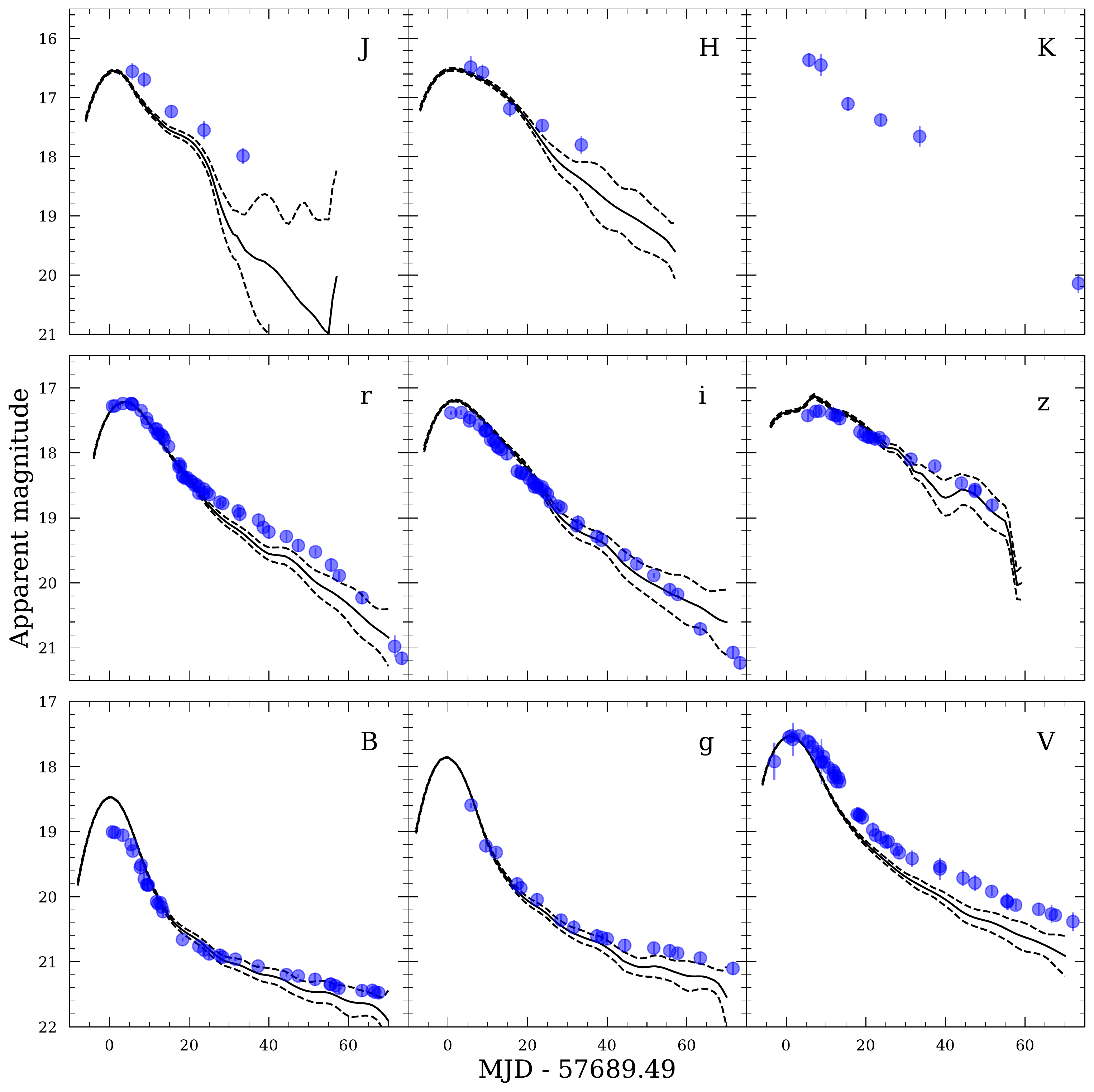}
\caption{Multiband \mbox{SN~2016hnk} \textsc{SNooPy} fits using a 91bg template and with a 10\% contribution of a light echo starting at +15 days until the end of the light curve. Fits in all band are reasonably better than those without a LE shown in Figure \ref{fig:snoopy}.}
\label{fig:snoopyLE}
\end{figure}

Another possible explanation for the light excess in the SN~2016hnk light curve after +15 days past maximum would be an intervening dust cloud close to the line-of-sight that is producing a light echo (LE; \citealt{2015ApJ...806..134M}).
To test this possibility, we constructed a simple model accounting for different contributions of the LE to the total observed light, in the form,
\begin{equation}
f_{\rm SN}^X (t) = f_{\rm obs}^X (t)- \alpha \, f_{\rm LE}^X \, S(t, t_{\rm delay}, t_{\rm end}),
\end{equation}
where $f_{\rm SN}^X$ is the actual emitted flux from the SN in the $X$ band, $f_{\rm obs}^X$ is the flux we observed in the same band, $f_{\rm LE}^X$ is the re-emitted flux from the cloud, $\alpha$ is a multiplicative factor that  corresponds to the contribution of the LE with respect to the actual emission from the SN (and which we assumed to be band independent) in the sense that $\alpha$=0 means no contribution at all and $\alpha$=1 corresponds to the extreme case of perfect reflection, resulting in $f_{\rm SN}$=$f_{\rm LE}$, and $S$ is a step function that has a value of one for $t_{\rm delay}$ < $t$ < $t_{\rm end}$ and zero at other epochs.

To estimate $f_{\rm LE}^X$ we assumed that early epochs are free of LE contribution and performed a \textsc{SNooPy} fit (fixing $t_{\rm max}$ and $R_V$ to the same values found on Section \ref{sec:phot}) but only up to +15 days past maximum. Then, the peak flux of the best-fit model was taken as $f_{\rm LE}^X$ for each band independently, and used to generate a series of LE-free light curves with different combinations of: $\alpha$ parameters from 0.02 to 0.14, in steps of 0.02; $t_{\rm delay}$ from 5 to 30 days in steps of 5 days; and $t_{\rm end}$ from 40 days on in steps of 5 days. See Figure \ref{fig:LE} as an example.
Those sets of light curves were fitted with \textsc{SNooPy} keeping the assumptions on the $t_{\rm max}$ and $R_V$, but extending the fit to the whole time extent.
The best fit, shown in Figure \ref{fig:snoopyLE}, was for a combination of parameters ($\alpha$, $t_{\rm delay}$, $t_{\rm end}$) = (0.10, 15, 80). Note that the best $t_{\rm end}$ value is right after the last photometric point, meaning that the LE is contributing flux at least until the epoch when we stopped our follow up campaign. 
The model also provided the best estimates of $s_{BV}$= 0.380 $\pm$ 0.032 and $E(B-V)$ = 0.458 $\pm$ 0.064 mag, the former being a bit lower than the reported in the main text, and the latter being consistent with the value found using CDMagic.

Although light curve fits improve significantly when this simple model of the LE is removed from observations, if this excess is due to a light echo, we should further see a mixture of spectral features in the spectra. 
At latter epochs when the LE is affecting the observations (>+15 d in our model), we should see spectral features from earlier epochs that are being reflected and that are contaminating the light directly received by the observer from the SN. 
This contamination is strongly dependent on the configuration of the ISM around the SN but, again, in the simplest scenario (single cloud at a certain distance from the SN) this would correspond to a mixture of the spectrum at the actual epoch and a spectrum at an early epoch equal to the delayed time. 

\begin{figure}
\includegraphics[trim=0.5cm 0cm 0.5cm 0.8cm,clip=true,width=\columnwidth]{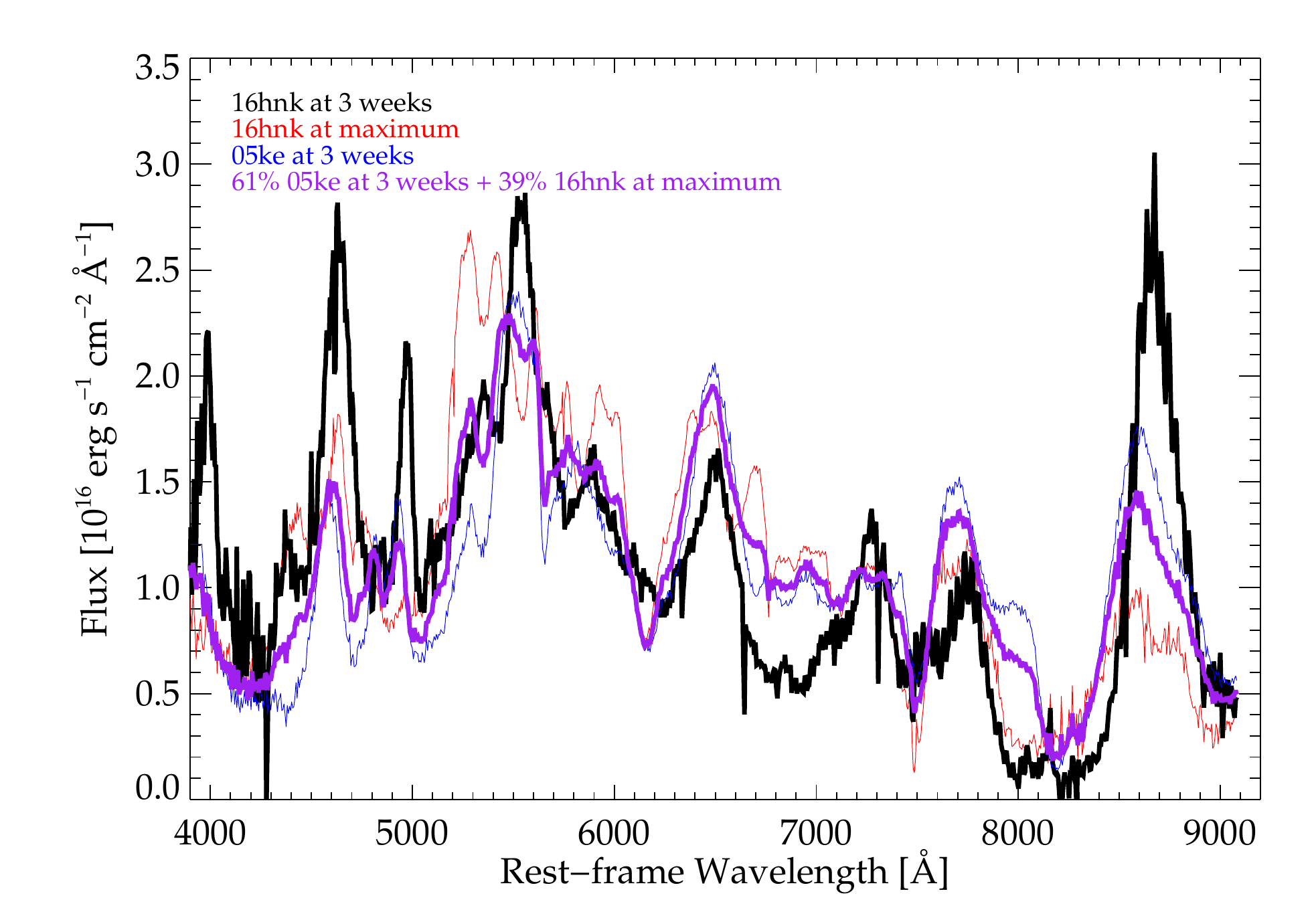}
\caption{Simulation of a light echo contaminated spectrum at three week past maximum, based on a fiducial subluminous spectrum at +3 weeks past maximum from SN~2005ke (contributing 61\% to the total), and adding 39\% of the SN~2016hnk spectrum at maximum light, which represents the contamination from the light echo. Comparing this LE-affected spectrum (in purple) to the observed SN~2016hnk (in black) we do not see any feature that definitely confirms that the latter is in fact affected by any feature reflected from earlier epochs.}
\label{fig:lecomp}
\end{figure}

To test if these features are present at late epochs we took the spectrum of a normal 1991bg-like SNIa, SN~2005ke at +23.6 days, and add up incremental contributions of SN~2016hnk spectrum at maximum light, which represents the contamination from the light echo. 
The resulting mixed spectrum is compared to the observed spectrum of SN~2016hk at +20.3 days to check if the mixed features from those two spectra reproduce those seen at late epochs in SN~2016hnk.
Starting from no contribution from the LE, and increasing in steps of 1\% up to 50\% contribution to the total light, we find the most similar spectrum for a 39\% LE contribution (based on $\chi^2$ minimization).
In this best case, shown in Figure \ref{fig:lecomp}, we do not find a convincing resemblance between the two spectra (observed, in black, and simulated, in purple) that confirms the hypothesis of a LE causing those features.
For this reason, we do not consider the light echo explanation for the late-time light excess, and favor the multiscattering approach detailed in the main text.

\section{Spectral modeling technique} \label{app:model}

The code is a 1D Monte Carlo radiation transport code that follows the propagation of photon packets through a SN atmosphere. The photon packets can undergo Thomson scattering and line absorption. Photon packets have two fates: they either re-enter the photosphere in a process known as back scattering or escape the SN atmosphere after possibly undergoing scattering or line absorption.  A modified nebular approximation is used to account for non-local thermodynamical equilibrium effects. The code makes use of the Schuster-Schwarzschild approximation, which assumes that radiative energy is emitted from an inner blackbody. This has the advantage that the radiation transport below the photosphere is not required to be known, but can cause excess flux in the red ($>$6500 \AA) past maximum light. We note that the values we obtain for $\vph$ will in most cases be lower than those measured from the individual ions in the spectra in Section \ref{sec:specanal}, as the lines are produced in the ejecta above the photosphere. 
For further details on the abundance stratification code and method used in this work see \citet{1993A&A...279..447M,1999A&A...345..211L,2000A&A...363..705M,Stehle05}.

\section{Photometry, local sequence, and spectral log}

\begin{table}\small
\centering
\caption{ASASSN $V$-band photometry and upper limits.}
\label{tab:SN2016hnk_ph_asassn}
\begin{tabular}{lcc} 
\hline\hline
\textbf{MJD} & \textbf{Epoch} & \textbf{$V$} \\
\hline
57652.435 & -36.46 & $>$17.72 \\
57653.253 & -35.65 & $>$18.00 \\
57655.145 & -33.79 & $>$18.31 \\
57660.479 & -28.54 & $>$18.36 \\
57663.160 & -25.90 & $>$17.87 \\
57664.204 & -24.88 & $>$18.33 \\
57668.340 & -20.81 & $>$17.57 \\
57669.340 & -19.82 & $>$17.94 \\
57670.382 & -18.80 & $>$18.01 \\
57671.476 & -17.72 & $>$17.97 \\
57672.324 & -16.88 & $>$17.91 \\
57673.521 & -15.71 & $>$17.83 \\
57685.513 &  -3.91 & $>$16.74 \\
57686.507 &  -2.93 & 17.92 (0.290) \\
57689.498 &  +0.02 & 17.25 (0.210) \\
57691.130 &  +1.62 & 17.58 (0.250) \\
57698.282 &  +8.66 & 17.92 (0.340) \\
\hline
\end{tabular}
\end{table}

\begin{table}\small
\centering
\caption{ATLAS $c$- and $o$-band photometry and upper limits.}
\label{tab:SN2016hnk_ph_atlas}
\begin{tabular}{lcc|lcc} 
\hline\hline
\textbf{MJD} & \textbf{Epoch} & \textbf{$c$} & \textbf{MJD} & \textbf{Epoch} & \textbf{$o$} \\
\hline
57688.474 &  -0.99 & 17.69 (0.07) &  57643.603 & -45.15 & $>$19.20     \\       
57688.492 &  -0.97 & 17.91 (0.07) &  57643.617 & -45.14 & $>$19.53     \\       
57688.506 &  -0.96 & 17.77 (0.07) &  57643.633 & -45.12 & $>$19.71     \\       
57688.523 &  -0.94 & 17.68 (0.07) &  57671.530 & -17.66 & $>$20.18     \\       
57688.542 &  -0.92 & 17.81 (0.08) &  57671.545 & -17.65 & $>$19.09     \\       
57696.448 &  +6.86 & 17.83 (0.07) &  57671.560 & -17.64 & $>$20.50     \\       
57696.463 &  +6.87 & 17.70 (0.06) &  57680.492 &  -8.85 & 18.74 (0.29) \\ 
57696.482 &  +6.89 & 17.65 (0.06) &  57680.507 &  -8.83 & 19.06 (0.41) \\ 
57696.489 &  +6.90 & 17.82 (0.07) &  57680.526 &  -8.81 & 18.67 (0.31) \\ 
57696.509 &  +6.92 & 17.83 (0.07) &  57680.533 &  -8.81 & 18.37 (0.24) \\ 
57699.356 &  +9.72 & 17.91 (0.13) &  57680.551 &  -8.79 & 19.32 (0.49) \\ 
57699.374 &  +9.74 & 18.08 (0.10) &  57700.433 &  +10.78 & 17.80 (0.14) \\ 
57699.394 &  +9.76 & 17.92 (0.09) &  57700.462 &  +10.81 & 17.53 (0.09) \\ 
57699.408 &  +9.77 & 17.79 (0.08) &  57700.477 &  +10.82 & 17.45 (0.08) \\ 
57699.422 &  +9.78 & 17.90 (0.12) &  57700.497 &  +10.84 & 17.45 (0.10) \\  
57716.402 & +26.50 & 18.48 (0.13) &  57704.387 &  +14.67 & 17.49 (0.16) \\ 
57716.417 & +26.51 & 18.95 (0.21) &  57704.406 &  +14.69 & 17.59 (0.18) \\ 
57716.433 & +26.53 & 18.89 (0.23) &  57704.415 &  +14.70 & 18.06 (0.27) \\ 
57716.449 & +26.54 & 18.42 (0.40) &  57704.436 &  +14.72 & 17.76 (0.21) \\ 
57716.465 & +26.56 & 19.03 (0.71) &  57704.443 &  +14.72 & 18.04 (0.28) \\ 
57744.335 & +53.99 & 19.81 (0.38) &  57712.443 &  +22.60 & 17.73 (0.15) \\ 
57744.348 & +54.00 & 20.80 (0.70) &  57736.351 &  +46.13 & 18.17 (0.25) \\ 
57744.388 & +54.04 & 19.46 (0.29) &  57736.364 &  +46.14 & 18.88 (0.42) \\ 
57756.336 & +65.80 & 20.39 (0.50) &  57736.376 &  +46.15 & 19.27 (0.57) \\ 
57756.346 & +65.81 & 20.57 (0.59) &  57736.389 &  +46.17 & 18.39 (0.34) \\ 
57780.272 & +89.35 & 20.23 (0.47) &  57743.343 &  +53.01 & 19.61 (0.40) \\  
57780.285 & +89.37 & 20.71 (0.71) &  57743.356 &  +53.02 & 18.96 (0.23) \\ 
57780.318 & +89.40 & 20.52 (0.63) &  57743.369 &  +53.03 & 19.14 (0.27) \\ 
57784.309 & +93.33 & 20.16 (0.52) &  57743.382 &  +53.05 & 18.73 (0.20) \\  
\hline
\end{tabular}
\end{table}

\onecolumn

\begin{table}\scriptsize
\centering
\caption[Photometry for \mbox{SN~2016hnk}. Epochs are given in the rest frame.]{Photometry for \mbox{SN~2016hnk}. Epochs are given in the rest frame.} \label{tab:SN2016hnk_ph} 
\begin{tabular}{ccccccccccc} 
\hline\hline \textbf{MJD} & \textbf{Epoch} & \textbf{$B$} & \textbf{$V$} & \textbf{$u$}& \textbf{$g$} &  \textbf{$r$} &  \textbf{$i$} & \textbf{$z$} & \textbf{Source} \\ \hline
\hline \hline
57690.200 &   +0.71 & 19.003 (0.088) & 17.544 (0.054) &  $\cdots$  &     $\cdots$      & 17.282 (0.046) & 17.384 (0.030) &     $\cdots$      &  fl03        \\ 
57690.815 &   +1.31 & 19.014 (0.079) & 17.526 (0.061) &  $\cdots$  &     $\cdots$      & 17.278 (0.037) &     $\cdots$      &     $\cdots$      &  fl14        \\ 
57692.830 &   +3.29 & 19.054 (0.098) & 17.526 (0.041) &  $\cdots$  &     $\cdots$      & 17.241 (0.037) & 17.380 (0.039) &     $\cdots$      &  fl06        \\
57694.920 &   +5.36 & 19.197 (0.092) & 17.607 (0.088) &  $\cdots$  &     $\cdots$      & 17.243 (0.038) & 17.455 (0.082) & 17.426 (0.088) &  IO:O        \\
57694.930 &   +5.36 &     $\cdots$      &     $\cdots$      &  $\cdots$  &     $\cdots$      & 17.240 (0.063) & 17.508 (0.059) &     $\cdots$      &  fl06        \\ 
57695.270 &   +5.70 & 19.296 (0.068) & 17.628 (0.034) &  $\cdots$  & 18.591 (0.038) & 17.257 (0.033) &     $\cdots$      &     $\cdots$      &  fl15        \\
57696.130 &   +6.54 &     $\cdots$      & 17.690 (0.082) &  $\cdots$  &     $\cdots$      &     $\cdots$      &     $\cdots$      &     $\cdots$      &  FLI         \\ 
57696.970 &   +7.37 &     $\cdots$      &     $\cdots$      & >18.832 &     $\cdots$      &     $\cdots$      &     $\cdots$      & 17.361 (0.089) &  fl16        \\ 
57697.205 &   +7.60 & 19.549 (0.108) & 17.802 (0.082) &  $\cdots$  &     $\cdots$      &     $\cdots$      &     $\cdots$      &     $\cdots$      &  FLI         \\
57697.435 &   +7.83 & 19.509 (0.089) & 17.768 (0.047) &  $\cdots$  &     $\cdots$      & 17.351 (0.035) & 17.571 (0.039) &     $\cdots$      &  fl12        \\
57697.835 &   +8.22 &     $\cdots$      &     $\cdots$      & >18.957 &     $\cdots$      &     $\cdots$      &     $\cdots$      & 17.358 (0.063) &  fl06        \\ 
57698.175 &   +8.56 & 19.726 (0.181) & 17.929 (0.093) &  $\cdots$  &     $\cdots$      &     $\cdots$      &     $\cdots$      &     $\cdots$      &  FLI         \\ 
57698.755 &   +9.13 & 19.812 (0.090) & 17.842 (0.065) &  $\cdots$  &     $\cdots$      & 17.471 (0.047) & 17.663 (0.053) &     $\cdots$      &  fl06        \\
57699.000 &   +9.37 & 19.821 (0.026) & 17.927 (0.088) &  $\cdots$  & 19.215 (0.081) & 17.534 (0.046) & 17.659 (0.045) &     $\cdots$      &  fl14        \\ 
57699.210 &   +9.58 & 19.820 (0.093) &     $\cdots$      &  $\cdots$  &     $\cdots$      &     $\cdots$      & 17.653 (0.073) &     $\cdots$      &  FLI         \\
57700.195 &  +10.55 &     $\cdots$      & 18.017 (0.091) &  $\cdots$  &     $\cdots$      &     $\cdots$      & 17.801 (0.111) &     $\cdots$      &  FLI         \\ 
57700.890 &  +11.23 &     $\cdots$      &     $\cdots$      &  $\cdots$  &     $\cdots$      & 17.634 (0.100) & 17.796 (0.101) & 17.402 (0.089) &  IO:O        \\ 
57701.270 &  +11.60 & 20.074 (0.100) & 18.110 (0.098) &  $\cdots$  &     $\cdots$      & 17.632 (0.050) & 17.830 (0.051) &     $\cdots$      &  fl04        \\
57701.630 &  +11.96 & 20.106 (0.026) & 18.094 (0.109) &  $\cdots$  & 19.318 (0.075) & 17.709 (0.069) &     $\cdots$      &     $\cdots$      &  fl11        \\
57701.785 &  +12.11 &     $\cdots$      &     $\cdots$      & >19.174 &     $\cdots$      &     $\cdots$      &     $\cdots$      & 17.425 (0.058) &  fl16        \\ 
57701.870 &  +12.19 &     $\cdots$      & 18.138 (0.099) &  $\cdots$  &     $\cdots$      & 17.706 (0.024) & 17.902 (0.043) &     $\cdots$      &  IO:O        \\ 
57702.235 &  +12.55 & 20.089 (0.093) & 18.231 (0.087) &  $\cdots$  &     $\cdots$      &     $\cdots$      & 17.923 (0.097) &     $\cdots$      &  FLI         \\ 
57702.405 &  +12.72 &     $\cdots$      &     $\cdots$      & >19.542 &     $\cdots$      &     $\cdots$      &     $\cdots$      & 17.430 (0.080) &  fl12        \\ 
57702.580 &  +12.89 & 20.158 (0.094) & 18.168 (0.089) &  $\cdots$  &     $\cdots$      & 17.724 (0.048) &     $\cdots$      &     $\cdots$      &  fl11        \\ 
57702.900 &  +13.21 & 20.228 (0.085) & 18.232 (0.041) &  $\cdots$  &     $\cdots$      & 17.751 (0.033) & 17.948 (0.035) & 17.475 (0.095) &  IO:O        \\
57703.225 &  +13.53 &     $\cdots$      &     $\cdots$      &  $\cdots$  &     $\cdots$      & 17.796 (0.134) &     $\cdots$      &     $\cdots$      &  fl04        \\
57704.320 &  +14.60 &     $\cdots$      &     $\cdots$      &  $\cdots$  &     $\cdots$      & 17.899 (0.086) & 18.014 (0.092) &     $\cdots$      &  fl04        \\ 
57706.860 &  +17.10 &     $\cdots$      &     $\cdots$      &  $\cdots$  &     $\cdots$      & 17.952 (0.090) &     $\cdots$      &     $\cdots$      &  fl06        \\ 
57706.890 &  +17.13 &     $\cdots$      &     $\cdots$      &  $\cdots$  & 19.570 (0.102) & 17.993 (0.101) & 18.083 (0.101) &     $\cdots$      &  fl16        \\
57707.290 &  +17.53 &     $\cdots$      & 18.487 (0.115) &  $\cdots$  &     $\cdots$      & 17.983 (0.084) &     $\cdots$      &     $\cdots$      &  fl15        \\ 
57707.815 &  +18.05 & 20.397 (0.030) & 18.491 (0.117) &  $\cdots$  & 19.620 (0.104) & 18.101 (0.102) & 18.101 (0.102) &     $\cdots$      &  fl06        \\ 
57708.030 &  +18.26 &     $\cdots$      & 18.501 (0.084) &  $\cdots$  &     $\cdots$      & 18.117 (0.048) & 18.113 (0.062) & 17.540 (0.088) &  IO:O        \\
57708.885 &  +19.10 &     $\cdots$      &     $\cdots$      &  $\cdots$  &     $\cdots$      & 18.122 (0.024) & 18.121 (0.025) & 17.584 (0.079) &  IO:O        \\ 
57709.870 &  +20.07 &     $\cdots$      &     $\cdots$      &  $\cdots$  &     $\cdots$      & 18.162 (0.030) & 18.181 (0.046) & 17.611 (0.088) &  IO:O        \\ 
57710.180 &  +20.37 &     $\cdots$      &     $\cdots$      &  $\cdots$  &     $\cdots$      & 18.176 (0.097) &     $\cdots$      &     $\cdots$      &  fl04        \\ 
57710.265 &  +20.46 &     $\cdots$      &     $\cdots$      & >19.462 &     $\cdots$      &     $\cdots$      &     $\cdots$      & 17.613 (0.062) &  fl03        \\ 
57710.850 &  +21.03 &     $\cdots$      &     $\cdots$      &  $\cdots$  &     $\cdots$      & 18.218 (0.022) & 18.227 (0.032) & 17.617 (0.089) &  IO:O        \\ 
57711.205 &  +21.38 &     $\cdots$      &     $\cdots$      &  $\cdots$  &     $\cdots$      &     $\cdots$      & 18.281 (0.095) &     $\cdots$      &  fl04        \\ 
57711.250 &  +21.42 &     $\cdots$      & 18.671 (0.112) &  $\cdots$  &     $\cdots$      & 18.205 (0.086) & 18.240 (0.092) &     $\cdots$      &  fl03        \\ 
57711.790 &  +21.96 &     $\cdots$      &     $\cdots$      &  $\cdots$  &     $\cdots$      &     $\cdots$      &     $\cdots$      & 17.639 (0.090) &  fl06        \\ 
57711.870 &  +22.04 & 20.476 (0.030) & 18.734 (0.105) &  $\cdots$  & 19.762 (0.102) & 18.308 (0.057) & 18.285 (0.067) &     $\cdots$      &  fl14        \\ 
57711.875 &  +22.04 &     $\cdots$      &     $\cdots$      &  $\cdots$  &     $\cdots$      & 18.234 (0.028) & 18.254 (0.041) &     $\cdots$      &  IO:O        \\ 
57712.845 &  +22.99 &     $\cdots$      &     $\cdots$      &  $\cdots$  &     $\cdots$      & 18.315 (0.029) & 18.303 (0.033) & 17.620 (0.066) &  IO:O        \\ 
57713.160 &  +23.30 & 20.521 (0.083) & 18.760 (0.072) &  $\cdots$  &     $\cdots$      & 18.262 (0.059) & 18.281 (0.062) &     $\cdots$      &  fl04        \\ 
57713.865 &  +24.00 &     $\cdots$      &     $\cdots$      &  $\cdots$  &     $\cdots$      & 18.301 (0.029) & 18.341 (0.033) & 17.672 (0.084) &  IO:O        \\ 
57714.505 &  +24.63 & 20.566 (0.087) & 18.810 (0.095) &  $\cdots$  &     $\cdots$      & 18.326 (0.061) & 18.369 (0.066) &     $\cdots$      &  fl11        \\ 
57715.155 &  +25.27 &     $\cdots$      & 18.805 (0.123) &  $\cdots$  &     $\cdots$      &     $\cdots$      & 18.459 (0.084) &     $\cdots$      &  FLI         \\ 
57717.175 &  +27.26 & 20.581 (0.097) & 18.892 (0.083) &  $\cdots$  &     $\cdots$      & 18.410 (0.053) & 18.511 (0.066) &     $\cdots$      &  fl04        \\
57717.850 &  +27.92 & 20.603 (0.090) & 18.929 (0.059) &  $\cdots$  & 19.995 (0.080) & 18.425 (0.042) & 18.529 (0.053) &     $\cdots$      &  fl14        \\
57720.820 &  +30.84 &     $\cdots$      &     $\cdots$      &  $\cdots$  &     $\cdots$      &     $\cdots$      &     $\cdots$      & 17.905 (0.053) &  fl16        \\ 
57721.065 &  +31.09 & 20.626 (0.028) & 18.990 (0.126) &  $\cdots$  & 20.074 (0.115) & 18.507 (0.095) & 18.729 (0.101) &     $\cdots$      &  fl14        \\ 
57722.210 &  +32.21 &     $\cdots$      &     $\cdots$      &  $\cdots$  &     $\cdots$      & 18.542 (0.113) & 18.694 (0.115) &     $\cdots$      &  fl04        \\ 
57726.820 &  +36.75 &     $\cdots$      &     $\cdots$      &  $\cdots$  &     $\cdots$      &     $\cdots$      &     $\cdots$      & 17.993 (0.057) &  fl06        \\ 
57726.840 &  +36.77 & 20.704 (0.032) &     $\cdots$      &  $\cdots$  & 20.164 (0.102) & 18.604 (0.101) & 18.841 (0.102) &     $\cdots$      &  fl16        \\ 
57728.050 &  +37.96 &     $\cdots$      & 19.070 (0.099) & >19.978 & 20.178 (0.053) & 18.677 (0.074) & 18.879 (0.047) &     $\cdots$      &  ALFOSC\_FASU\\  
57728.130 &  +38.04 &     $\cdots$      & 19.095 (0.174) &  $\cdots$  &     $\cdots$      &     $\cdots$      &     $\cdots$      &     $\cdots$      &  FLI         \\ 
57729.545 &  +39.43 &     $\cdots$      &     $\cdots$      &  $\cdots$  & 20.192 (0.104) & 18.723 (0.102) &     $\cdots$      &     $\cdots$      &  fl11        \\ 
57733.885 &  +43.70 & 20.796 (0.032) & 19.185 (0.125) &  $\cdots$  & 20.258 (0.102) & 18.766 (0.086) & 19.021 (0.094) & 18.205 (0.061) &  fl06        \\ 
57736.910 &  +46.68 & 20.810 (0.034) & 19.230 (0.127) &  $\cdots$  &     $\cdots$      & 18.850 (0.101) & 19.102 (0.102) & 18.291 (0.092) &  fl16        \\ 
57741.155 &  +50.86 & 20.845 (0.101) & 19.305 (0.077) &  $\cdots$  & 20.284 (0.084) & 18.907 (0.056) & 19.201 (0.039) & 18.463 (0.086) &  fl04        \\
57744.930 &  +54.57 & 20.897 (0.033) & 19.388 (0.128) &  $\cdots$  &     $\cdots$      &     $\cdots$      &     $\cdots$      &     $\cdots$      &  fl16        \\ 
57745.155 &  +54.79 & 20.895 (0.043) & 19.393 (0.122) &  $\cdots$  & 20.310 (0.103) & 19.017 (0.104) & 19.313 (0.097) &     $\cdots$      &  fl03        \\
57746.170 &  +55.79 & 20.911 (0.040) &     $\cdots$      &  $\cdots$  &     $\cdots$      &     $\cdots$      &     $\cdots$      &     $\cdots$      &  fl14        \\ 
57747.105 &  +56.71 & 20.936 (0.086) & 19.419 (0.076) &  $\cdots$  & 20.332 (0.082) & 19.099 (0.079) & 19.347 (0.072) &     $\cdots$      &  fl04        \\
57752.870 &  +62.39 & 20.959 (0.096) & 19.453 (0.093) &  $\cdots$  & 20.378 (0.082) & 19.250 (0.101) & 19.563 (0.102) &     $\cdots$      &  fl14        \\
57755.520 &  +64.99 & 20.957 (0.051) &     $\cdots$      &  $\cdots$  &     $\cdots$      &     $\cdots$      &     $\cdots$      &     $\cdots$      &  fl12        \\ 
57756.155 &  +65.62 & 20.974 (0.052) & 19.488 (0.134) &  $\cdots$  &     $\cdots$      &     $\cdots$      &     $\cdots$      &     $\cdots$      &  fl04        \\
57757.120 &  +66.57 & 20.980 (0.099) & 19.498 (0.077) &  $\cdots$  &     $\cdots$      &     $\cdots$      &     $\cdots$      &     $\cdots$      &  fl03        \\ 
57761.100 &  +70.49 &     $\cdots$      &     $\cdots$      &  $\cdots$  & 20.470 (0.105) & 19.496 (0.168) & 19.676 (0.103) &     $\cdots$      &  fl04        \\ 
57761.485 &  +70.86 &     $\cdots$      & 19.544 (0.138) &  $\cdots$  &     $\cdots$      &     $\cdots$      &     $\cdots$      &     $\cdots$      &  fl12        \\ 
57762.877 &  +72.24 &     $\cdots$      &     $\cdots$      &  $\cdots$  &     $\cdots$      & 19.540 (0.101) & 19.718 (0.101) &     $\cdots$      &  fl06        \\ 
57799.780 & +108.55 &     $\cdots$      &     $\cdots$      &  $\cdots$  &     $\cdots$      & 20.458 (0.332) &     $\cdots$      &     $\cdots$      &  fl16        \\ 
\hline
\end{tabular}
$ $\\
Notes: 
fl03, LCOGT node at Cerro Tololo Inter-American Observatory;
fl04, LCOGT node at Cerro Tololo Inter-American Observatory;
fl06, LCOGT node at SAAO;
fl11, LCOGT node at Siding Spring Observatory;
fl12, LCOGT node at Siding Spring Observatory;
fl14, LCOGT node at SAAO;
fl15, LCOGT node at Cerro Tololo Inter-American Observatory;
fl16, LCOGT node at SAAO;
FLI, DEMONEXT;
IO:O, Liverpool Telescope;
ALFOSC\_FASU, Nordic Optical Telescope.
\end{table}

\begin{table*}
\centering
\caption{Near-infrared Photometry for \mbox{SN~2016hnk}. Epochs from peak are given in the rest frame.} \label{tab:SN2016hnk_ph_nir}
\begin{tabular}{lrcccl}
\hline\hline 
\textbf{ID} & \textbf{Epoch} & \textbf{$J$} & \textbf{$H$} & \textbf{$K$} & \textbf{Source} \\ 
\hline
57695.193 &  +5.62 & 16.555 (0.139) & 16.483 (0.186) & 16.366 (0.124) & SOFI   \\
57698.163 &  +8.55 & 16.694 (0.132) & 16.573 (0.136) & 16.448 (0.191) & SOFI   \\
57705.030 & +15.30 & 17.059 (0.112) & 17.005 (0.137) & 16.915 (0.121) & NOTCAM \\
57713.220 & +23.36 & 17.320 (0.156) & 17.239 (0.100) & 17.141 (0.104) & SOFI   \\
57722.957 & +32.95 & 17.658 (0.137) & 17.495 (0.154) & 17.358 (0.172) & NOTCAM \\
57762.930 & +72.29 & 19.995 (0.176) & 19.193 (0.161) & 18.604 (0.158) & NOTCAM \\
\hline
\end{tabular}
$ $\\
Notes: 
SOFI, New Technology Telescope;
NOTCAM, Nordic Optical Telescope.
\end{table*}

\twocolumn

\begin{sidewaystable*}\footnotesize
\centering
\caption{Journal of spectroscopic observations for \mbox{SN~2016hnk}.}
\label{tab:spectropcopy}
\begin{tabular}{lccccccccc} 
\hline\hline
\textbf{UT date} & \textbf{MJD} & \textbf{Epoch} & \textbf{Exposure}& \textbf{Airmass} &  \textbf{Telescope} &  \textbf{Spectrograph} & \textbf{Grism} & \textbf{Range} & \textbf{Resolution} \\
\hline
2016-10-29 06:51:58.975 & 57690.29 &   +0.80 &  900                         & 1.30  & HILTNER & OSMOS        & VPH Blue         & 3960-6880  & 1600              \\
2016-10-30 06:03:08.180 & 57691.25 &   +1.74 & 1200                         & 1.19  & SOAR    & Goodman      & G400       & 3600-7040  & 1850              \\
2016-10-31 00:20:51.571 & 57692.01 &   +2.49 & 2600                         & 1.26  & SALT    & RSS          & PG0900       & 3500-9400  & 900              \\
2016-11-01 01:52:29.692 & 57693.08 &   +3.54 & 1800                         & 1.30  & NOT     & ALFOSC       & Gr\#4         & 3400-9400  & 700              \\
2016-11-02 03:21:06.248 & 57694.14 &   +4.59 &  600                         & 1.09  & NTT     & EFOSC2       & Gr\#13        & 3640-8930  & 2400              \\
2016-11-02 21:32:08.720 & 57694.90 &   +5.33 & 1500                         & 1.74  & LT      & SPRAT        & Wasatch600    & 4020-6950  & 1300              \\
2016-11-04 01:57:23.700 & 57696.08 &   +6.49 & 12x150                       & 1.245 & GEMINI  & FLAMINGOS-2  & G5801         & 9900-18000 & 900              \\
2016-11-04 15:10:05.940 & 57696.63 &   +7.04 & 2100                         & 1.30  & LJT     & YFOSC        & G3            & 3500-9170  & 2000              \\
2016-11-05 01:59:35.148 & 57697.08 &   +7.48 & 1800                         & 1.35  & NTT     & EFOSC2       & Gr\#11+Gr\#16 & 3340-9650  & 3600              \\
2016-11-06 02:05:03.429 & 57698.09 &   +8.47 & 3240                         & 1.19  & NTT     & SOFI         & Gr\#BG        & 9340-16480 & 1000              \\
2016-11-07 00:05:48.541 & 57699.00 &   +9.37 & 1800                         & 1.30  & NOT     & ALFOSC       & Gr\#4         & 3400-9700  & 700              \\
2016-11-07 02:06:58.316 & 57699.09 &   +9.46 & 1800                         & 1.15  & NTT     & EFOSC2       & Gr\#11+Gr\#16 & 3340-9980  & 3600              \\
2016-11-09 01:34:17.864 & 57701.07 &  +11.41 & 1900                         & 1.38  & LT      & SPRAT        & Wasatch600    & 4020-8000  & 1300              \\
2016-11-09 17:21:16.187 & 57701.72 &  +12.04 & 2100                         & 1.30  & LJT     & YFOSC        & G3            & 3500-9160  & 2000              \\
2016-11-09 23:05:53.854 & 57701.96 &  +12.28 & 2850                         & 1.26  & LT      & SPRAT        & Wasatch600    & 5170-7900  & 1300              \\
2016-11-11 00:59:54.332 & 57703.04 &  +13.34 & 2850                         & 1.33  & LT      & SPRAT        & Wasatch600    & 5220-7920  & 1300              \\
2016-11-18 02:00:24.765 & 57710.08 &  +20.27 & 1800x2                       & 1.09  &NTT      & EFOSC2       & Gr\#11+Gr\#16 & 3340-9980  & 3600              \\
2016-11-28 04:39:23.638 & 57720.19 &  +30.22 & 3600+2x1800                  & 1.74  &NTT      & EFOSC2       & Gr\#13        & 3640-9220  & 2400              \\
2016-12-03 03:46:45.336 & 57725.16 &  +35.11 & 2700x3                       & 1.41  &NTT      & EFOSC2       & Gr\#13        & 3640-9230  & 2400              \\
2016-12-04 07:22:48.000 & 57726.31 &  +36.24 & 1200                         & 1.73  & MMT     & Blue-Channel & G300         & 4440-8490  & 740              \\
2016-12-06 19:49:23.524 & 57728.83 &  +38.72 & 3600                         & 1.60  & NOT     & ALFOSC       & Gr\#4         & 3400-9700  & 700              \\
2016-12-19 04:05:09.167 & 57741.17 &  +50.87 & 2700x2                       & 1.84  & NTT     & EFOSC2       & Gr\#13        & 3640-9220  & 2400              \\
2016-12-28 03:49:27.464 & 57750.16 &  +59.72 & 2700+1900                    & 2.12  & NTT     & EFOSC2       & Gr\#13        & 3630-9220  & 2400              \\
2017-01-06 01:14:34.082 & 57759.05 &  +68.47 & 2700x2                       & 1.30  & NTT     & EFOSC2       & Gr\#13        & 3640-9230  & 2400              \\
2017-10-16 03:21:08.439 & 58042.14 & +347.07 & 3060(UV)+3000(Opt)+3000(NIR) & 1.16  & VLT     & X-Shooter    & $\cdots$         & 3000-25000 & 4100/6500/5600 \\
2017-10-17 03:25:50.270 & 58043.14 & +348.06 & 3060(UV)+3000(Opt)+3000(NIR) & 1.13  & VLT     & X-Shooter    & $\cdots$         & 3000-25000 & 4100/6500/5600 \\
2017-10-17 04:35:12.328 & 58043.19 & +348.11 & 3060(UV)+3000(Opt)+3000(NIR) & 1.05  & VLT     & X-Shooter    & $\cdots$         & 3000-25000 & 4100/6500/5600 \\
2017-11-10 02:35:30.919 & 58067.11 & +371.65 & 3060(UV)+3000(Opt)+3000(NIR) & 1.07  & VLT     & X-Shooter    & $\cdots$         & 3000-25000 & 4100/6500/5600 \\
2017-11-10 03:46:08.985 & 58067.16 & +371.70 & 3060(UV)+3000(Opt)+3000(NIR) & 1.05  & VLT     & X-Shooter    & $\cdots$         & 3000-25000 & 4100/6500/5600 \\
2017-11-15 04:31:21.627 & 58072.19 & +376.65 & 3060(UV)+3000(Opt)+3000(NIR) & 1.12  & VLT     & X-Shooter    & $\cdots$         & 3000-25000 & 4100/6500/5600 \\
\hline\end{tabular}
$ $\\
HILTNER, Hiltner Telescope 2.4m,
SOAR, Southern Astrophysical Research Telescope 4.1m,
SALT, Southern African Large Telescope 11m,
NOT, Nordic Optical telescope 2.6m,
NTT, New Technology Telescope 3.58m,
LT, Liverpool Telescope 2.0m,
GEMINI, Gemini South Telescope 8.1m,
LJT, Lijiang Telescope 2.4m,
MMT, MMT Telescope 6.5m,
VLT, Very Large Telescope 8.2m.                      
\end{sidewaystable*}

\begin{table*}\scriptsize
\caption{Measured velocities (in units of 10$^3$ km s$^{-1}$). Epochs are measured from the $B$-band maximum brightness.}
\label{tab:vel}
\centering        
\begin{tabular}{ccccccccc}
\hline\hline 
Epoch  &\ion{Ca}{ii} H\&K &    \ion{Fe}{ii} &  \ion{S}{ii} W1     &   \ion{S}{ii} W2     & \ion{Si}{ii}~5972\AA & \ion{Si}{ii}~6355\AA &     \ion{O}{i}   &  \ion{Ca}{ii}  NIR     \\
\hline
 +0.80 &             & 12.74(0.51) & 5.89(0.16) & 5.51(0.19) & 7.83(0.16) & 10.23(0.18) &  $-$        &  $-$        \\
 +1.74 & 18.14(0.49) & 12.46(0.14) & 5.61(0.05) & 5.16(0.06) & 7.42(0.20) & 10.02(0.12) &  $-$        &  $-$        \\
 +2.49 & 17.47(0.82) & 12.33(0.11) & 5.26(0.07) & 5.01(0.06) & 7.40(0.10) &  9.59(0.07) & 11.21(0.04) & 15.34(0.07) \\
 +3.54 & 17.46(1.78) & 12.67(0.14) & 4.97(0.13) & 4.66(0.12) & 6.68(0.31) &  9.51(0.20) & 11.07(0.04) & 15.03(0.07) \\
 +4.59 & 17.73(1.84) & 12.73(0.36) & 4.85(0.25) & 4.60(0.37) & 6.70(0.64) &  9.42(0.38) & 11.18(0.09) & 14.93(0.22) \\
 +5.33 &  $-$        & 13.04(0.87) & 4.81(0.85) & 4.55(1.07) & 5.73(0.27) &  9.34(0.79) & 11.06(0.12) &  $-$        \\
 +7.04 & 16.92(1.48) & 13.34(0.31) & 4.62(0.21) & 4.39(0.36) & 5.34(0.36) &  9.02(0.22) & 10.91(0.09) & 14.52(0.15) \\
 +7.48 & 17.28(0.55) & 13.47(0.14) & 4.08(0.14) & 4.12(0.26) & 4.84(0.61) &  8.76(0.29) & 10.90(0.06) & 14.82(0.10) \\
 +9.37 & 16.33(1.55) & 17.19(0.25) & 3.89(0.15) & 3.81(0.35) & 4.88(0.27) &  8.71(0.21) & 10.74(0.05) & 14.72(0.07) \\
 +9.46 & 16.53(1.83) & 17.52(0.17) & 3.64(0.12) & 3.57(0.21) & $-$        &  8.64(0.23) & 10.83(0.04) & 14.76(0.06) \\
+11.41 &  $-$        & 13.73(1.73) & 3.69(1.30) & 3.16(1.72) & $-$        &  8.48(1.01) & 10.53(0.14) &  $-$        \\
+12.04 & 15.72(1.19) & 13.76(0.62) & 3.41(0.35) & 3.22(0.52) & $-$        &  8.34(0.26) & 10.61(0.11) & 14.37(0.14) \\
+12.28 &  $-$        &  $-$        & 3.02(1.06) & 3.12(1.41) & $-$        &  8.04(0.68) & 10.80(0.34) &  $-$        \\
+13.34 &  $-$        &  $-$        & 2.33(1.47) & 2.99(1.22) & $-$        &  8.04(0.38) & 10.49(0.16) &  $-$        \\
+20.27 & 15.51(1.88) & 13.27(0.50) & $-$        & $-$        & $-$        &  7.09(0.23) & 10.10(0.07) & 14.22(0.12) \\
+30.22 & 14.57(0.84) & 10.04(1.28) & $-$        & $-$        & $-$        &  6.52(0.25) & 10.37(0.20) & 14.01(0.17) \\
+35.11 & 14.52(0.36) &  9.43(1.54) & $-$        & $-$        & $-$        &  6.08(0.20) &  9.76(0.19) & 13.55(0.08) \\
+36.24 &  $-$        &  8.61(1.24) & $-$        & $-$        & $-$        &  5.73(0.57) & 10.26(0.21) &  $-$        \\
+38.72 & 15.42(1.80) &  8.72(1.42) & $-$        & $-$        & $-$        &  5.76(0.21) & 10.27(0.07) & 13.66(0.07) \\
+50.87 & 14.50(1.56) &  7.25(1.20) & $-$        & $-$        & $-$        &  5.30(0.34) & 11.59(0.67) & 13.15(0.24) \\
+59.72 & 13.88(1.52) &  5.54(1.77) & $-$        & $-$        & $-$        &  5.05(1.41) & 10.03(0.82) & 12.79(0.19) \\
+68.47 & 12.09(1.25) &  4.77(1.97) & $-$        & $-$        & $-$        &  $-$        &  9.64(2.85) & 12.59(0.42) \\
\hline
\end{tabular}
\end{table*}

 \begin{table*}\scriptsize
\caption{Measured pseudo-equivalent widths (in \AA). Epochs are measured from the $B$-band maximum brightness.}
\label{tab:pew}
\centering        
\begin{tabular}{ccccccccc}
\hline\hline 
Epoch &\ion{Ca}{ii} H\&K &    \ion{Fe}{ii} &  \ion{S}{ii} W1     &   \ion{S}{ii} W2     & \ion{Si}{ii}~5972\AA & \ion{Si}{ii}~6355\AA &     \ion{O}{i}   &  \ion{Ca}{ii}  NIR         \\
\hline
 +0.80 &    $-$        & 263.17(13.86) & 10.34(2.01) & 34.26(3.85) & 66.39(6.45) & 127.92(4.78)  &     $-$      &    $-$        \\
 +1.74 & 120.05(14.80) & 281.95(4.54)  & 12.27(0.78) & 34.37(1.51) & 61.48(3.10) & 132.88(3.16)  &     $-$      &    $-$        \\
 +2.49 & 119.88(12.61) & 272.21(2.86)  & 10.92(0.75) & 25.15(1.15) & 53.96(1.47) & 126.71(1.12)  & 110.08(2.37) & 418.58(5.14)  \\
 +3.54 & 125.70(11.11) & 267.05(4.59)  & 12.38(1.58) & 24.22(2.40) & 50.08(5.67) & 132.54(4.21)  & 117.13(3.26) & 496.48(4.12)  \\
 +4.59 & 121.70(11.54) & 259.03(9.63)  & 10.95(3.64) & 14.29(6.56) & 45.46(6.68) & 136.51(7.87)  & 120.91(7.96) & 515.12(16.60) \\
 +5.33 &    $-$        & 248.69(37.93) & 12.69(6.52) & 13.13(7.30) & 42.64(6.10) & 140.62(8.82)  & 123.50(5.85) &    $-$        \\
 +7.04 & 118.09(19.43) & 264.50(10.92) & 16.57(3.44) &  9.44(4.85) & 36.85(5.67) & 141.86(8.92)  & 132.73(6.14) & 526.95(11.37) \\
 +7.48 & 119.84(17.19) & 269.35(5.41)  & 16.31(2.10) &  8.97(2.98) & 30.96(4.45) & 148.26(5.46)  & 130.95(5.47) & 533.60(7.44)  \\
 +9.37 & 104.04(12.02) & 332.98(16.58) & 14.97(2.89) &  6.31(2.54) & 28.58(6.58) & 156.34(7.80)  & 130.80(4.11) & 533.86(5.69)  \\
 +9.46 & 106.07(13.34) & 326.13(13.35) & 15.86(2.33) &  5.79(1.74) &    $-$      & 152.67(5.51)  & 126.98(2.79) & 523.19(4.20)  \\
+11.41 &    $-$        &    $-$        & 13.18(3.77) &  5.80(1.07) &    $-$      & 170.60(11.55) & 115.76(8.71) &    $-$        \\
+12.04 &  99.36(19.81) & 360.20(25.19) & 16.16(5.83) &  5.97(4.38) &    $-$      & 183.31(11.39) & 121.58(7.75) & 531.55(10.94) \\
+12.28 &    $-$        &    $-$        & 16.04(3.70) &  5.75(1.89) &    $-$      & 175.90(11.35) & 115.14(9.23) &    $-$        \\
+13.34 &    $-$        &    $-$        & 13.14(3.69) &  6.03(1.52) &    $-$      & 194.76(15.53) & 115.96(8.84) &    $-$        \\
+20.27 & 104.64(14.39) & 381.58(13.18) &    $-$      &    $-$      &    $-$      & 207.95(5.89)  & 112.86(4.87) & 563.62(8.96)  \\
+30.22 & 108.01(10.98) & 420.72(26.65) &    $-$      &    $-$      &    $-$      & 181.10(13.57) & 103.47(6.88) & 605.20(12.11) \\
+35.11 & 105.92(11.02) & 403.08(12.29) &    $-$      &    $-$      &    $-$      & 168.78(7.64)  & 101.44(6.73) & 602.40(6.01)  \\
+36.24 &    $-$        & 389.52(23.55) &    $-$      &    $-$      &    $-$      & 166.72(10.50) &  90.80(5.09) &    $-$        \\
+38.72 &  92.52(18.40) & 392.41(14.21) &    $-$      &    $-$      &    $-$      & 168.73(5.28)  &  93.18(8.02) & 607.72(3.96)  \\
+50.87 &  90.27(10.21) & 369.78(19.98) &    $-$      &    $-$      &    $-$      & 143.79(13.15) &  85.50(7.27) & 640.54(17.37) \\
+59.72 & 129.27(11.52) & 361.46(11.38) &    $-$      &    $-$      &    $-$      & 125.17(14.53) &  79.72(6.44) & 657.21(14.31) \\
+68.47 & 122.63(18.45) & 306.01(20.03) &    $-$      &    $-$      &    $-$      &    $-$        &  74.19(7.12) & 650.72(33.03) \\
\hline
\end{tabular}
\end{table*}

\end{appendix}

\end{document}